\documentclass[a4paper, 12pt, openany]{book}
\usepackage{amssymb,amsthm,amsmath,latexsym}
\usepackage{amscd,epsfig}

\let\oldbibliography\thebibliography
\renewcommand{\thebibliography}[1]{%
  \oldbibliography{#1}%
  \setlength{\itemsep}{0pt}%
}

\def\C{{\mathbb{C}}}
\def\R{{\mathbb{R}}}
\def\P{{\mathbb{P}}}
\def\E{{\mathbb{E}}}
\def\M{{\mathbb{M}}}
\def\PT{{\mathbb{PT}}}
\def\PS{{\mathbb{PS}}}
\def\db{\overline{\partial}}
\def\hpi{\hat{\pi}}
\def\bpi{\bar{\pi}}
\def\homega{\hat{\omega}}
\def\ap{{A'}}
\def\bp{{B'}}
\def\cp{{C'}}
\def\ubar{\bar{u}}
\def\vbar{\bar{v}}

\DeclareMathOperator{\tr}{tr}
\DeclareMathOperator{\End}{End}

\newtheorem{theorem}{Theorem}

\begin{document}
\thispagestyle{empty}
\vspace*{1cm}
\begin{center}
{\Huge\bf Aspects of Yang-Mills Theory in Twistor Space}
\vskip 2cm

\epsfig{file=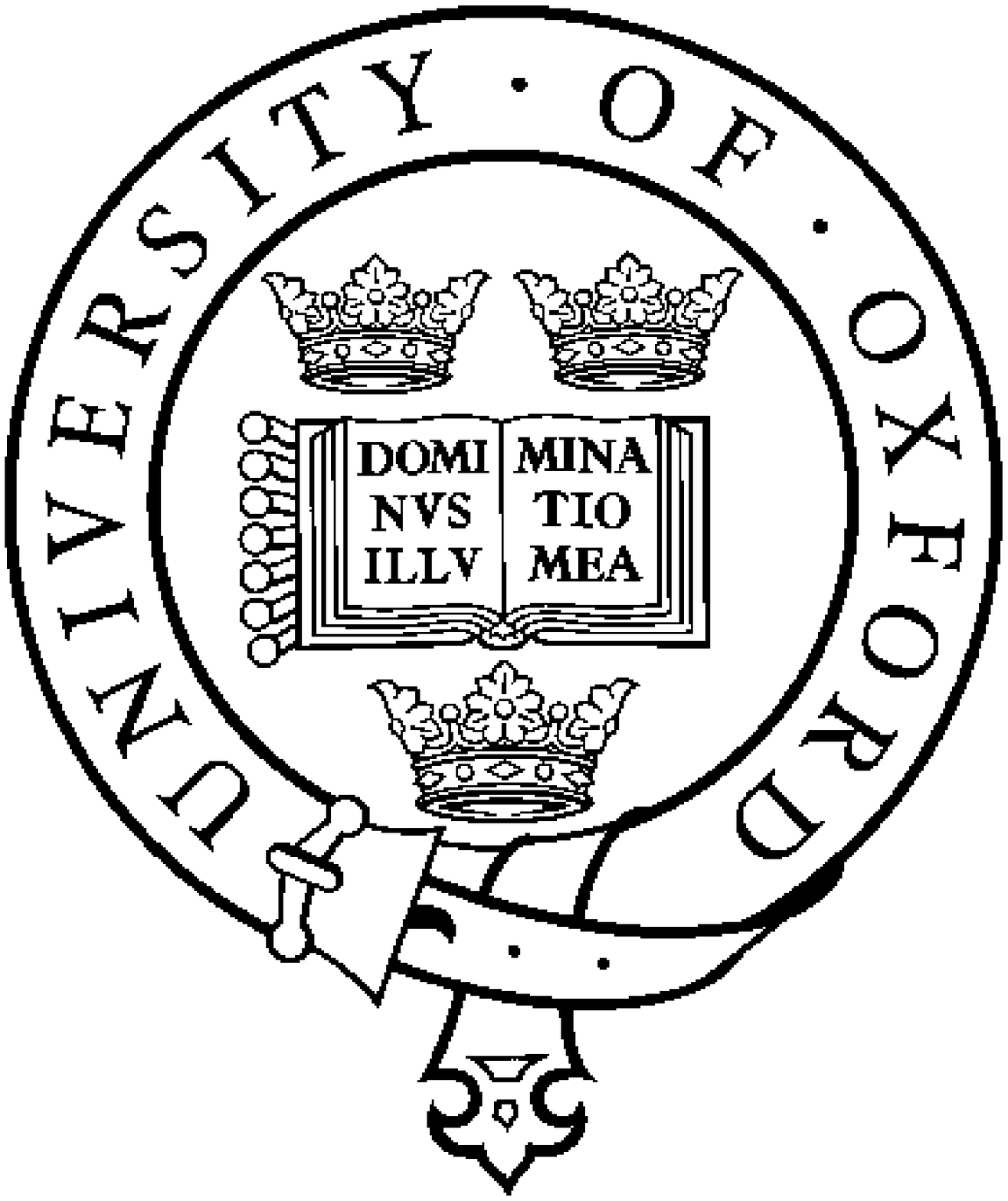, width=3.3cm, height=4cm}

\vskip 2cm

{\Large Wen Jiang}

{\Large Oriel College

University of Oxford}

\vskip 3cm

{\large A thesis submitted for the degree of}

{\large\it Doctor of Philosophy}

{\large 2007}

\end{center}

\newpage

\thispagestyle{empty}
\begin{center}
	{\Large\bf Acknowledgements}
\end{center}

This thesis is only possible because of the support and assistance from many people. I would like to thank Lionel Mason foremost, for generously sharing his ideas with me and providing me with invaluable counsel throughout the gestation of this work. Many thanks are due to my supervisor, Philip Candelas, for his patient guidance and unstinting help during my time at Oxford.

Also indispensable to my research were discussions with my colleagues, especially Rutger Boels, Vincent Bouchard and David Skinner, from whom I learned a great deal about QFT, twistors, string theory and much besides.

The Mathematical Institute and Oriel College created a stimulating atmosphere for me to work and live in and importantly, supplied many teaching opportunities that greatly enriched my D.Phil.\ experience.

I cannot thank my parents and my wife enough for their love and constant support, without which this thesis could not have been written.

Finally, I acknowledge generous funding from the EPSRC that freed me from any financial worries as a student.

\newpage

\thispagestyle{empty}
\begin{center}
	{\Large\bf Aspects of Yang-Mills Theory in Twistor Space}
	
	\vspace{0.5cm}
	
	{\large Wen Jiang}
	
	{\large Oxford University D.Phil.\ Thesis}
	
	\vspace{0.8cm}
	
	{\Large\bf Abstract}
	
	\vspace{0.5cm}
\end{center}

This thesis carries out a detailed investigation of the action for pure Yang-Mills theory which L.\ Mason formulated in twistor space. The rich structure of twistor space results in greater gauge freedom compared to the theory in ordinary space-time. One particular gauge choice, the CSW gauge, allows simplifications to be made at both the classical and quantum level.

The equations of motion have an interesting form in the CSW gauge, which suggests a possible solution procedure. This is explored in three special cases. Explicit solutions are found in each case and connections with earlier work are examined. The equations are then reformulated in Minkowski space, in order to deal with an initial-value, rather than boundary-value, problem. An interesting form of the Yang-Mills equation is obtained, for which we propose an iteration procedure.

The quantum theory is also simplified by adopting the CSW gauge. The Feynman rules are derived and are shown to reproduce the MHV diagram formalism straightforwardly, once LSZ reduction is taken into account. The three-point amplitude missing in the MHV formalism can be recovered in our theory. Finally, relations to Mansfield's canonical transformation approach are elucidated.

\newpage
\thispagestyle{empty}
\mbox{}

\newpage

\tableofcontents


\chapter{Introduction}\label{chap:intro}

Yang-Mills, or non-abelian gauge, theory is now the accepted framework for modelling  fundamental particles. (An account of its long and interesting history may be found in \cite{dawn}). The idea in a physical context first originated in Hermann Weyl's attempt to unify gravity and electromagnetism. A precise formulation was discovered by Yang and Mills \cite{yangmills} in 1954, but only in the context of nuclear isospin. Its pertinence to the fundamental weak and strong interactions was gradually realised through the 1960's and 70's, culminating in the standard model of particle physics, described in terms of a gauge theory with symmetry group $SU(3)\times SU(2)\times U(1)$. This model has passed numerous experimental tests. Indeed so successful has it been that a discovery of any phenomenon beyond the standard model would be regarded as a major breakthrough.

The motivation for this thesis comes from the $SU(3)$ sector describing the strong force, or QCD, although the precise choice of the gauge group is mostly unimportant in our model. QCD is a rich and elegant theory whose structure is still not completely understood. An instance of this is the simplicity of tree-level gluon scattering amplitudes, which had been a mystery ever since Parke and Taylor \cite{Parke86} first conjectured the general form of the so-called MHV amplitudes in 1986. Traditional Feynman diagram techniques for computing these amplitudes involve a huge number of diagrams and yet, the final answers turn out to be remarkably compact.

The tree-level scattering amplitude of gluons all of the same helicity vanishes. So does that of $n-1$ gluons of one helicity and one gluon of the opposite helicity. The simplest non-vanishing amplitudes involve two gluons of one helicity, with the rest of the opposite helicity. These are called maximally helicity-violating (MHV) amplitudes. In spinor notation (explained later and also in \cite{Witten03}, \cite{Dixon96}), when the $i^\text{th}$ and $j^\text{th}$ of $n$ gluons have positive helicity and the rest negative helicity\footnote{It is unfortunate that there is no uniform convention for assigning helicity in the literature. We will stick to the rule that a field with primed indices, e.g.\ the self-dual part of curvature, has positive helicity.}, the tree-level scattering amplitude is essentially given by
\begin{equation}
A_n= \frac{(\lambda_i\cdot \lambda_j)^4}{(\lambda_1\cdot \lambda_2) (\lambda_2 \cdot \lambda_3) \cdots (\lambda_n\cdot \lambda_1)},\label{eq:MHV}
\end{equation}
where the group theory factors are omitted. The number of Feynman diagrams, however, increases factorially with $n$ and runs into hundreds even for $n=6$ and more than 10 million for $n=10$. It is therefore surprising that there is one simple formula applicable for any $n$.

Even though the formula \eqref{eq:MHV} was soon proved using a recursion relation by Berends and Giele \cite{Berends87}, it did not dispel the suspicion that deeper reasons are responsible for this phenomenon. Nair \cite{Nair88} first suggested in 1988 an alternative formalism in which the simplicity of the MHV amplitudes was a direct consequence of the underlying model. Even though his theory was not fully-fledged and though only the MHV amplitudes were explicitly derived, several important ingredients in subsequent developments already emerged in this paper.

First, a null momentum vector $p_\mu$ in Minkowski space can be written as a matrix $p_{AA'}=  \sigma^\mu_{AA'} p_\mu$, where $\sigma_\mu$'s are the Pauli matrices. Then $\det (p_{AA'})= p_\mu p^\mu=0$ and so $p_{AA'}$ is decomposable as $p_{AA'}= \bar{\lambda}_A \lambda_\ap$, for some two-component complex spinor $\lambda_\ap$. Thus, the direction of a null vector is determined by $\lambda_\ap$ modulo scale, that is, a point on $\C\P^1 \cong S^2$, the celestial sphere. These $\lambda$'s are precisely those appearing in the amplitude \eqref{eq:MHV} and may therefore be suspected to play a fundamental role, possibly as part of the space on which a model is defined. If one recalls that Euclidean twistor space is precisely Euclidean space-time\footnote{We pass over the question of signature for now. This thorny issue is discussed further in Chapter \ref{chap:quant}.} plus the set of null directions $\lambda$ at each point, then this provides one motivation for re-formulating Yang-Mills theory in twistor space.

The second ingredient has to do with an important result in twistor theory, namely the Penrose transform, which, roughly speaking, relates functions of homogeneity $-2k$ on twistor space to space-time fields of helicity $k-1$. The details will be reviewed in the introduction. For now, observe that functions of different homogeneities are naturally bundled together by use of supersymmetry. If four fermionic parameters $\eta^i$, $i=1,\ldots,4$, are added to the ordinary twistor space $\C\P^3$, the result is super-twistor space $\C\P^{3|4}$,
\[
\C\P^{3|4}= \C^{4|4} / \sim, \qquad \text{where } (Z^\alpha, \eta^i) \sim (c Z^\alpha, c \eta^i), \text{ for any } c\in \C^*.
\]
A function $\Phi$ on $\C\P^{3|4}$ is, more or less by definition, expressible as
\begin{equation}
\Phi = A + \eta^i \chi_i + \frac{1}{2} \eta^i \eta^j \phi_{ij} + \frac{1}{3!} \epsilon_{ijkl} \eta^i \eta^j \eta^k \tilde{\chi}^l + \eta_1 \eta_2 \eta_3 \eta_4 B,\label{eq:fermionexpansion}
\end{equation}
where the fields $(A,\chi,\phi,\tilde{\chi},B)$ have homogeneities $(0,-1,-2,-3,-4)$ and hence correspond to space-time fields of helicities $(-1, -\frac{1}{2}, 0, +\frac{1}{2}, 1)$. These are precisely the field content of the ${\cal N}=4$ super-Yang-Mills multiplet. Since additional supersymmetry does not affect tree-level scattering amplitudes, Nair's, and later Witten's, investigation into the MHV amplitudes was focused on the ${\cal N}=4$ extension of Yang-Mills theory. Supersymmetry will, however, not play a role in this thesis but we take this opportunity to point out how $(\lambda_i \cdot \lambda_j)^4$ in the numerator of Equation \eqref{eq:MHV} arises naturally in supersymmetric theories.

Let the part of $\Phi$ depending on four $\eta$'s be denoted by $\Phi^{(4)}$ and introduce the `incidence relation' $\eta^i= \theta^{iA'} \pi_\ap$. Then by virtue of properties of fermionic integration,
\[
\int d^8 \theta \Phi^{(4)}(\pi_i) \Phi^{(4)}(\pi_j) = (\pi_i\cdot\pi_j)^4 B(\pi_i) B(\pi_j).
\]
Thus the factor $(\pi_i\cdot\pi_j)^4$ is associated with two fields of helicity 1, as it should be, given the result \eqref{eq:MHV}.

Lastly, it remains to find a way to interpret the $1/(\lambda_k\cdot \lambda_{k+1})$ factors in the denominator in Equation \eqref{eq:MHV}. Given the homogeneous coordinate $\pi_\ap$ on $\P^1$, one can pick an inhomogeneous coordinate $z$ by taking $\pi_2=1$ and $z=\pi_1$. Then
\[
\frac{1}{\pi\cdot\pi'}= \frac{1}{\pi_1\pi_2'-\pi_2\pi_1'}= \frac{1}{z-z'}
\]
in inhomogeneous coordinates, which may be recognised as the correlator of two free fermions on the Riemann sphere. To be more precise, the theory of free fermions on the Riemann sphere has action
\begin{equation}
\int \beta\db \alpha \ d^2 z.\label{eq:fermionaction}
\end{equation}
The correlator of $\alpha$ and $\beta$ is Green's function of the $\db$-operator. Therefore, as explained in the appendix,
\begin{equation}
\langle \alpha(z) \beta(z') \rangle = \frac{1}{z-z'}.\label{eq:fermioncorr}
\end{equation}

In a combination of these ideas, Nair considered a Wess-Zumino-Witten model on $\C\P^1$ defined by a current algebra, which encodes \eqref{eq:fermioncorr}. The MHV amplitude can then be generated by calculating a correlation function. It is clear, however, that this model is far from complete. Non-MHV amplitudes are not easily accommodated and there is no direct link to Yang-Mills theory. It was some 15 years before Witten conjectured a duality between a topological string theory and perturbative gauge theory \cite{Witten03}, which would fully explain the simplicity of QCD scattering amplitudes.

Witten had the insight of recognising the ingredients described above as features of a string theory, in particular, the topological B-model with target space $\C\P^{3|4}$. The B-model (see \cite{Witten:1991zz}) comes from taking an ${\cal N}=2$ supersymmetric field theory in two dimensions and `twisting' it by changing the spins of various operators using certain R-symmetries. The condition that the R-symmetries be anomaly-free means that the target space must be a Calabi-Yau manifold for the B-model to be well defined. The super-twistor space $\C\P^{3|4}$ is indeed Calabi-Yau, since its canonical bundle is trivialised by the non-vanishing section
\[
\Omega= \frac{1}{4!} \epsilon_{\alpha\beta\gamma\delta} Z^\alpha dZ^\beta dZ^\gamma dZ^\delta d\eta^1 d\eta^2 d\eta^3 d\eta^4.
\]
This expression in homogeneous coordinates is well defined on $\C\P^{3|4}$ because the usual rule that $\int d\eta \> \eta=1$ forces $d\eta^i\mapsto c^{-1} d\eta^i$ under a scaling $(Z^\alpha, \eta^i) \mapsto (cZ^\alpha, c\eta^i)$. The triviality of the canonical bundle implies that the first Chern class of the tangent bundle is zero and, hence, the manifold is Calabi-Yau.

The basic field in this theory is a $(0,1)$-form $\cal A$, which may be expanded in precisely the same form as Equation \eqref{eq:fermionexpansion}, thus reproducing the fields in ${\cal N}=4$ Yang-Mills theory. The fact that both the Calabi-Yau condition and the correspondence with the ${\cal N}=4$ multiplet lead to the super-twistor space $\C\P^{3|4}$ make Witten's theory especially compelling.

The mechanism that provides something akin to the correlator \eqref{eq:fermioncorr} is the coupling of open strings to Euclidean D-instantons. Without going into details (please refer to Witten's paper \cite{Witten03} or Cachazo and Svrcek's lectures \cite{Cachazo:2005ga}), Equation \eqref{eq:fermionaction} becomes the effective action of the D1-D5 and D5-D1 strings
\[
\int_C \beta (\db+ {\cal A}) \alpha.
\]
After choosing appropriate scattering wave-functions, the derivation of the MHV formula from a correlator proceeds along similar lines to Nair's calculations. 

Witten's theory goes much further in that he conjectures a correspondence between the perturbative expansion of ${\cal N}=4$, $U(N)$ gauge theory with the D-instanton expansion of the B-model in $\C\P^{3|4}$. Evidence for this is provided by observations that the gluon scattering amplitudes seem to be supported on holomorphic curves in twistor space. For example, the formula \eqref{eq:MHV} for the MHV amplitude depends solely on $\lambda_i$'s, but not $\bar{\lambda}_i$'s (where $p_i= \lambda_i \bar{\lambda}_i$), except through the momentum conservation delta function. Let us express the amplitude as
\[
\delta^{(4)}(\sum p_i) f(\lambda_j)= \int d^4 x \exp \left( ix^{AA'} \sum \lambda_{iA'} \bar{\lambda}_{iA} \right) f(\lambda_j),
\]
by re-writing the delta function. If one naively treats $\lambda$ and $\bar{\lambda}$ as independent (justified in split space-time signature) and Fourier transform with respect to $\bar{\lambda}_i$'s, one finds the amplitude to be 
\begin{align*}
\int \prod_k d^2 \bar{\lambda}_k \exp \left( i \bar{\lambda}_{kA} \mu_k^A \right) \int d^4 x \prod_i & \exp \left( ix^{AA'} \lambda_{iA'} \bar{\lambda}_{iA} \right) f(\lambda_j)\\
&= \int d^4 x \prod_k \delta^{(2)}(\mu_{kA} + x_{AA'}\lambda_k^\ap) f(\lambda_j),
\end{align*}
in $(\mu_A, \lambda_\ap)$ space. The crucial point is that the same $x$ appears in the delta function for any $k$ (index for the external particles) and so the integrand vanishes unless all pairs $(\mu_k,\lambda_k)$ lie on the subset defined by
\begin{equation}
\mu_A + x_{AA'} \lambda^\ap=0.\label{eq:mhvvertexeqn}
\end{equation}
This, being a linear equation, defines a straight line in $(\mu_A,\lambda_\ap)$ space (viewed as twistor space) for each $x$. Hence, the MHV amplitudes are said to localize on (genus 0) degree 1 curves, i.e.\ straight lines, in twistor space. The amplitude is thus expressed as an integral over the moduli space of lines in $\C\P^3$. It may seem natural to interpret these holomorphic curves as the world-sheets of strings and the integral as over all possible positions of these strings. Witten was prompted to consider the B-model with D-instantons, even though alternative models emerged subsequently, e.g. Berkovits's open twistor string \cite{Berkovits:2004hg}.

This view-point immediately generalises to other amplitudes, where $q$ gluons have different helicities to the rest. From scaling considerations, the string theory suggests that the $l$-loop amplitude of such a configuration should localize on holomorphic curves of degree $d=q-1+l$ and genus $g\leq l$. Since algebraic conditions after Fourier transform correspond to differential constraints before the transform, it follows that the scattering amplitudes satisfy certain differential equations, which are easier to check than doing the Fourier transform directly. This was verified for certain tree-level and one-loop amplitudes in Witten's paper and in subsequent work by him and others \cite{Cachazo:2004zb, Cachazo:2004by}.

Assuming this string-gauge theory correspondence, one could compute Yang-Mills amplitudes by performing the integral over the moduli space of holomorphic curves of appropriate degree and genus. Roiban \emph{et.\ al.}\ \cite{Roiban:2004yf,Roiban:2004vt} did this for some tree-level amplitudes with non-MHV configurations and found agreement with known results. The fact that only genus 0 curves, i.e.\ rational maps of $\P^1$ into $\P^3$, are involved makes the moduli space integral tractable.  However, a suitable parametrization and measure are difficult to implement on moduli spaces of non-zero genus curves, rendering the string theory ineffective for practical calculations of amplitudes.

A more serious problem with Witten's conjecture is that the closed string sector seems to give a version of ${\cal N}=4$ conformal supergravity \cite{Berkovits:2004jj}, which is generally regarded as unphysical. The supergraviton interactions cannot be easily decoupled from Yang-Mills fields when the number of loops is non-zero, thus suggesting that Witten's model will not exactly reproduce super-Yang-Mills loop amplitudes, despite its success at tree-level.

Even though the conjectured duality between twistor string theory and Yang-Mills only seems to be valid in the planar limit, it has inspired much progress in calculations of QCD scattering amplitudes. One prominent example is the use of MHV diagrams. As pointed out above, MHV amplitudes localize on subsets of twistor space defined by Equation \eqref{eq:mhvvertexeqn}.  These lines, or copies of $\P^1$, correspond to the points $x$ appearing the equation, according to the standard relation between twistor space and space-time (more details in Chapter \ref{chap:action}). While non-MHV tree amplitudes entail localization on curves of degree $d>1$, calculations suggested that the `connected' picture, in which the instanton is a connected curve of degree $d$, is equivalent\footnote{Gukov \emph{et. al.}\ \cite{Gukov:2004ei} provided an argument for this equivalence by reasoning that the moduli space integrals in both pictures localize on the same subset -- the subset of intersecting lines.} to the `totally disconnected' picture, in which the curve is made up of $d$ straight lines, labelled by different $x$'s. Open strings then stretch between these disjoint degree 1 instantons. If suitably inspired, one might be led to regard MHV \emph{amplitudes}, represented by $d=1$ instantons, as local interaction \emph{vertices}. Cachazo, Svrcek and Witten \cite{CSW04} proposed a diagrammatic scheme, akin to the usual Feynman diagrams, in which one links up these `MHV vertices' with $1/p^2$ propagators in accordance with the following rules.

\begin{itemize}
	\item Only legs representing gluons of opposite helicities may be joined.
	\item An internal line carries a momentum $P$ that is determined by momentum conservation and generally not null. Define a spinor $\lambda^P_\ap= P_{AA'} \eta^A$, with an arbitrarily chosen spinor $\eta^A$.
	\item For every diagram, write down the Parke-Taylor formula \eqref{eq:MHV} for each vertex, using the spinor $\lambda^P_\ap$ defined above for any internal leg. Multiply these together with a $1/P^2$ factor for each internal line.
	\item Sum over all possible tree diagrams.
\end{itemize}

\begin{figure}[htb]
\vspace{0.3cm}
\begin{center}
\epsfig{file=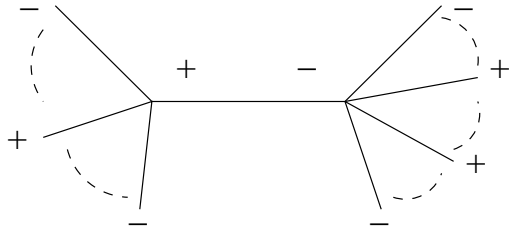}
\label{fig:mhvdiagram}
\caption{\emph{An example of an MHV diagram, in which two MHV vertices are joined by one propagator.}}
\end{center}
\vspace{0cm}
\end{figure}

This scheme, known as the CSW rules, was verified for a number of configurations and used to work out other tree-level amplitudes. A proof was provided by Risager in \cite{Risager:2005vk}. Using properties of massless field theories, he showed that the tree-level CSW rules followed from the so-called BCF recursion relations, which also came out of work stimulated by Witten's twistor string theory. This, however, is only the story at tree-level. Brandhuber \emph{et.\ al.}\ \cite{Brandhuber:2004yw} extended the CSW prescription to include loops, so that one joins up MHV vertices with propagators in the same way as before, but allows loops to appear. The loop momenta are integrated over, just as in ordinary quantum field theory. The results for certain one-loop amplitudes were found to agree with known expressions obtained using traditional techniques.

All evidence then points to MHV diagrams being valid for loop amplitudes as well as trees. It is natural to speculate whether an alternative formulation of Yang-Mills theory exists, which has the CSW rules as the usual Feynman rules. One of the purposes of this thesis is to address this question. Before describing the approach taken in this thesis, we will mention for comparison the work of Mansfield, Morris, \emph{et.\ al.}\ (\cite{Mansfield:2005yd}, \cite{Ettle:2006bw} and \cite{Ettle:2007qc}), which also exploits an alternative version of the Yang-Mills Lagrangian. The idea is to put the ordinary Lagrangian ${\cal L}[A]$ in light-cone gauge, which eliminates two unphysical degrees of freedom, leaving the two helicity states represented by $A_+$ and $A_-$. Now ${\cal L}\propto \|F\|^2=\|dA+A\wedge A\|^2$ and schematically,
\[
{\cal L}={\cal L}_{+-}+{\cal L}_{++-}+{\cal L}_{--+}+{\cal L}_{++--},
\]
where the subscripts indicate the number and type of $A_\pm$'s in each term. In the MHV prescription\footnote{Our convention of using mostly negative helicity configurations in MHV vertices is different from that of Mansfield and others.}, a vertex has two positive helicity gluons with the rest negative. The third term does not have this property and one could try to remove it by finding new fields $B_\pm$, such that this term is absorbed into the quadratic, non-interacting term:
\[
\tilde{\cal L}_{+-} [B] = {\cal L}_{+-} [A] + {\cal L}_{--+} [A].
\]
This is only plausible because the two terms on the right hand side would only give rise to tree diagrams with all but one external particle of negative helicity. These produce zero amplitudes (on shell) and so, classically, seem to represent a free theory. Imposing the restriction that the transformation be canonical allows one to express $A$ as an infinite series in $B$ \cite{Ettle:2006bw} and the Yang-Mills Lagrangian is found to be of the form
\[
{\cal L}=\tilde{\cal L}_{+-} [B] + \tilde{\cal L}_{++-} [B] + \tilde{\cal L}_{++--} [B] + \tilde{\cal L}_{++---} [B] + \cdots,
\]
that is, an infinite sum of terms involving MHV combinations of $B_\pm$. The fields $B_\pm$ can be used to represent gluon states in amplitude calculations, by the Equivalence Theorem (modulo subtleties \cite{Ettle:2007qc}). The CSW rules may then be seen to follow from this new Lagrangian.

\vspace{0.5cm}

Having summarised the recent history of twistor string theory and developments stemming from it, we hope to have set the stage for our approach to understanding perturbative Yang-Mills theory.

With hindsight, twistor space seems to be a natural setting for reformulating Yang-Mills theory. As noted above, the ordinary Yang-Mills Lagrangian contains a cubic term that, classically, does not contribute to interactions and amplitudes involving gluons all of the same helicity vanish. This means that the self-dual or anti-self-dual sector of the theory is somehow free at the classical level. This is in fact suggested by a result in twistor theory dating back to the 70's. The Penrose-Ward transform \cite{Ward:1977ta} relates anti-self-dual (ASD) gauge bundles on space-time to essentially `free' holomorphic bundles on twistor space. To be more precise, given a gauge bundle with connection $A$ on space-time, the condition that the curvature $F_A$ of $A$ be ASD is a first order, non-linear PDE in $A$ --- a special case of the Yang-Mills equation. An ASD curvature form $F_A$ vanishes on self-dual null two-planes in complexified space-time and the bundle is consequently integrable on these planes. By translating the bundle to the space of these planes, i.e.\ twistor space, Ward showed that the result was a holomorphic bundle with only topological constraints (being trivial on lines in $\C\P^3$).

One insight of Witten's is that this relation could be exhibited at the level of actions. For $B$ a self-dual two-form in space time, consider an action
\begin{equation}
\int B\wedge F_A.\label{eq:chalmersintro}
\end{equation}
One of its equations of motion is $(\text{self-dual part of }F_A)=0$, implying that the bundle is ASD. On the twistor side, let $a$ and $b$ be (0,1)-forms and $(\db+a)$ a partial connection defined through $a$. The action
\begin{equation}
\int b\wedge f_a
	\label{eq:holocsintro}
\end{equation}
leads to the equation of motion $f_a\equiv \db a + a\wedge a=0$, implying that the underlying bundle is holomorphic. Thus, via the Penrose-Ward correspondence, there appears to be a relation between the ASD part of the Chalmers-Siegel action in space-time \eqref{eq:chalmersintro} and a kind of holomorphic Chern-Simons theory in twistor space \eqref{eq:holocsintro}.

The use of variational principles in twistor theory is in fact rather new. One reason is that (massless) fields on space-time have often been translated into \v Cech cohomology classes on twistor space in the usual form of the Penrose transform; and \v Cech cohomology classes, with their definition in open patches, etc., do not seem to lend themselves to calculus of variations. However, the isomorphism between \v Cech and Dolbeault cohomologies means that differential forms could be used instead. These, of course, fit much more easily into a variational framework. The focus of this thesis is on the Dolbeault point of view.

To go beyond the ASD sector and obtain interactions of the full Yang-Mills theory, one natural idea is to extend the relationship between the actions \eqref{eq:chalmersintro} and \eqref{eq:holocsintro}. It is known that the complete Chalmers-Siegel action
\[
\int B\wedge F_A + \frac{\epsilon}{2} \int B\wedge B
\]
describes full Yang-Mills theory perturbatively, so one could try to translate the second term into twistor language and find a twistor Yang-Mills Lagrangian of the form,
\[
\int b\wedge f_a + \frac{\epsilon}{2} I[a,b].
\]
This was successfully carried out by Mason \cite{Mason05}. This thesis examines consequences of Mason's Lagrangian, at the classical and the quantum level.

After briefly reviewing the construction of the twistor Yang-Mills theory in Chapter \ref{chap:action}, we investigate the equations of motion that arise from the action in Chapter \ref{chap:class}. The basic fields are $a$ and $b$, which are (0,1)-forms on a three-dimensional complex manifold. A gauge of axial type will be used in which one of the three components of each field is set to zero. In other words, we pick a basis of (0,1)-forms $\{e^0,e^1,e^2\}$, express $a$ as $a=\sum a_i e^i$ (similarly for $b$) and demand that
\begin{equation}
	a_2=b_2=0.
	\label{eq:cswgaugeintro}
\end{equation}
Since only two components remain of each form, the $b\wedge a \wedge a$ term in the action \eqref{eq:holocsintro} disappears. The effect is similar to removing the ${\cal L}_{--+}$ term in the Yang-Mills Lagrangian $\cal L$ as mentioned above, although in that case, it is achieved by gauging followed by a transformation of fields. Because of motivation from the MHV diagram formalism, the condition in Equation \eqref{eq:cswgaugeintro} will be called the CSW gauge.

The equations of motion take the following special form in this gauge,
\begin{align}
\db_2 a_0 &=0, \qquad \db_2 a_1= J_1[a_0, b_0],\notag\\
\db_2 b_0 &=0, \qquad \db_2 b_1= J_2[a_0, b_0],\label{eq:eomintro}
\end{align}
where $J_1$ and $J_2$ are functionals of $a_0$ and $b_0$ only. There are also two constraint equations, not involving $\db_2$, which are compatible with Equations \eqref{eq:eomintro}. This prompts one to consider Equations \eqref{eq:eomintro} as evolution equations with constraints. If one treats $\db_2$ as a real operator, $\partial_t$ for some `time' parameter $t$, one can imagine specifying initial values for the fields on some hypersurface, which satisfy the constaints, and then evolving the fields under $\db_2$. The result would be a solution of the full system of equations. Since it can (and will) be shown that solutions of Equations \eqref{eq:eomintro} are in one-to-one correspondence with solutions of the full Yang-Mills equation, the foregoing procedure would lead to a solution of the Yang-Mills equation. To investigate its feasibility, we first look at three special cases.

When the gauge group is abelian, the equations can be solved by virtue of the $\db$-Poincar\'e lemma. When we look for solutions corresponding to ASD fields in space-time, we deduce that the equations reduce to the Sparling equation, which has been obtained before by Sparling, Newman \emph{et.\ al.}\ \cite{Newman80}, albeit by different arguments. As an example, the case of an ASD 1-instanton is solved, providing a smooth representation in CSW gauge, which is extremely simple in form and appears not to have been known. However, when self-dual solutions are considered, the situation becomes much less clear. A representation of an SD 1-instanton in CSW gauge is found, but the computations involved are very lengthy, even though the mathematics is straightforward. In addition, an iteration procedure is proposed, which approximates a solution by taking terms of increasing order in $t$.

The complex nature of $\db_2$ means that the proposal above cannot be easily achieved. Setting up the theory in Minkowski space enables us to make $\db_2$ a real operator, but complications spoil the form of Equations \eqref{eq:eomintro}. Even though a solution procedure remains out of reach (perhaps unsurprising, given the complexity of the Yang-Mills equation), we obtain an interesting reformulation of Yang-Mills equation, which echoes earlier work by Kozameh, Mason and Newman \cite{Kozameh:1992ss}.

Chapter \ref{chap:quant} is concerned with the quantum theory of the twistor Lagrangian, returning to the original motivation for our work. In addition to Mason's first paper \cite{Mason05}, articles by Boels, Mason and Skinner \cite{Boels:2007qn} \cite{Boels:2006ir} \cite{Boels:2007gv} have also examined various quantum aspects of Yang-Mills theory in twistor space. This thesis contains a detailed investigation, addressing some questions that have not been dealt with. In particular, we propose a diagrammatic representation of the twistor Feynman rules, thus leading to an explicit derivation of CSW formalism. The relationship between twistor space correlators and space-time scattering amplitudes is elucidated. In the process, the `missing' $--+$ amplitude is recovered.

Our approach has the advantage of being natural in some sense. Once the relationship between the space-time ASD Chalmers-Siegel action \eqref{eq:chalmersintro} and the holomorphic Chern-Simons action in twistor space \eqref{eq:holocsintro} is recognised, translating the full Chalmers-Siegel action to twistor space is the obvious next step. When the action is expressed as a series in the basic fields $a$ and $b$, the single interaction term $I[a_0,b_0]$ is seen to contain the whole tower of MHV vertices\footnote{Here $a_0$ and $b_0$ denote components of the (0,1)-forms $a$ and $b$ up the $\P^1$ fibres.}
\[
a_0\cdots a_0 b_0 a_0\cdots a_0 b_0 a_0 \cdots a_0.
\]
The fields $a_0$ and $b_0$ are not \emph{a priori} the minus and plus helicity states. However, there is a simple formula relating the potential $A_{AA'}$ in space-time to $a$ in twistor space. If we make a convenient choice of polarization vectors and apply the LSZ reduction formalism, we find that $a_0$ and $b_0$ suffice as representatives of the helicity states. In the case of only three external particles, relations between the external momenta result in a subtlety when applying LSZ reduction. A tree diagram, which is killed by LSZ reduction at first sight, is actually non-zero and gives the $--+$ amplitude. This bears some resemblance to Ettle \emph{et.\ al.}'s proposal \cite{Ettle:2007qc} that `Equivalence Theorem evasion' is the source of the missing amplitude. However, as we explain in Chapter \ref{chap:quant}, the precise mechanisms seem to be different.

In Section \ref{sec:mhvlagrangian}, the canonical transformation approach is compared to the twistor formulation in detail. As already mentioned, only two components of the Yang-Mills fields, $\cal A$ and $\bar{\cal A}$, remain in light-cone quantization. These are canonically transformed into new fields $\cal B$ and $\bar{\cal B}$, in terms of which the interaction vertices takes an explicit MHV form. The transformation was defined implicitly by Mansfield in \cite{Mansfield:2005yd} and explicit formulae for $\cal A$ in terms of $\cal B$ were later derived by Ettle and Morris in \cite{Ettle:2006bw}. When we interpret the light-cone fields in the twistor framework, we find that $\cal B$ and $\bar{\cal B}$ essentially correspond to the fields $a_0$ and $b_0$ in our formalism. The expression of $\cal A$ as a series in $\cal B$ is simply the expansion of (a component of) $A_{AA'}$ in $a_0$, which can be derived from general properties of the objects involved. There is no need to solve any recursion relations as in \cite{Ettle:2006bw}. One difference between the two Lagrangian-based theories is the form of the external states that is used in amplitude calculations. In the canonical transformation approach, the new fields $\cal B$ and $\bar{\cal B}$ are used as the $\pm$ helicity states, facilitated by the `Equivalence Theorem'. However, to recover the `missing' three-point tree amplitude (and also the $(++++)$ loop amplitude), it is necessary to supplement the MHV vertices in the Lagrangian with the so-called `MHV completion vertices', which come from higher-order terms in the expansion of $\cal A$ in $\cal B$. On the twistor side, the same external states as for ordinary tree-level calculations can be used to find the `missing' amplitude. The subtlety lies in kinematic relations between the momentum spinors. Given the essential equivalence between the twistor and light-cone formalisms, the loop amplitudes should be retrievable from the twistor theory. Preliminary calculations show that the box one-loop diagram with four external particles of $-$ helicity\footnote{Owing to different conventions of assigning helicities, our all-minus loop is the all-plus loop in \cite{Ettle:2007qc}.} survives LSZ reduction. If we dimensionally regularise in the sense of treating twistor space as $\R^{4-2\epsilon} \times \P^1$, then the resulting momentum space integral has been shown to reproduce the know formula for the amplitude in \cite{Ettle:2007qc}. However, still open is the intriguing question whether there is a regularisation scheme appropriate to twistor space, which obviates the need to appeal to space-time techniques such as unitarity cut and dimensional regularisation.

\vspace{0.5cm}

We would like mention that the proposal in this thesis is not the only possible twistor-based reformulation of Yang-Mills. The asymmetry between right- and left-handed fields (essentially between $a_0$ and $b_0$) is apparent, when the classical equations are examined for self-dual and anti-self-dual fields, and when a choice is made in taking MHV vertices to involve mostly negative-helicity external particles. This disadvantage may be addressed by constructing a theory in ambitwistor space, as Mason and Skinner have done in \cite{MasonSkinner}. Despite the complexity of ambitwistor space compared to twistor space, further investigations might reveal an action-based explanation for the Britto-Cazhazo-Feng-Witten recursion relations \cite{BCFW}, which build up tree amplitudes by combining $+--$ and $-++$ amplitudes.  Encouraging evidence in this direction is provided by Hodges, who has shown that gluon tree amplitudes are naturally expressed as twistor diagrams in an ambidextrous formalism \cite{Hodges:2005bf} \cite{Hodges:2005aj}. The methods in this work, applied to the Lagrangian in \cite{MasonSkinner}, might elucidate the origin of Hodges's twistor diagrams.

\chapter{Twistor Yang-Mills Action}\label{chap:action}

In this chapter we briefly review an alternative action for Yang-Mills theory, which Lionel Mason introduced on Euclidean twistor space \cite{Mason05}. Sections \ref{sec:twistor} and \ref{sec:ward} will establish notation and explain necessary results from twistor theory  \cite{Atiyah,Woodhouse}. The motivation for our Lagrangian is outlined in Section \ref{sec:chalmers}. The enlarged gauge freedom is discussed in Section \ref{sec:gauge} and two particular gauge choices are introduced. One, the CSW gauge, allows many simplifications in the classical and quantum theories. The other, the harmonic gauge, can be used to provide a direct link between the twistor and space-time formulations of Yang-Mills theory.

\section{Twistor space of $\R^4$}\label{sec:twistor}
\subsection{Spinors and twistors}\label{subsec:spinors}

Twistor theory originated in a paper \cite{Penrose67} by Penrose in 1967 and has been successfully applied to a wide range of mathematical and physical problems. Although the theory is most often associated with Minkowski space, this thesis stays mostly in the framework of Euclidean twistors, as utilised, for example, by Atiyah, Hitchin and Singer \cite{AHS78} in their work on the instanton moduli space.

One way of motivating the introduction of twistors is via spinors. The local isomorphism (2-fold covering) $SU(2)\times SU(2)\stackrel{\sim}{\to} SO(4)$ means that $\R^4$, which is topologically trivial, is a spin manifold. Its tangent bundle decomposes into a tensor product of the `primed' and `unprimed' spin bundles:
\[
T\R^4 \cong S\otimes S'.
\]
Practically, this means replacing each $SO(4)$ index by a pair of $SU(2)$ indices, e.g.\ $x^\mu\mapsto x^{AA'}$, where $A,A'=1,2$. A concrete realization is
\begin{equation}
	x^{A\ap}=\frac{1}{\sqrt{2}}
	\left(
\begin{array}{ccc}
	x^0+ix^1 & x^2+ix^3\\
	-x^2+ix^3 & x^0-ix^1
\end{array}
\right),
	\label{eq:xmutoxaa}
\end{equation}
which is the usual relation of $\R^4$ with quaternions represented as matrices. (This differs from the usual expression of $x^{AA'}$ as $\sigma_\mu^{AA'}x^\mu$ using Pauli matrices. The current form is more adapted to Euclidean space as we explain below.) The Euclidean metric $g_{\mu\nu}$ may be replaced by $g_{AA'BB'}=\epsilon_{AB} \epsilon_{A'B'}$, where $\epsilon$'s are the usual anti-symmetric symbols\footnote{Our convention is that $\varepsilon_{12}=1$ and $\varepsilon_{21}=-1$.}. These are used to raise and lower spin indices analogous to the role of $g_{\mu\nu}$. However, as the $\epsilon$'s are anti-symmetric, the `direction' of contraction is important. The north-westerly convention will be adopted:
\[
\alpha_A=\alpha^B \varepsilon_{BA}\quad \text{and}\quad \beta^\ap=\varepsilon^{\ap \bp} \beta_\bp.
\]
It may then be checked from Equation \eqref{eq:xmutoxaa} that $\|x\|^2\equiv x_\mu x^\mu=x_{AA'} x^{AA'}$ as expected.

Owing to the 2-dimensional nature of the spinors, there is a very useful relation
\begin{equation}\label{eq:antisym}
	\alpha_\ap\beta_\bp-\alpha_\bp\beta_\ap= (\alpha\cdot\beta)\ \varepsilon_{\ap\bp},
\end{equation}
where $\alpha\cdot\beta\equiv \alpha_\cp \beta^\cp$. An instance of the simplifications that arise is the decomposition of differential forms. The gauge field $F_{\mu\nu}$ in Yang-Mills theory is a 2-form that can be expressed in spinor notation as
\begin{equation}
F_{AA'BB'}=F_{AB}\epsilon_{A'B'}+F_{A'B'}\epsilon_{AB},	
	\label{eq:fdecomp}
\end{equation}
where $F_{AB}=\frac{1}{2} F_{ABA'}^{\qquad A'}$ and $F_{A'B'}=\frac{1}{2} F_{A'B'A}^{\qquad A}$ are symmetric. They correspond precisely to the anti-self-dual and self-dual part of $F_{AA'BB'}$. An anti-self-dual (ASD) curvature is defined to have vanishing $F_{A'B'}$, and as a consequence, for any spinor $\pi_\ap$ and ASD $F$,
\[
\pi^\ap \pi^\bp F_{ABA'B'}=0.
\]
This suggests that an ASD vector bundle is flat when restricted to certain directions, except that, of course, $\pi_\ap$ does not live in $\R^4$! If, however, one enlarges the base space to $\PT'\equiv\R^4\times\P^1=\{(x^{AA'},\pi_\bp)\}$, one can conceivably exploit the `integrability' condition above. This story will be taken up again once we have set up the necessary structures in the space $\PT'$.

The most important feature of $\PT'$ is that it is naturally a subset of the complex manifold $\C\P^3$, enabling us to deal with holomorphic as well as differential-geometric objects. Let us introduce projective co-ordinates $Z^\alpha=(\omega^A,\,\pi_\ap)$ on $\C\P^3$, which is, by definition, the projective twistor space $\PT$, and define the `incidence relation'
\begin{equation}\label{eq:incidence}
	\omega^A=x^{A\ap}\pi_\ap.
\end{equation}
Then the inclusion of $\PT'$ in $\PT$ is given by 
\[
(x^{AA'},\pi_\bp)\mapsto (\omega^A=x^{A\ap}\pi_\ap,\pi_\bp),
\]
which identifies $\PT'$ as the subset of $\PT$ in which $\pi_\ap\neq 0$.

The real Euclidean space $\E=\R^4$ can be picked out through the following conjugation defined on spinors.
\begin{equation}\label{eq:conj}
	(\omega^A)=\left(\begin{array}{c}
	\alpha \\ \beta
\end{array}\right)\mapsto (\hat\omega^A)=\left(\begin{array}{c}
	-\bar\beta \\ \bar\alpha
\end{array}\right),\quad
(\pi_\ap)=\left(\begin{array}{c}
	\gamma \\ \delta
\end{array}\right)\mapsto (\hat\pi_\ap)=\left(\begin{array}{c}
	-\bar\delta \\ \bar\gamma
\end{array}\right).
\end{equation}
Note that $\widehat{\widehat{\omega}}^A=-\omega^A$, $\omega_A\> \hat{\omega}^A = |\alpha|^2+|\beta|^2$, and similarly for $\pi_\ap$. Crucially, the conjugation induced on $x^{AA'}$ leaves it invariant: $x^{A\ap}=\hat{x}^{A\ap}$. Thus this operation is called Euclidean conjugation, as distinct from the Minkowskian conjugation to be encountered later in Chapter \ref{chap:class}.

The space $\PT'$ is a trivial $\P^1$-bundle over $\E$, the projection in twistor co-ordinates being
\begin{align}
	pr:\ & \PT' \to \E\>;\label{eq:proj}\\
	 & (\omega^A,\pi_\ap)\mapsto x^{A\ap}= \frac{\omega^A\hat{\pi}^\ap - \hat{\omega}^A\pi^\ap}{\pi\cdot\hat{\pi}}.\notag
\end{align}
Its fibre over $x^{A\ap}$ may be seen to be $L_x:=\{(x^{A\ap}\pi_\ap,\> \pi_\ap);\ \pi_\ap\in \P^1\}$, with the aid of Equation \eqref{eq:antisym}. The $\hat{}$ conjugation in Equation \eqref{eq:conj} fixes $\E$ and restricts to the antipodal map in each $\P^1$ fibre.

On the other hand, there is also a natural projection $pr':\>\PT'\to \P^1$ mapping $[\omega^A, \pi_\ap]$ to $[\pi_\ap]$, the fibres being $\C^2$. This shows that $\PT'$ is equivalent to the total space of ${\cal O}(1)\oplus {\cal O}(1)\to \P^1$, as $\omega^A$ has the same weight as $\pi_\ap$.

These two fibrations of $\PT'$ are summarised by the following diagram
\begin{equation}
\begin{CD}
@. \C^2 @.\\
@. @VVV @.\\
\P^1@>>> \PT' @>pr>> \E\\
@. @Vpr'VV @.\\
@. \P^1 @.
\end{CD}
\label{diag:fib}
\end{equation}
The horizontal fibration in fact extends to a fibration $\P^1\to\PT\to S^4$, where $S^4$ is regarded as the Euclidean compactification of $\R^4$. The fibre over the point $\infty\in S^4$ is $\{[\omega^A,\,0]\}$, precisely the subset of $\PT$ that we omit on restricting to $\PT'$.

\subsection{Differential forms on projective space}\label{subsec:diffforms}

The easiest way of specifying functions and forms on a projective space $\P^N$ is to use homogeneous coordinates, briefly denoted by $Z^i$, $i=0,\ldots,N$. To see how this works, recall that $\P^N$ may be described as the quotient space of $V^*\equiv \C^{N+1}\backslash \{0\}$ by the $\C^*$ group action $c: Z \mapsto c Z$. Vector bundles $E\to \P^N$ appearing in this thesis all have an interpretation as the quotient space of a bundle $E'\to V^*$ by certain $\C^*$-action. For example, the line bundle ${\cal O}(n)\to \P^N$ can be represented by
\[
\C\times V^*/\{(z,Z)\sim (c^n z,cZ)\}.
\]
A section of ${\cal O}(n)$ is then equivalent to a section of $\C\times V^*\to V^*$ which is invariant under the $\C^*$-action. But a section of $\C\times V^*\to V^*$ is given by $Z\mapsto (f(Z),Z)$ for a function $f$ on $V^*$. The requirement of invariance under $\C^*$-action means that $Z$ and $cZ$ are both mapped to the same orbit of the action, i.e.
\[
(f(cZ),cZ) \sim (f(Z), Z).
\]
It follows that $f(cZ)=c^n f(Z)$ and, hence, a section of ${\cal O}(n)$ is given by a function $f(Z)$ of holomorphic weight $n$ in $Z$. Similarly, differential forms or vector fields on $\P^N$ are given by forms or vector fields on $V^*$ which are invariant under the $\C^*$ scaling. E.g.\ a general 1-form on $V^*$
\[
\sum_i f_i(Z)dZ^i + g_i(Z)d\bar{Z}^i
\]
descends to $\P^N$ if $f_i(cZ)=c^{-1} f_i(Z)$ and $g_i(cZ)=\bar{c}^{-1} g_i(Z)$. All forms and vector fields in this thesis are all written in terms of homogeneous coordinates, having holomorphic weights, so that they naturally take values in ${\cal O}(n)$ for some $n$. This is not merely for convenience, as we will explain in the context of the Penrose transform.

Under the identification $\PT'\cong \E\times\P^1$, we may specify a basis for $(0,1)$-forms on $\PT'$ by\footnote{The awkward factors of $\pi\cdot\hpi$ appearing in subsequent formulae are there to make everything have holomorphic weight only.}
\begin{equation}\label{eq:formbasis}
	e^0=\frac{\hpi^\ap d\hpi_\ap}{(\pi\cdot\hat{\pi})^2},\quad\quad e^A= \frac{\hat{\pi}^\ap dx^A_{\ \ap}}{\pi\cdot\hat{\pi}},
\end{equation}
with its dual basis of (0,1)-vector fields
\begin{equation}\label{eq:vecbasis}
	\db_0=(\pi\cdot\hpi)\pi_\ap \frac{\partial}{\partial \hpi_\ap},\quad\quad \db_A=\pi^\ap \frac{\partial}{\partial x^{A\ap}},
\end{equation}
where, by Equation \eqref{eq:proj},
\[
e^A= \frac{\omega^A \hpi^\ap d\hpi_\ap}{(\pi\cdot\hat{\pi})^2} - \frac{d\homega^A}{\pi\cdot \hpi} \quad \text{ and }\quad \db_A= -(\pi\cdot\hpi) \frac{\partial}{\partial \homega^A}.
\]
Referring to the last paragraph, we see that these indeed define (0,1)-forms and (0,1)-vector fields on $\PT'\subset \P^3$. From now on, we fix these sets of bases and talk about the components $a_A$, $f_{0A}$, etc.\ of forms on $\PT'$ without further comment. Note that the discussion leading to Equation \eqref{eq:dbarpi} in the Appendix shows that we have the expected relation in $\PT'$,
\[
\db= e^0\>\db_0 + e^A\>\db_A.
\]

We will use the notation $\hat{\Omega}$ for the $(0,3)$-form of holomorphic weight $-4$
\[
\hat{\Omega}=e^0\wedge e^1\wedge e^2
\]
and also $\Omega=\varepsilon_{\alpha\beta\gamma\delta}Z^\alpha dZ^\beta dZ^\gamma dZ^\delta$, the standard $(3,0)$-form of weight 4. Its complex conjugate is $(\pi\cdot\hat{\pi})^4\,\hat{\Omega}$. Then
\[
\frac{1}{2\pi i} \Omega \wedge \hat{\Omega}= d^2 \pi d^4 x
\]
is the (unweighted) volume form on $\PT'$. Here $d^2 \pi$ is the normalised volume form on $\P^1$ as defined in Equation \eqref{eq:d2pi}.

It must be pointed out that the ${\partial}/{\partial\hpi_\ap}$ in Equation \eqref{eq:vecbasis} is the $(0,1)$-vector field on the $\P^1$ factor in $\E\times\P^1$, so that $\partial x^{A\ap}/\partial\hpi_\ap\> =0$. It should not be confused with the vector field associated with the $\pi_\ap$ coordinate on $\PT'$, which we will not use. The two are related as follows
\[
\left.\frac{\partial}{\partial\hpi_\ap}\right|_{\E\times\P^1} = \left.\frac{\partial}{\partial\hpi_\ap}\right|_{\PT'}+ x^{A\ap} \frac{\partial}{\partial \hat{\omega}^A}.
\]

\subsection{Penrose transform}\label{subsec:penrose}

The Penrose transform was mentioned in the introduction as providing a one-to-one correspondence between massless fields on space-time and cohomology classes in twistor space. The precise statement is that the (C\v ech or Dolbeault) cohomology group
\[
H^1(\PT', {\cal O}(-n-2))
\]
is isomorphic to the set
\[
\{ \text{free massless fields on }\R^4 \text{ of helicity }n/2 \}.
\]
In spinor language, a free massless field of helicity $n/2$ is a function symmetric in $|n|$ indices (primed if $n>0$; unprimed if $n<0$), which satisfies
\begin{equation}
\left\{ 
\begin{array}{ll}
 \partial^{AA'} \phi_{A'\cdots B'}=0 &\mbox{ if $n>0$}, \\
 \square \phi=0 &\mbox{ if $n=0$}, \\
 \partial^{AA'} \phi_{A\cdots B}=0 &\mbox{ if $n<0$}.
\end{array} 
\right.\label{eq:fieldeqns}
\end{equation}
Several ways exist of exhibiting this correspondence. The paper by Eastwood \emph{et.\ al.}\ \cite{Eastwood} contains a comprehensive proof using several complex variables. We will follow Woodhouse \cite{Woodhouse} and focus on the Dolbeault point of view.

The case of right-handed fields, that is, $n\geq 0$, is simpler. Given a (0,1)-form $\beta$ on $\PT'$ with values in ${\cal O}(-n-2)$, the function
\begin{equation}
	\phi_{A'\cdots B'}(x)= \frac{1}{2\pi i}\int_{L_x} \pi_\ap \cdots \pi_\bp \beta \wedge \pi_\cp d\pi^\cp
	\label{eq:fieldfromform}
\end{equation}
is an $n$-index symmetric spinor field, which satisfies the field equations \eqref{eq:fieldeqns} because $\beta$ is $\db$-closed. To get an idea why this formula works, let us look at the case $n=1$,
\[
\partial_{AA'} \phi^\ap = \frac{1}{2\pi i} \partial_{AA'} \int \pi^\ap \beta\wedge (\pi d\pi) = \frac{1}{2\pi i} \int \db_A \beta\wedge (\pi d\pi) = \int \db_A \beta_0 d^2\pi,
\]
where we recalled the definition $\db_A=\pi^\ap\partial_{AA'}$ and noted that the components $\db_A\beta_B e^B\wedge (\pi d\pi)$ integrate to 0 over $\P^1$. Since $\beta$ is $\db$-closed,
\[
\db\beta = (\db_0 \beta_A - \db_A \beta_0) e^0\wedge e^A + \db_A \beta_B e^A\wedge e^B = 0.
\]
Hence, $\db_A \beta_0=\db_0 \beta_A$ and $\partial_{AA'} \phi^\ap= \int \db_0 \beta_A d^2 \pi=0$ by Stoke's theorem. The higher spin cases are virtually the same, except for more $\pi_\ap$'s to keep track of (c.f.\ \cite{Woodhouse}).

Note that the relationship between homogeneity and helicity comes from the requirement that the integrand be of weight 0, viz.\ well-defined on $L_x$, the $\P^1$ fibre over $x$. 

In the opposite direction, if $\phi_{A'\cdots B'}$ is a free massless field of helicity $n/2\geq 0$, the (0,1)-form
\begin{equation}
	\beta_\phi = (n+1) \phi_{A'\cdots B'} \frac{\hpi^\ap \cdots \hpi^\bp}{(\pi\cdot\hpi)^{n}}\ e^0 + \partial_{AC'} \phi_{A'\cdots B'} \frac{\hpi^\ap \cdots \hpi^\bp \hpi^\cp}{(\pi\cdot\hpi)^{n+1}}\ e^A
	\label{eq:formfromfield}
\end{equation}
is $\db$-closed and represents a class in $H^1(\PT', {\cal O}(-n-2))$. Substituting $\beta_\phi$ into the integral in Equation \eqref{eq:fieldfromform} will reproduce $\phi$, thanks to the identity \eqref{eq:multipiint} in the Appendix. This is in fact the motivation behind the definition of $\beta_\phi$ --- it suggests the $\beta_0$ component and one can then write down an appropriate $\beta_A$ to make $\beta_\phi$ $\db$-closed.

Left-handed fields $\phi_{A\cdots B}$ are more complicated to deal with, because the correspondence has to be phrased in terms of potentials modulo gauge freedom. A spinor field $\psi_{AB'\cdots C'}$, symmetric in $n-1$ indices $B',\ldots, C'$, is a potential for $\phi$ if
\[
\phi_{AB\cdots C}= \partial_{(BB'} \cdots \partial_{CC'} \psi_{A)}^{B'\cdots C'},
\]
where the symmetrization is only over the unprimed indices. The field equation is satisfied if
\[
\partial^{A(A'} \psi_A^{B'\cdots C')} =0.
\]
There is the freedom of adding to $\psi$ a term of the form $\partial_{A(A'} \xi_{B'\cdots C')}$. The (0,1)-form defined by
\[
\beta_\psi = \psi_A^{A'\cdots C'} \pi_\ap \cdots \pi_\cp e^A,
\]
where $e^A$ is the (0,1)-form introduced in Equation \eqref{eq:formbasis}, is $\db$-closed thanks to the field equation. Changing the potential $\psi$ by a term $\partial_{A(A'} \xi_{B'\cdots C')}$ amounts to the addition of a $\db$-exact term $\db (\xi^{B' \cdots C'} \pi_\bp \cdots \pi_\cp)$, leaving the cohomology class unaffected.

For completeness, let us also mention Penrose's original contour integral formulae for solving the massless field equations. If $f(Z^\alpha)$ is a (holomorphic) function on twistor space, of homogeneity $-n-2$ ($n\geq 0$), then
\begin{equation}
	\phi_{A'\cdots B'} (x) = \frac{1}{2\pi i} \oint \pi_\ap \cdots \pi_\bp f(Z) \pi_\cp d\pi^\cp,
	\label{eq:penroseformula1}
\end{equation}
is a free massless field of helicity $n/2$. Despite its similarity to formula \eqref{eq:fieldfromform}, this integral is supposed to be a contour integral in the Riemann sphere $L_x$. The reason that this definition satisfies the field equation is also different. The important property to note is that, for a holomorphic function $f(Z)$,
\[
\frac{\partial}{\partial x^{AA'}} f = \pi_\ap \frac{\partial f}{\partial \omega^A}.
\]
Therefore,
\[
\nabla_A^{A'} \phi_{A'\cdots B'} (x) = \frac{1}{2\pi i} \oint \pi_\ap \cdots \pi_\bp \pi^\ap \frac{\partial f}{\partial \omega^A} \pi_\cp d\pi^\cp =0,
\]
since $\pi_\ap \pi^\ap=0$.

A somewhat similar formula for an $n$-index left-handed field is
\begin{equation}
	\phi_{A\cdots B} = \frac{1}{2\pi i} \oint \frac{\partial}{\partial \omega^A} \cdots \frac{\partial}{\partial \omega^B} f(Z) \pi_\cp d\pi^\cp,
	\label{eq:penroseformula2}
\end{equation}
where $f(Z)$ is now of weight $n-2$. The field equation is satisfied for the same reason as above, ($\frac{\partial}{\partial \omega^A} \frac{\partial}{\partial \omega_A} =0$ this time)
\[
\nabla_\ap^A \phi_{A\cdots B} = \frac{1}{2\pi i} \oint \frac{\partial}{\partial \omega^A} \cdots \frac{\partial}{\partial \omega^B} \pi_\ap \frac{\partial}{\partial \omega_A} f(Z) \pi_\cp d\pi^\cp =0.
\]

The specification of the contour in the integral and the choice of $f$ are unclear as the formulae stand. Note first of all that $f$ needs only be defined in a neighbourhood of the contour and it could be modified by adding a function holomorphic inside or outside the contour. This freedom is best dealt with by treating $f$ as an element of the \v Cech cohomology group $H^1(\PT', {\cal O}(-2h-2))$. Given a (Leray) cover $(U_i)$ of $\PT'$, let $f_{i\cdots k}$ denote a section of ${\cal O}(-2h-2)$ defined in $U_i \cap \cdots \cap U_k$. Then an element of $H^1(\PT', {\cal O}(-2h-2))$ is represented by a collection of holomorphic sections $(f_{ij})$, satisfying $f_{ij}+ f_{jk}+ f_{ki}=0$, modulo the freedom of adding $g_i - g_j$. (A more detailed exposition may be found in \cite{Huggett} or \cite{WardWells}.) These \v Cech cohomology classes could be related to $\db$-closed (0,1)-forms as follows. Choose a partition of unity $(\rho_i)$ subordinate to the cover $(U_i)$ (i.e.\ the $\rho_i$'s are smooth, with support inside $U_i$ and $\sum_i \rho_i=1$). Define $f_i = \sum_j f_{ij} \rho_j$ in $U_i$ and it follows from the cocycle condition on $(f_{ij})$ that
\[
f_i - f_j = \sum_k f_{ik} \rho_k - \sum_k f_{jk} \rho_k = \sum_k f_{ij} \rho_k = f_{ij}.
\]
Therefore, $\db f_i - \db f_j = \db f_{ij} =0$, so the $\db f_i$'s agree on overlaps and give a globally defined (0,1)-form $\alpha$ (obviously $\db$-closed). This is the desired \v Cech representative. An addition of $g_i-g_j$ to $f_{ij}$ amounts to an addition of $-\db(\sum g_i \rho_i)$ to $\alpha$.

In the opposite direction, given a Dolbeault representative $\alpha$, we solve the equation $\db f_i= \alpha$ in $U_i$, which is possible by the assumed topological triviality of $U_i$ and by the $\db$-Poincar\'e lemma. Define $f_{ij}= f_i- f_j$, then evidently $\db f_{ij}=0$ on the overlap $U_i \cap U_j$ and $f_{ij}$ satisfies the cocycle condition. This is the \v Cech representative corresponding to $\alpha$. Note that changing $\alpha$ by $\db g$ does not change $f_{ij}$, but there is a freedom of choosing $f'_i= f_i + g_i$ as the local solution to $\db f_i = \alpha$, for any holomorphic $g_i$. This would add $g_i - g_j$ to $f_{ij}$.

We will mostly use the Dolbeault formalism in this work. If necessary, the procedure above could be used to pass between the two cohomologies. Since the twistor space $\PT'$ can be covered by just two open sets, $\{\pi_1\neq 0\}$ and $\{\pi_2 \neq 0\}$, one could choose an explicit partition of unity, as in \cite{Witten03}, for example.

\section{Instantons and the Ward correspondence}\label{sec:ward}

One major strength of twistor theory is that it allows one to reformulate differential-geometric questions in terms of holomorphic data. For example, the classic Ward correspondence between instanton bundles on $\E=\R^4$ (vector bundles with an anti-self-dual connection) and holomorphic bundles on $\PT'$ arises as follows. A (necessarily trivial) vector bundle $E\to \E$ with connection $A$ can be pulled back to a bundle $pr^*E\to\PT'$, and the pull-back of the connection 1-form is
\begin{align*}
pr^*A &= A_{AA'} dx^{AA'}= A_{AA'} dx^A_\bp \left(\frac{\pi^\ap \hpi^\bp - \pi^\bp \hpi^\ap}{\pi\cdot \hpi} \right) \\ 
&= \pi^\ap A_{AA'}e^A+ \hpi^\ap A_{AA'} \hat{e}^A.
\end{align*}
If we let $a\in \Omega^{0,1}(\End(pr^*E))$ be the $(0,1)$-component of $pr^*A$,
\[
a=\pi^\ap A_{A\ap} e^A,
\]
we can define a `partial connection' on $pr^*E$
\[
\db_a:=\db+a=e^0\db_0+e^A(\db_A+\pi^\ap A_{A\ap}),
\]
which allows differentiation of sections in anti-holomorphic directions, i.e.\ $\db_a$ is a covariant derivative from $\Gamma(E)\to \Omega^{0,1}(\PT')\otimes \Gamma(E)$. The curvature $f_a\in \Omega^{0,2}(\End(pr^*E))$ of this partial connection is
\begin{equation}\label{eq:fF}
f_a=\db_a^2=\db a+a\wedge a=\frac{1}{2} \pi^\ap \pi^\bp F_{A\ap B\bp} e^A\wedge e^B.
\end{equation}
When the connection $A$ is anti-self-dual (ASD), $F_{A\ap B\bp}=\varepsilon_{\ap\bp} \tilde{F}_{AB}$ and $f_a=\db_a^2=0$, so that $\db_a$ is a genuine $\db$-operator. By a special case of the Newlander-Nirenberg theorem (c.f.\ \cite{Donaldson}), the bundle $pr^*E\to\PT'$ has a holomorphic structure, where the holomorphic sections are precisely those annihilated by $\db$.

In the other direction, suppose a topologically trivial bundle $E'\to\PT'$ is equipped with a partial connection $\db_a$, whose curvature is purely horizontal, i.e.\ $f_a$ has no component in the $e^0$-direction,
\begin{equation}
	\db_0\lrcorner f_a=0.
	\label{eq:conditiononf}
\end{equation}
One can construct a bundle $E\to\E$ with connection $A$, which is ASD when $f_a=\db_a^2=0$. Take $E$ simply to be the trivial bundle $\E\times\C^k$ and define a connection $A$ as follows. The restriction of $E'$ to each $\P^1$ in $\PT'$ has a holomorphic structure, as the (0,2)-form $\db_a^2$ vanishes on the 1-dimensional complex manifold $\P^1$. Therefore, $E'\to \PT'$ may be trivialized globally with a basis of sections $\{s_i\}$ which are holomorphic when restricted to each fibre $L_x$, i.e.
\begin{equation}\label{eq:holsi}
\db_0\lrcorner \db_a s_i=0.	
\end{equation}
In this frame, $\db_a s_i=\tilde{a}_{ji}s_j$ for some matrix of 1-forms $\tilde{a}_{ji}$ that have no component in the $e^0$-direction because of the condition \eqref{eq:holsi}. Therefore, $\tilde{a}_{ji}=\tilde{a}_{ji\, A}e^A$, using the basis of (0,1)-forms defined in Equation \eqref{eq:formbasis}. Since $e^A$ has weight $-1$ in $\pi_\ap$, $\tilde{a}_{ji\, A}$ has weight 1 in $\pi_\ap$. Now,
\[
\db_0\lrcorner f_a= \db_0\lrcorner \db_a^2 = \db_0 \tilde{a}_{ji\,A},
\]
so that the requirement $f_{a\,0A}=0$ implies that $\tilde{a}_A$ is holomorphic in $\pi_\ap$. Since it is also of weight 1 in $\pi$, an extension of Liouville's theorem shows that
\begin{equation}\label{eq:aA}
\tilde{a}_{A}=\pi^\ap A_{A\ap}(x)
\end{equation}
for some matrix of 1-forms $A_{A\ap}$ depending on $x$ only. This provides the required connection on $E$.

For later use in perturbation theory, we express $A$ directly in terms of $\db_a$. Let us define
\begin{align}
H(x,\pi):\ \C^k &\longrightarrow E'_{(x,\pi)};\notag\\
 (c_i) &\longmapsto \sum_i c_i\, s_i(x,\pi),\label{eq:Hdef}
\end{align}
which makes explicit the role of $H$ as a frame holomorphic up the fibres, characterized by $\db_0\lrcorner \db H=0$. Its properties are summarised in the Appendix \ref{sec:formulae}. Now, $\pi^\ap\partial_{A\ap}\lrcorner\> \db_a H = H \tilde{a}_A$ and, by Equation \eqref{eq:aA} and the integral formula \eqref{eq:piint1},
\begin{equation}
	A_{A\ap}(x) =-2\int_{L_x} H^{-1} (\pi^\bp \partial_{A\bp} \lrcorner\> \db_a H)\> \frac{\hat{\pi}_\ap}{\pi\cdot\hat{\pi}}\> d^2 \pi. \label{eq:Ainta}
\end{equation}
Note that the conjugation by $H$ gives an ordinary matrix-valued integrand, which is then possible to integrate entry by entry. The expression for $\db_a$ in this frame is
\begin{equation}
	H^{-1}\db_a H= e^0 \db_0 + e^A \pi^\ap \nabla_{AA'}.
	\label{eq:dbainH}
\end{equation}
The frame $H(x,\pi)$ is not uniquely determined, but there is only freedom to change it to $H(x,\pi)\gamma(x)$, where $\gamma$ is a matrix independent of $\pi$ because it has to be globally holomorphic on each $\P^1$ fibre. The end effect is merely a gauge transformation of $A$. Further, we deduce from Equations \eqref{eq:fF} and \eqref{eq:fdecomp} that when $\db_a^2=0$, the curvature of $A$ is ASD.

We have in fact shown a little more than the Ward correspondence. We have related vector bundles with a connection over $\E$ to vector bundles over $\PT'$ with a partial connection satisfying the condition \eqref{eq:conditiononf}. In the next section, action principles will be formulated for connections in bundles over $\E$ and $\PT'$ and the correspondence explained above will relate solutions of  the Yang-Mills equation to extrema of the twistor action.

{\bf Note.} For the sake of practicality, we will often implicitly fix a frame, so that all the fields are treated as matrix-valued. The condition on $H$ becomes $(\db_0 +a_0)H=0$ and Equation \eqref{eq:Ainta} reads
\[
A_{A\ap}(x) =-2\int_{L_x} H^{-1} (\db_A + a_A) H\> \frac{\hat{\pi}_\ap}{\pi\cdot\hat{\pi}}\> d^2 \pi.
\]

\section{Chalmers-Siegel action in twistor space}\label{sec:chalmers}

In four dimensions, the field strength 2-form of a gauge theory splits into a self-dual (SD) and an anti-self-dual (ASD) part,
\[
F=F^++F^-, \quad \text{with }\ *F^{\pm}=\pm F^{\pm}.
\]
As explained in the Introduction, many pieces of evidence show that the ASD sector is special. From a physical point of view, particle scattering amplitudes in this sector all vanish at tree level, indicating that the theory is classically `free'. This is somewhat reflected mathematically by the Ward correspondence in the last section, which shows that ASD bundles turn into holomorphic bundles on twistor space without further differential constraints. It is therefore reasonable to consider an action for the theory which represents perturbation around the `free' ASD sector, as first proposed by Chalmers and Siegel \cite{Chalmers}.

The usual Yang-Mills action for a connection $A$ on a vector bundle $E$ is
\[
S=\int \| F\|^2 =\int \|F^+\|^2 +\|F^-\|^2
\]
Perturbation theory will not be affected if the topological invariant (a constant multiple of the second Chern class)
\[
-\int \tr F^2 = \int \|F^+\|^2 -\|F^-\|^2,
\]
is added to the action. The resulting integral, $\int \|F^+\|^2$, takes the minimum value 0 if the field is ASD. We may introduce an auxiliary field $B$, an $\End(E)$-valued self-dual 2-form, and define a new action
\begin{equation}\label{eq:csaction}
	S[A,B]=\int\tr B\wedge F - \frac{\epsilon}{2} \int\tr B\wedge B.
\end{equation}
Note that ASD and SD forms are orthogonal in the sense that $\tr A\wedge B=0$ if $A$ is ASD and $B$ self-dual. This follows from
\[
\tr A\wedge B = \tr A\wedge *B = \tr *A\wedge B = -\tr A\wedge B.
\]
Therefore, the first term in Equation \eqref{eq:csaction} is really just $\int\tr B\wedge F^+$.

The relation of \eqref{eq:csaction} to the original action may be seen through the equations of motion, from varying $B$ and $A$ respectively:
\begin{equation}
F^+=\epsilon B\quad \text{and}\quad D_A\, B=0.
	\label{eq:cseqsmotion}
\end{equation}
This implies in turn that $D_A \, F^+=0$ and so
\[
D_A *F=D_A(F^+-F^-)= D_A (2F^+-F)=-D_A F=0,
\]
where $D_A \, F=0$ by the Bianchi identity. Hence $F$ satisfies the full Yang-Mills equation. When the parameter $\epsilon=0$, $F^+=0$, that is, $F$ is ASD. Thus, $\epsilon$ acts as a perturbation parameter away from the ASD sector.

Witten pointed out that the ASD Chalmers-Siegel action, i.e.\ \eqref{eq:csaction} with $\epsilon=0$, has a reformulation in twistor space. In the set-up of Section \ref{sec:ward}, a bundle $E$ over $\R^4$ is pulled back to a bundle $E'$ over $\PT'$ and the connection $A$ corresponds to a partial connection $a$, an $\End(E')$-valued (0,1)-form. The translation of $F_A$ is then naturally the curvature of $\db_a$, the (0,2)-form $f_a=\db a+ a\wedge a$. But what about the self-dual 2-form $B$? The Penrose transform in Section \ref{subsec:penrose} suggests an answer. A self-dual 2-form on space-time $B=B_{A'B'}\epsilon_{AB}$ is a spin 1 field, and corresponds to a (0,1)-form of homogeneity $-4$ under the transform. Therefore, one is led to consider an action of the form $\int b\wedge f_a$ over $\PT'$. However, the integrand, as it stands, is a (0,3)-form and cannot be integrated over a 3-dimensional complex manifold. This is easily remedied by using the holomorphic 3-form $\Omega$ introduced in Section \ref{subsec:diffforms}. The action,
\begin{equation}
	S_{\text{ASD}}[a,b]=\frac{1}{2\pi i} \int_{\PT'} \tr (b\wedge f_a) \wedge \Omega,
	\label{eq:asdtwistorcs}
\end{equation}
where $b\in \Omega^{0,1}(\End(E')\otimes {\cal O}(-4))$, is then well defined. Its equations of motion are
\[
\db_a^2=f_a=0\quad \text{ and }\quad \db_a b=0,
\]
which imply that $E'$ has a holomorphic structure (and $b$ is $\db$-closed in that structure).

This alternative point of view on the Ward transform has the advantage that it could be extended to include non-ASD fields. To reproduce the full Chalmers-Siegel action, Mason \cite{Mason05} introduced an interaction which is essentially the lift of the second term in \eqref{eq:csaction} to twistor space. His approach, which is probably the most direct one, is to use a non-abelian extension of the linear Penrose transform to express the self-dual 2-form $B$ in terms of the $(0,1)$-form $b$. When these forms are not $\End(E)$-valued, their relationship is given in Equation \eqref{eq:fieldfromform}, so that one might consider writing
\[
B_{A'B'}= \frac{1}{2\pi i} \int \pi_\ap \pi_\bp\ b\wedge (\pi d\pi).
\]
In order to make sense of the integral of an $\End(E')$-valued form, we would like to choose a trivialization of the bundle that is, in some sense, the most natural. It has been remarked before in Section \ref{sec:ward} that the restriction of $E'$ to each $\P^1$ fibre necessarily has a holomorphic structure. When $a$ is small, as in perturbation theory, the restricted bundle is holomorphically trivial. Therefore, we take the frame $H$ in Equation \eqref{eq:Hdef} and define
\begin{equation}
	B_{A'B'}(x)= \int_{L_x} \pi_\ap \pi_\bp H^{-1}\> b_0\> H\ d^2\pi.
	\label{eq:Bfromb0}
\end{equation}
Conjugation by $H$ produces an ordinary matrix of forms which one is allowed to integrate as in Equation \eqref{eq:Ainta}. The field $B$ is then determined by $a$ and $b$ up to \emph{space-time} gauge transformations.

The lift of the interaction term $\int \tr B\wedge B$ to twistor space is then
\begin{align}
	I[a,b]=\int_\E \int_{L_x \times L_x} \tr\> \Bigl( (H^{-1} b_0 H)\bigr|_{(x,\pi_1)} & (H^{-1} b_0 H)\bigr|_{(x,\pi_2)} \Bigr) \notag\\
	 &\times (\pi_1\cdot\pi_2)^2\ d^2\pi_1 d^2 \pi_2 d^4 x.\label{eq:interactionterm}
\end{align}
Note that this only depends on the components $a_0$ and $b_0$ of the fields. The components $b_A$ evidently do not appear and the dependence on $a_0$ enters through the frame $H$, which must satisfy $(\db_0+a_0)H=0$.

In sum, the total action is
\begin{equation}\label{eq:totalaction}
	S[a,b]=S_{\text{ASD}}[a,b] - \frac{\epsilon}{2}\>I[a,b].
\end{equation}

The Yang-Mills action acquires additional gauge freedom on being transferred from space-time to twistor space. Transformations of $a$ and $b$ that leave the action invariant are
\begin{equation}\label{eq:gaugefreedom}
	a\mapsto a'= h^{-1} a h+ h^{-1} \db h,\qquad b\mapsto b'= h^{-1} b h +\db_{a'} \chi,
\end{equation}
where $h$ and $\chi$ are arbitrary smooth sections of $\End(E')\to \PT'$, homogeneous of degree 0 and $-4$ respectively in $\pi$. The change effected by $h$ is of course a standard gauge transformation. The $\db_{a'} \chi$ term leaves $S_\text{ASD}$ invariant by Bianchi's identity $\db_{a'} f_{a'}=0$, while Equation \eqref{eq:hconjdba0}
\[
H^{-1} (\db_0 +a_0) \chi H= \db_0 (H^{-1} \chi H)
\]
shows that it contributes a total divergence to the integrand in $I$.

The interaction term can be written slightly differently if we introduce the following definition,
\begin{equation}
	K_{12}= H(x,\pi_1) \frac{1}{\pi_1\cdot\pi_2} H^{-1}(x,\pi_2).
	\label{eq:defK}
\end{equation}
It is explained in Appendix \ref{sec:formulae} that $K_{12}$ is Green's function for the operator $\db_0+a_0$ and that its variation with respect to $a_0$ is given by
\begin{equation}
	\delta K_{12}=-\int_{L_x} K_{13} \delta a_0^{(3)} K_{32}\ d^2\pi_3,
	\label{eq:varKwrta0}
\end{equation}
where the shorthand $a_0^{(k)}=a_0(x,\pi_k)$ is used. The term $I[a_0,b_0]$ becomes
\begin{equation}
	I[a_0,b_0]= -\int_\E \int_{L_x\times L_x} \tr \left( b_0^{(1)} K_{12} b_0^{(2)} K_{21} \right) \ (\pi_1\cdot \pi_2)^4 \ d^2\pi_1 d^2 \pi_2 d^4 x.
	\label{eq:Iab}
\end{equation}

The equations of motion of the twistor action $S_\text{T}[a,b]$ are as follows. Varying with respect to $b$ gives
\begin{equation}
	f_a=\epsilon \pi^\ap \pi^\bp H B_{A'B'} H^{-1} e^1\wedge e^2,
	\label{eq:eomfromb}
\end{equation}
where $B_{A'B'}$ is defined by Equation \eqref{eq:Bfromb0}. The equation from varying $a$ is more complicated,
\begin{equation}
	\db_a b\, (x,\pi_3)= \epsilon \mathop{\int}_{L_x\times L_x} K_{32}b_0^{(2)} K_{21} b_0^{(1)} K_{13} \> (\pi_1 \cdot \pi_2)^4 \ d^2\pi_1 d^2\pi_2 \ e^1\wedge e^2.
	\label{eq:eomfroma}
\end{equation}
These equations will be examined in more detail in the next chapter. One key property to note for now is that the right-hand sides have no component in the $e^0$-direction.

\vspace{0.5cm}

The equivalence of the twistor and space-time actions is provided by the following result, proved in \cite{Mason05}.

\begin{theorem}
There is a one-to-one correspondence between extrema of the action $S_\text{T}[a,b]$ and those of the Chalmers-Siegel action $S_\text{YM}[A,B]$, modulo gauge equivalence. Moreover, the values of the actions agree at the extrema.
\end{theorem}

The material introduced in Sections \ref{sec:twistor} and \ref{sec:ward} allows us to sketch a proof of this theorem, closely related to the discussion of the Ward correspondence. First note that any triple $(E,A,B)$ on $\E$ lifts to $(E',a,b)$ on $\PT'$, such that $S_\text{T}[a,b]=S_\text{YM}[A,B]$. It was explained how to obtain $E'=\text{pr}^*E$ and $a=\pi^\ap A_{AA'} e^A$, so that
\begin{equation}
	\db_a= e^0 \db_0 + e^A \pi^\ap \nabla_{AA'},
	\label{eq:afromA}
\end{equation}
where $\nabla_{AA'}=\partial_{AA'}+A_{AA'}$ is the covariant derivative in $E$. The (0,1)-form $b$ is given by a generalisation of Equation \eqref{eq:formfromfield},
\begin{equation}
	b = 3 B_{A'B'} \frac{\hpi^\ap \hpi^\bp}{(\pi\cdot\hpi)^2}\ e^0 + \nabla_{AC'} B_{A'B'} \frac{\hpi^\ap \hpi^\bp \hpi^\cp}{(\pi\cdot\hpi)^3}\ e^A,
	\label{eq:bfromB}
\end{equation}
where $\partial_{AC'}$ has been replaced by $\nabla_{AC'}$, which acts in the standard way on the $\End(E)$-valued form $B$. With $a$ and $b$ thus defined, it is straightforward to check that they are solutions of Equations \eqref{eq:eomfromb} and \eqref{eq:eomfroma} if $A$ and $B$ satisfy the Yang-Mills equation, i.e.
\begin{equation}
	\nabla^{AA'}B_{A'B'}=0, \qquad F_{A'B'}=\epsilon B_{A'B'}.
	\label{eq:fullym}
\end{equation}
Here $F_{A'B'}$ is the self-dual part of the curvature $F(A)$, c.f.\ Equation \eqref{eq:fdecomp}.

In the other direction, $(E',a,b)$ on $\PT'$ corresponds to $(E,A,B)$ on $\E$ if $a$ and $b$ satisfy their equations of motion. Since the interaction term in the twistor action is only dependent on $a_0$ and $b_0$, the equation of motion from varying $b_A$ implies that $f_{0A}=0$, i.e.\ $f_a$ has no $e^0$ component. Therefore, the condition \eqref{eq:conditiononf} is satisfied and, as explained in Section \ref{sec:ward}, we can obtain from $a$ a connection $A$ on a bundle $E$ over $\E$. The field $B$ can then be defined by Equation \eqref{eq:Bfromb0}. 

If the expressions of $f_a$ in terms of $F_A$, \eqref{eq:fF}, and $b$ in terms of $B$, \eqref{eq:bfromB}, are substituted into the integral $S_\text{ASD}[a,b]$, it can be seen to equal $\int B\wedge F$, thanks to the identity \eqref{eq:multipiint}. The interaction terms, $I[a,b]$ and $I[B]$, are equal by construction. Therefore, the value of $S_T[a,b]$ agrees with that of $S_{YM}[A,B]$, whenever the fields on the two spaces are related. Hence, extrema of $S_T$ must correspond to extrema of $S_{YM}$ and solutions of the twistor equations of motion give rise to solutions of the full Yang-Mills equation.

This can also be checked directly. The relation \eqref{eq:dbainH} implies that $f_a = \pi^\ap \pi^\bp H F_{A'B'} H^{-1} e^1\wedge e^2$. Comparing with Equation \eqref{eq:eomfromb}, we have $F_{A'B'}=\epsilon B_{A'B'}$. The other equation in \eqref{eq:fullym} is obtained in a similar fashion to the calculation below Equation \eqref{eq:fieldfromform}. The fact that $\db_a b$ has no $e^0$ component means that $\db_0 H^{-1} b_A H= \pi^\ap \nabla_{AA'} H^{-1} b_0 H$. Then
\[
\nabla_{AA'} B^{A'B'}= \int_{L_x} \pi^\bp \pi^\ap \nabla_{AA'} H^{-1} b_0 H \ d^2\pi= \int \db_0 \left( \pi^\bp H^{-1} b_A H \right) d^2\pi=0.
\]

It only remains to show that the correspondence is one-to-one modulo gauge equivalence. For this, we use the result from \cite{Woodhouse} that the gauge freedom up the fibres can be completely fixed by demanding that $a$ has no $e^0$ component and $b$ is harmonic on each $\P^1$ fibre, as in Equations \eqref{eq:afromA} and \eqref{eq:bfromB}. (More on this in the next section.) If we only work with such representatives in twistor space, it is then easy to see that the correspondence outlined above is one-to-one modulo \emph{space-time} gauge transformations.

\section{Gauge conditions}\label{sec:gauge}

One of the advantages of a twistorial formulation of Yang-Mills theory is the greater freedom in choosing a gauge. The focus of this thesis is on the Cachazo-Svrcek-Witten (CSW) gauge \cite{CSW04}. This is an axial-type gauge available only in twistor space, which dramatically simplifies computations of gluon amplitudes. This section will describe the features of this gauge and also mention another gauge condition --- the harmonic gauge, which will provide a direct link between the twistor and the space-time formulations.

To define the CSW gauge, it is necessary to choose a reference spinor $\eta^A$, normalised for convenience, i.e.\ $\eta_A \hat{\eta}^A=1$. This gives a point $N^\alpha=(\hat{\eta}^A,0)$ in the $\infty$ fibre of $\PT\to S^4$. Through any point $Z^\alpha\in \PT'$, there is a straight (complex) line $z\mapsto Z^\alpha+z N^\alpha= (\omega^A+z\hat{\eta}^A,\,\pi_\ap)$ connecting $Z^\alpha$ and $N^\alpha$. Lines of this form produce a holomorphic foliation of $\PT'$ (and a foliation of each $\C^2$ fibre of $\PT'\to \P^1$, the vertical fibration in diagram \eqref{diag:fib}). We will then define the CSW gauge for a $(0,1)$-form $a$ to be: $a=0$ when restricted to these lines. Equations \eqref{eq:incidence} and \eqref{eq:formbasis} show that a $(0,1)$-form $a=a_0 e^0 + a_A e^A$ restricts to $a_A \eta^A d\bar{z}$. Therefore, the CSW gauge is
\begin{equation}\label{eq:csw}
	\eta^A a_A= \eta^A b_A=0.
\end{equation}
This is analogous to the usual light-cone gauge condition $n^\mu A_\mu=0$, where $n^\mu$ is a null vector. If we choose a basis in which $\eta^A=(0,1)$, the gauge condition on the fields is then simply
\[
a_2=b_2=0.
\]

The CSW gauge has the advantage of reducing the components of the fields from 3 to 2 and thus eliminating the cubic $b\wedge a\wedge a$ term in the ASD action \eqref{eq:asdtwistorcs}. This will lead to simplifications in the classical and quantum theories, as demonstrated in Chapters \ref{chap:class} and \ref{chap:quant}.

\vspace{0.5cm}

In the proof of the theorem in Section \ref{sec:chalmers}, we mentioned that the gauge freedom owing to the $\P^1$ fibres can be completely fixed, leaving only the possibility of space-time transformations. This is done by first demanding that $a_0=0$, which is possible for any `small' enough $a$. To be more precise, the restriction of any complex vector bundle to $\P^1$ has a holomorphic structure, since the (0,2)-form $\db_a^2|_{\P^1}$ is necessarily zero. It is well-known that a holomorphic bundle on $\P^1$ is of the form $\bigoplus_{i=1}^n {\cal O}(k_i)$ for some integers $k_i$. The set of $a$ such that all $k_i=0$ in the holomorphic structure given by $\db_a$, is then an open subset containing $a=0$ in the space of all possible $a$'s. A small perturbation from $a=0$ will again lead to a holomorphically trivial bundle, and a frame will exist in which $a_0=0$.

If a gauge transformation $a\mapsto h^{-1} a h + h^{-1} \db h$ preserves the property $a_0=0$, we must have $\db_0 h=0$. For any $x\in \R^4$, the function $\pi\mapsto h(x,\pi)$ is then a holomorphic function on $\P^1$ and so must be constant by Liouville's theorem. Hence, $h$ is a function of $x$ only, i.e.\ a space-time transformation.

The condition to be imposed on $b\in \Omega^{0,1}(\End(E')\otimes {\cal O}(-4))$ is that it is harmonic on restriction to $L_x$ for any $x$. We use the standard round metric on $\P^1$,
\[
g = \frac{\pi_\ap d\pi^\ap \odot \hpi_\bp d\hpi^\bp}{(\pi \cdot \hpi)^2},
\]
and define a Hodge star * as usual. Since $\db b|_{L_x}$ necessarily vanishes, $b|_{L_x}$ being harmonic is equivalent to
\begin{equation}
	\db_0^{\>*} b|_{L_x} =0.
	\label{eq:gaugeconditiononb}
\end{equation}
Expressing the above in components, we have
\[
\partial_0 \left( (\pi\cdot\hpi)^2 b_0 \right) = 0,
\]
where $\partial_0$ is the conjugate of $\db_0$, namely $\partial/\partial z$ in local co-ordinates. Recall that $b_0$ is homogeneous of degree $-2$ in $\pi$, so that $(\pi\cdot\hpi)^2 b_0$ is homogeneous of degree 2 in $\hpi$. On the other hand, the equation above says that $(\pi\cdot\hpi)^2 b_0$ is holomorphic in $\hpi$. Hence, it has to be of the form $B_{A'B'}(x) \hpi^\ap \hpi^\bp$, and so
\begin{equation}
	b_0= \frac{B_{A'B'}(x) \hpi^\ap \hpi^\bp}{(\pi\cdot\hpi)^2},
	\label{eq:b0B}
\end{equation}
for some function $B_{A'B'}$ depending only on $x$. If $b$ is changed by adding $\db_a \chi$, then $\db_0 \chi$ must also be of the form \eqref{eq:b0B}, for some $\tilde{B}_{A'B'}$, say. But by the identity \eqref{eq:multipiint},
\[
\tilde{B}_{A'B'}(x)= 3 \int_{L_x} \frac{\tilde{B}_{C'D'} \pi_\ap \pi_\bp \hpi^{C'} \hpi^{D'}}{(\pi\cdot\hpi)^2} d^2\pi = 3 \int_{L_x} \db_0 (\pi_\ap \pi_\bp \chi) d^2\pi =0.
\]
Hence, $\db_0 \chi=0$. It follows that $\chi$ is a holomorphic function in $\pi$ of homogeneity $-4$ and, therefore, must vanish.

The end result is that the only gauge freedom left is
\[
a\mapsto h^{-1}(x)\, a\, h(x)+ h^{-1}(x) \db h(x), \qquad b\mapsto h^{-1}(x)\, b\, h(x),
\]
which is precisely the transformation induced by a space-time transformation $(A,\,B)\mapsto (h^{-1}Ah+ h^{-1} \partial h,\> h^{-1} B h)$.

Let us see what the action $S_\text{T}[a,b]$ looks like, when expressed using these representatives. First, $a_0=0$ means that we can take the frame $H$ to be trivial. Since $b_0$ is given by \eqref{eq:b0B}, the interaction term $I[a_0,b_0]$ becomes simply $\int B_{A'B'} B^{A'B'} d^4x$. A few lines of calculations show that
\[
S_\text{ASD}[a,b]= \int_{\PT'} \left( b^A \db_0 a_A + b_0(\db_A a^A + a_A a^A) \right) d^2 \pi d^4 x.
\]
The field $b_A$ only appears linearly in the gauge-fixed action $S_\text{T}$ and so can be regarded as a Lagrange multiplier. Integrating it out leaves a delta function forcing $\db_0 a_A=0$. Since $a_A$ has weight 1 in $\pi$, it must be of the form $a_A= \pi^\ap A_{AA'}(x)$. Inserting this and \eqref{eq:b0B} into $S_\text{ASD}$ gives $\int B\wedge F_A$, after making use of the formula \eqref{eq:multipiint} again. We therefore recover the Chalmers-Siegel action \eqref{eq:csaction}.

The harmonic gauge is convenient for the quantum theory, too. If we consider the condition \eqref{eq:gaugeconditiononb} on an infinitesimal gauge transformation,
\[
a \to a + \db_a \xi,\qquad b\to b+ \db_a \chi + [\xi, b],
\]
we find that
\[
\db^*\db \xi |_{L_x}=0\quad\text{ and }\quad \db^*(\db \chi + [\xi,b_0]) |_{L_x} =0.
\]
It follows that the Fadeev-Popov determinant will be independent of $a$ and $b$, as $b_0$ only appears `off the diagonal' in the equation above. As a consequence, ghosts associated with this gauge-fixing will decouple from physical amplitudes.

\chapter{Classical Aspects}\label{chap:class}

We saw in the last chapter that solutions of the twistor equations of motion are in one-to-one correspondence with solutions of the Yang-Mills equation. Therefore, understanding these equations in twistor space sheds light on the Yang-Mills equation, which is of considerable mathematical interest. Given its important role in the study of physics and geometry, a great deal of effort has been spent investigating the properties of its solutions.

Written out in components, the Yang-Mills equation is a coupled system of second-order PDE's in the connection 1-form $A$ on a vector bundle $E$ over Euclidean $\R^4$ or Minkowskian $\R^{3,1}$. (Assume the gauge group to be $SU(2)$ for simplicity.) As such, its structure is very complex and at the beginning, attention was focused on solving the special case of ASD Yang-Mills equation. Recall that the integral
\[
\frac{-1}{8\pi^2} \int \tr F\wedge F = \int (\|F^+\|^2- \|F^-\|^2) d^4 x
\]
is an integer topological invariant $k$. Therefore, within each topological class, the Yang-Mills functional
\[
-\int \tr F\wedge *F = \int (\|F^+\|^2+ \|F^-\|^2) d^4 x
\]
attains an absolute minimum when $F^+=0$ or $F^-=0$, i.e.\ when the connection is anti-self-dual (ASD) or self-dual (SD). The ASD Yang-Mills equation $*F=-F$ is a system of first-order PDE's, for which explicit solutions were found in the 1970's. These are called instantons. Atiyah, Drinfeld, Hitchin and Manin \cite{Atiyah:1978ri} invented a method in 1978 which allows one to construct all instantons of a given instanton number $k$. Their technique developed from the Ward correspondence outlined in Section \ref{sec:ward}, relating ASD bundles over $S^4$ to algebraic bundles over $\C\P^3$. As the latter objects were better understood, ADHM were able to formulate their celebrated construction by means of quaternion linear algebra. It is also well-known that, when the base space is allowed to be a general Riemannian manifold, the moduli space of these instantons is used in Donaldson theory to study the differential structure of the base manifold.

Subsequently, mathematicians naturally began to search for non-ASD or SD instantons, that is, solutions of the Yang-Mills equation which are not absolute minima of the action. These were elusive and for a decade or so, the evidence seems to suggest that these do not exist. For example, Bourguignon \emph{et.\ al.}\ \cite{Bourguignon} showed that, for an $SU(2)$ bundle over $S^4$, no local minima of the Yang-Mills functional exists apart from the ASD or SD solutions. Furthermore, if $\|F^+\|^2<3$ everywhere, then in fact $F^+=0$ and so the bundle is ASD. It was somewhat surprising when Sibner, Sibner and Uhlenbeck \cite{Sibner} demonstrated the existence of non-(anti-)self-dual critical points in 1989, by exploiting a link between equivariant gauge fields over $S^4$ and monopoles on hyperbolic 3-space. For a sample of the mathematical analysis that goes into these results, we refer the reader to a recent survey \cite{Tian}.

The story in Lorentzian signature is rather different in character. Here, one is more interested in the Cauchy problem, i.e.\ understanding the evolution of a gauge field from some initial data, given on a spatial slice, say. I.\ E.\ Segal \cite{Segal} proved a theorem in 1978 that ensures the existence of a solution for some finite time interval if the initial data have square integrable first and second derivatives. This was extended to global existence by Eardley and Moncrief \cite{Eardley}, who exploited the gauge invariance of the equations to derive important estimates bounding the non-linear terms. The same well-posedness result was also proved by Klainerman and Machedon \cite{Klainerman} without the condition on smoothness of the initial data. The proofs are all highly technical because of the difficulty in deriving appropriate estimates for non-linear terms in the equation. Typically, one begins with equations in the temporal gauge, makes use of another gauge (Cronstrom or Coulomb) for some special property, and reverts to the original gauge at the end.

From another angle, the Yang-Mills equation can be dimensionally reduced to various simpler equations, for example, to an ODE by removing space dependence \cite{Steeb}; or to an equation in two variables by assuming spherical symmetry \cite{Kawabe}. These can be investigated with a variety of algebraic and numerical techniques. It appears that, contrary to the ASD case, these reductions from the full Yang-Mills equation are not integrable and exhibit chaotic behaviour.

\vspace{0.5cm}

We have seen that the Yang-Mills equation displays interesting phenomena in Euclidean and Minkowski space. The first section of this chapter takes our twistor Yang-Mills action and looks at its equations of motion in the CSW gauge. These equations have an suggestive form, which we explore in some simple cases in Section \ref{sec:specialcases}. In Section \ref{sec:mink}, the framework is adapted to Minkowski space and a link to an earlier work of Kozameh, \emph{et.\ al.}\ is examined. It is hoped that our reformulation may lead to alternative approaches to the investigations mentioned above.

\section{Equations of motion}\label{sec:eom}

The CSW gauge $a_2=b_2=0$ eliminates two field components and makes the $S_\text{ASD}$ part of the twistor action \emph{quadratic}. The gauge-fixed action is
\[
S[a,b]=\int \tr\> (b_1 \db_2 a_0- b_0\db_2 a_1)\, d^2\pi\, d^4x - \frac{\epsilon}{2}\> I[a_0,b_0].
\]
Varying the components of $a$ and $b$, we find
\begin{align}
	\db_2 a_0 &=0, \qquad \db_2 a_1 = -\frac{\epsilon}{2}\> \frac{\delta I[a_0,b_0]}{\delta b_0},\label{eq:eom0}\\
	\db_2 b_0 &=0, \qquad \db_2 b_1 = -\frac{\epsilon}{2}\> \frac{\delta I[a_0,b_0]}{\delta a_0}.\label{eq:eom1}
\end{align}
However, we have to remember to add equations obtained from varying the original action with respect to $a_2$ and $b_2$:
\begin{equation}\label{eq:eom2}
	\db_0 a_1-\db_1 a_0+ [a_0,a_1]=0, \qquad \db_0 b_1+ [a_0,b_1]-\db_1 b_0 - [a_1,b_0] = 0.
\end{equation}

Note that these form a first-order system of 4 functions, in constrast to the Yang-Mills equation, which is a second-order system in 3 functions after gauge-fixing. The way the fields couple in the equations is special and explored in detail below. It must also be pointed out that part of the simplicity arises because the fields appearing here depend on 3 complex (6 real) variables, compared to 4 real variables in ordinary Yang-Mills theory. Note also that there is some non-locality in $\pi$ in the $\delta I$ terms, which involve integrals over $L_x$ or $L_x\times L_x$.

Since only the $\db_2$ derivatives appear on the left-hand sides of Equations \eqref{eq:eom0} and \eqref{eq:eom1}, one could think of them as $\db_2$-evolution equations, while treating \eqref{eq:eom2} as constraints. This is of course only valid if the constraints are preserved by evolution, as will be checked below. However, this may be expected from the path integral point of view. The action \eqref{eq:totalaction} is linear in $a_2$ and $b_2$, which may therefore be treated as Lagrange multipliers. Integrating out these fields would lead to a theory with the gauge-fixed action above and consistent constraints \eqref{eq:eom2}.

In order to see that the constraints are consistent with the equations of motion, we need to show that the left-hand sides of \eqref{eq:eom2} vanish under $\db_2$, given that Equations \eqref{eq:eom0} and \eqref{eq:eom1} are satisfied. This amounts to showing 
\begin{equation}\label{eq:eom2a}
	\db_0 \frac{\delta I}{\delta b_0} + \left[a_0, \frac{\delta I}{\delta b_0} \right]=0
\end{equation}
and
\begin{equation}\label{eq:eom2b}
	\db_0 \frac{\delta I}{\delta a_0} + \left[a_0, \frac{\delta I}{\delta a_0} \right] + \left[b_0, \frac{\delta I}{\delta b_0} \right]=0.
\end{equation}

To simplify notation, we will temporarily write $a_i$ and $b_i$ for the values of $a_0$, $b_0$ at $\pi_i$. These should not be confused with the components of $a$, $b$. Indeed, only their ${0}^\text{th}$ components appear here. We have
\begin{equation}\label{eq:iterm2}
I=-\int_\E \int_{L_x\times L_x} \tr b_1 K_{12} b_2 K_{21} \> (\pi_1\cdot\pi_2)^4 \> d^2\pi_1\ d^2\pi_2 d^4 x,
\end{equation}
and so
\begin{equation}\label{eq:delidelb}
	\frac{\delta I}{\delta b(x,\pi_3)} =-2 \int_{L_x} K_{32} b_2 K_{23} (\pi_3\cdot \pi_2)^4\ d^2\pi_2.
\end{equation}
For the variation with respect to $a_0$, we use Equation \eqref{eq:varkdiff}
\[
\frac{\delta K_{12}{}^A_{\ B}}{\delta a_3 {}^C_{\ D}} = -K_{13}{}^A_{\ C} K_{32}{}^D_{\ B}\ ,
\]
which shows that
\begin{equation}\label{eq:delidela}
	\frac{\delta I}{\delta a(x, \pi_3)} = 2 \int_{L_x\times L_x} K_{32} b_2 K_{21} b_1 K_{13} (\pi_1\cdot \pi_2)^4\ d^2\pi_1 d^2\pi_2.
\end{equation}
Equation \eqref{eq:eom2a} is then immediate, since its left-hand side is $\db_{a_0} (\delta I/\delta b)$ and $\db_{a_0} K_{32}=\delta(\pi_3\cdot\pi_2)$, which gives 0 when multiplied by $(\pi_3\cdot\pi_2)^4$.

Similarly, the first two terms in Equation \eqref{eq:eom2b} combine to give
\begin{align*}
\db_{a_0} \left(\frac{\delta I}{\delta a_3}\right) = &  2\int_{L_x} b_3 K_{31} b_1 K_{13} (\pi_1\cdot \pi_3)^4\> d^2 \pi_1\\
&\qquad\qquad\qquad - 2 \int_{L_x} K_{32} b_2 K_{23} b_3 (\pi_3\cdot \pi_1)^4\> d^2\pi_2\\
= & -\left[ b_3, \frac{\delta I}{\delta b_3} \right],
\end{align*}
confirming Equation \eqref{eq:eom2b}.

We are therefore justified in treating the equations of motion as evolution equations \eqref{eq:eom0}, \eqref{eq:eom1} plus a set of constraints \eqref{eq:eom2}. The complex nature of $\db_2$ complicates the picture and means that these equations really comprise a boundary-value rather than initial-value problem. But let us imagine $\db_2=\partial/\partial t$ for some real time parameter $t$ and propose a possible solution procedure. Suppose the values of the fields are given on a hypersurface, e.g. $x^0=0$. The equations \eqref{eq:eom0} and \eqref{eq:eom1} imply that $a_0$ and $b_0$ are constant under $\partial_t$ and hence known at any time. Putting these values back into the equations gives the source term on the right-hand sides of the $a_1$ and $b_1$ equations. The functions $a_1$ and $b_1$ can then be obtained by integrating the source terms with respect to $t$. The constraints, if satisfied initially, are satisfied for all time. Thus, the values of $a$, $b$ in the CSW gauge are completely determined over $\PT'$ from the initial data.

The situation is actually more complicated, as we will see in the sections below. Recall the fibration $\C^2\to \PT'\to \P^1$, the vertical line in diagram \eqref{diag:fib}. For any given $\pi_\ap$, $\db_2=\eta^A \db_A=-(\pi\cdot\hpi)\eta^A (\partial/\partial \homega^A)$ defines an anti-holomorphic vector field in the $\C^2$-fibre over $\pi_\ap$. The equations, therefore, present an analytic continuation, rather than evolution, problem. One property, however, is the same in both cases. If $\db_2 f=0$ for some function $f$ which is zero on a hypersurface, then by the Identity Theorem for holomorphic function, $f=0$ everywhere. More generally, if (analytic) values of $f$ are given on a hypersurface, solving $\db_2 f=0$ amounts to analytically continuing $f$ to the rest of the space. Any solution is necessarily unique. However, in contrast to an evolution problem, it may happen that analytic continuation develops singularities from smooth initial data, c.f.\ continuing $1/(1+x^2)$ from the real line to $\C$. We have not the space to delve into this kind of subtlety in this work and will concentrate on formal aspects of the equations and their solutions.

\section{Special cases}\label{sec:specialcases}
We will examine below three special cases of the solution procedure outlined above: when the corresponding Yang-Mills field (i) is abelian, (ii) gives rise to anti-self-dual fields on space-time and (iii) leads to self-dual space-time fields. In fact, Case (ii) is the simplest of the three and has been investigated in the context of the Sparling equation \cite{Newman80,specialsolns}.

\subsection{Abelian fields}\label{subsec:abelian}
When the gauge group is abelian, all $\End(E)$-valued forms commute, so that the frame $H$ drops out and the term $I[a_0,b_0]$ becomes
\[
I[b_0]=\int b_0^{(1)} b_0^{(2)} \> (\pi_1\cdot\pi_2)^2 \> d^2\pi_1\ d^2\pi_2 d^4 x.
\]
The equations of motion are then $\db_2 b_1\propto \delta I/\delta a_0 =0$ and
\[
\db_2 a_1 = -\frac{\epsilon}{2}\> \frac{\delta I}{\delta b_0} = -\epsilon\int_{L_x} b_0^{(2)} (\pi\cdot\pi_2)^2 d^2\pi_2=\phi_{\ap\bp} (x) \pi^\ap \pi^\bp,
\]
where $\phi_{\ap\bp} (x) = -\epsilon\int_{L_x} b_0 \pi_\ap \pi_\bp d^2\pi$. On the other hand, the constraints reduce to
\[
\db_0 a_1 - \db_1 a_0=0 \quad \text{ and }\quad \db_0 b_1 - \db_1 b_0 =0.
\]
These equations, together with $\db_2 a_0 = \db_2 b_0 =0$, are all the equations we need to solve. Our point of view is that $a_0$, $b_0$, and so $\phi_{A'B'}$, are determined from initial data by analytic continuation. The task, then, is to express $a_1$ and $b_1$ in terms of $a_0$ and $b_0$.

If the initial data are not given in the CSW gauge, it is necessary to enforce the gauge choice, by finding a transformation $a\mapsto a+\db \xi$, $b\mapsto b+\db \chi$ such that $\db_2 \xi=-a_2$ and $\db_2 \chi=-b_2$. A physical problem may involve specified values of $A$ and $\partial A$ on a space-time hypersurface, which are then lifted to twistor space to give initial values of $a$ and $b$. If we demand that $\db_0 \xi=0$, which is possible as $\pi$ appears as a linear factor in the equation $\db_2 \xi =-a_2$, we can impose the additional condition that $\db_0 a_1 = \db_0 \db_1 \xi =0$ (and similarly $\db_0 b_1=0$) initially. It follows from the constraints that $\db_1a_0=\db_1b_0=0$ initially.

The only non-trivial evolution equation is the one for $a_1$. Note that by the first constraint and the equation $\db_2 a_0=0$,
\[
\db_2 \db_0 a_1=\db_2 \db_1 a_0= \db_1\db_2 a_0=0.
\]
Given that $\db_0 a_1=0$ initially, it must remain 0 for all time. The function $a_1$ is then holomorphic and of weight one in $\pi_\ap$, so $a_1= a_{1\ap}(x)\pi^\ap$. The $\db_2 a_1$ equation therefore becomes
\begin{equation}\label{eq:abeliana1eqn}
\eta^A \partial_{A(\ap} a_{1\bp)} = \phi_{\ap\bp},
\end{equation}
where the identification $\db_2=\eta^A\pi^\ap\partial_{A\ap}$ is recalled and ( ) denotes symmetrization of the primed indices.

We will solve for $a_1$ as follows. Since $\db_1b_0=\db_2b_0=0$, $b_0$ is a function of $\omega_A$ (and $\pi$, $\hpi$) only, so 
\[
\partial_{A\ap}b_0= \frac{\partial b_0}{\partial \omega^A} \pi_\ap,
\]
which implies that 
\begin{equation}\label{eq:phiabeqn}
\partial^\ap_A \phi_{\ap\bp} = -\epsilon\int \frac{\partial b_0}{\partial \omega^A} \pi^\ap \pi_\ap \pi_\bp d^2\pi =0.
\end{equation}
This condition in fact implies the existence of a function $\phi$ such that 
\begin{equation}\label{eq:potentialphi}
\phi_{\ap\bp}=\eta^A\partial_{A\ap}\eta^B\partial_{B\bp} \phi.
\end{equation}
To see this, let $z^\ap=\eta_A x^{A\ap}$, $\hat{z}^\ap=\hat{\eta}_A x^{A\ap}$ and $\hat{\partial}_\ap=\eta_A \partial^A_{\ap}$. The last notation is sensible since $\hat{\partial}_\ap z^\bp=0$ and $\hat{\partial}_\ap \hat{z}^\bp = \delta_\ap^\bp$. Equation \eqref{eq:phiabeqn} implies that
\[
\hat{\partial}_{[\ap} \phi_{\bp]\cp} =0.
\]
This may be interpreted as saying that the form $\phi_{\bp\cp} d\hat{z}^\bp$ is $\db$-closed over $\C^2$. By the $\db$-Poincar\'e lemma, there is a function $\phi_\cp$ such that $\phi_{\bp\cp}=\hat{\partial}_\bp \phi_\cp$. Furthermore, the symmetry of $\phi_{\bp\cp}$ in $\bp$ and $\cp$ means that
\[
\hat{\partial}_{[\bp} \phi_{\cp]}=0.
\]
The same reasoning as above leads to a function $\phi$ with the property in Equation \eqref{eq:potentialphi}. The `potentials' $\phi_\bp$ and $\phi$ may be given by an integral formula such as the one in \cite{Nickerson} or \cite{Nijenhuis}. We quote the formula in \cite{Nijenhuis}, which is given as follows. Suppose $\alpha=\alpha_1 d\bar{z}^1 + \alpha_2 d\bar{z}^2$ is a $\db$-closed (0,1)-form on a polycylinder $D\times D'\subset \C^2$. Let $T^1$ be the operator on functions
\[
T^1 f(z_1, z_2)= \int_{D} \frac{f(\zeta,z_2)}{z_1-\zeta} d^2\zeta,
\]
and let $T^2$ denote the analogous operator, where the integral is done in the second variable. Define another operator $S^1$ by the contour integral
\[
S^1 f(z_1, z_2)= \frac{1}{2\pi i} \oint_{\partial D} \frac{f(\zeta, z_2)}{\zeta-z_1} d\zeta.
\]
Then it may be checked that the function
\[
g= T^1 \alpha_1 + S^1 T^2 \alpha_2
\]
satisfies $\alpha =\db g$. The function $\phi_\bp$ can be found by applying this formula to $\phi_{A'B'}$, and then similarly $\phi$ is obtained from $\phi_\bp$.

We can finally solve Equation \eqref{eq:abeliana1eqn} as
\[
a_1=a_{1\ap}\pi^\ap=\eta^A\pi^\ap\partial_{A\ap} \phi = \db_2 \phi.
\]
More covariantly, since $\eta_A=(-1,0)$, we can write $a_A=-\eta_A \eta^B \pi^\bp \partial_{BB'} \phi$, which indeed satisfies the gauge condition $\eta^A a_A=0$.

It remains to deduce the connection $A_{A\ap}$ on space-time. The procedure as mentioned before requires transforming to a gauge in which $a_0=0$ and then writing the new $a_A$ as $\pi^\ap A_{A\ap}$. In the current abelian setting, gauge transformations of $a$ are of the form $a\to a+\db \xi$ and putting $a_0= 0$ amounts to solving $\db_0 \xi=-a_0$. From formulae given in the Appendix \ref{sec:formulae}, we have
\[
\xi=-\int_{L_x} a_0 \frac{\pi\cdot o}{\pi'\cdot o} \frac{1}{\pi\cdot\pi'} d^2\pi',
\]
where $o_\ap$ is an arbitrary spinor. Note that $a_0$ is a function of $\omega$, $\pi$ and $\hpi$ only and, similar to Equation \eqref{eq:phiabeqn},
\[
\db_A h= \pi^\ap \partial_{A\ap} h =\int_{L_x} \frac{\partial a_0}{\partial \omega^A} \frac{\pi\cdot o}{\pi'\cdot o} d^2\pi'.
\]
The component $a_A$ in the new gauge is then $\tilde{a}_A=a_A +\db_A \xi$. Recalling that $\db_A=\pi^\ap\partial_{A\ap}$ and $a_A=-\eta_A \eta^B\pi^\bp\partial_{B\bp} \phi$ as found above, we see that indeed $\tilde{a}_A=\pi^\ap A_{A\ap}(x)$, where
\[
A_{A\ap}(x)= -\eta_A \eta^B \partial_{B\ap} \phi(x) - o_\ap \int_{L_x} \frac{\partial a_0}{\partial \omega^A} \frac{1}{\pi\cdot o} d^2\pi.
\]

When the gauge group is $U(1)$, the curvature of $A$ simply gives familiar Maxwell fields. The self-dual part is
\[
F_{A'B'}= \partial_{A(A'} A^A_{B')}= - \eta^A \partial_{AA'} \eta^B \partial_{BB'} \phi = -\phi_{A'B'} = \epsilon \int_{L_x} b_0 \pi_\ap \pi_\bp\ d^2 \pi,
\]
and the ASD part
\[
F_{AB}= \partial_{A'(A} A^\ap_{B)} = - \int_{L_x} \frac{\partial}{\partial \omega^A} \frac{\partial}{\partial \omega^B} a_0\ d^2\pi.
\]
Comparing these equations to formulae \eqref{eq:penroseformula1} and \eqref{eq:penroseformula2}, we see that our formula for the potential reproduces Penrose's formulae for free massless fields. It also shows that the potential itself is a linear combination of a self-dual piece and an ASD piece, which is special to the abelian situation.

\subsection{Anti-self-dual solutions}\label{subsec:asd}
If a Yang-Mills field is ASD, the self-dual part $B$ of the curvature $F$ must be 0. In view of Equation \eqref{eq:Bfromb0}, we will focus our attention on the case $b_0=0$ everywhere in this section. Consequently, $I=0$ and all the fields vanish under $\db_2$.

Recall from Section \ref{sec:ward} that the curvature $F_{A\ap B\bp}$ is ASD if it is equal to $F_{AB} \epsilon_{\ap\bp}$, or equivalently, $\pi^\ap \pi^\bp F_{A\ap B\bp} = 0$ for any $\pi$; also, fixing a $\pi$, we can define a partial connection $\db_{aA}=\pi^\ap (\partial_{A\ap}+ A_{A\ap})$ on $\C^2=\{\omega^A=x^{A\ap}\pi_\ap \}$, which is simply the fibre of the vertical fibration in the diagram \eqref{diag:fib}. The anti-self-duality of $F$ implies that this partial connection is flat, i.e.\ $\db_a^2=0$. The vector bundle $pr^* E|_{\C^2}$ is then holomorphically trivial. In other words, there is a gauge transformation $G(x,\pi,\hpi)$ such that
\begin{equation}\label{eq:asdGeqn}
	\pi^\ap (\partial_{A\ap} G + A_{A\ap} G)=0.
\end{equation}
The left-hand side is just $a_A$ in the new gauge. Thus, in the case of an ASD field, a gauge may be chosen in which $a_A=0$.

In view of the above discussion, any initial data can be put in a form where
\[
a_1=a_2=0.
\]
From the constraint \eqref{eq:eom2}, $\db_1 a_0=0$ on the initial hypersurface. The field $b_1$ has to satisfy the other constraint equation, but it does not play any role in the equations for $a$ and will be ignored from the discussion below.

These initial conditions must in fact hold for all time, since everything vanishes under $\db_2$. The equations to be solved are, therefore,
\[
\db_2 a_0=0 \quad\text{ and }\quad \db_1 a_0=0,
\]
or simply $\db_A a_0=0$, which is satisfied by any function $a_0=a_0(\omega^A,\pi_\ap,\hpi_\ap)$, i.e.\ independent of $\homega^A$.

To see how such a function $a_0$ relates to a self-dual field on space-time, we make a gauge transformation so that $a_0$ becomes 0, that is, we take the inverse of the $G$ appearing in Equation \eqref{eq:asdGeqn}. Thus $G a_0 G^{-1} +G\db_0 G^{-1}=0$. When simplified, this is just the Sparling equation
\begin{equation}\label{eq:sparling}
	\db_0G=G a_0,
\end{equation}
which has been investigated extensively in the literature (see \cite{Newman80, specialsolns, Goldberg} and references therein). The potential $A_{A\ap}(x)$ is recovered as follows. First, $a_A=G\db_A G^{-1}=-\db_A G G^{-1}$ is holomorphic in $\pi_\ap$, for
\begin{align*}
\db_0(\db_A G G^{-1}) &= \db_A \db_0 G G^{-1}- \db_AG G^{-1} \db_0G G^{-1}\\
&= \db_A(Ga_0) G^{-1}- \db_A G a_0 G^{-1}\\
&= G (\db_A a_0) G^{-1},
\end{align*}
which vanishes since $\db_A a_0=0$. Therefore, $a_A$, being of weight one in $\pi_\ap$, must be of the form $\pi^\ap A_{A\ap}(x)$ and so $\pi^\ap A_{A\ap}= -\pi^\ap \partial_{A\ap} G G^{-1}$. Hence,
\[
A_{A\ap}= -\partial_{A\ap} G G^{-1}+ C_A \pi_\ap,
\]
for some function $C_A$. It can be found by using $\partial A(x)/\partial \pi=0$ and contracting it with combinations of $\pi$ and $\eta$. Letting $\partial_0=\frac{1}{\pi\cdot\hpi} \hpi\cdot\frac{\partial}{\partial \pi}$, the end result can be expressed as
\[
C_A= -\pi^\bp \partial_0 (\partial_{A\bp} G G^{-1}).
\]

{\bf 't Hooft ansatz}\nopagebreak

We will illustrate the above correspondence between $a_0$ and $A$ for the 't~Hooft ansatz of SU(2) instantons. As the gauge group is taken to be SU(2), or rather its complexification SL(2,$\C$), we may mix the index of $\omega^A$ with the gauge bundle index. More precisely, we take the unprimed spinor bundle as the underlying vector bundle. (The unprimed spinor bundle of a four-manifold is anti-self-dual if the metric is Einstein.) The conditions on $a^{\ A}_{0B}$ are that it be a function of $\omega^A$, $\pi_\ap$ and $\hpi_\ap$ of holomorphic weight two and $a_{0AB}$ must be symmetric in $A$, $B$ so as to take values in sl(2,$\C$). The simplest candidate is evidently
\begin{equation}\label{eq:1instantona0}
	a^{\ A}_{0B} = 2 \omega^A \omega_B,
\end{equation}
where the factor of 2 is inserted for later convenience. Any scaling of $a_0$ can be absorbed into a scaling of $\omega$, or the space-time coordinate $x$.

To determine $G$ in Equation \eqref{eq:sparling}, we essentially need two linearly independent sections $s$ such that $\db_0s-sa_0=0$. Searching for linear combinations of $\omega^A$ and $\homega^A$ gives the result
\begin{equation}\label{eq:1instantonG}
	G^A_B= \frac{1}{\sqrt{1+|x|^2}} \left( \delta^A_B + \frac{2\homega^A \omega_B}{\pi \cdot \hpi}\right),
\end{equation}
where the prefactor comes from demanding $\det G = \frac{1}{2}G_{AB} G^{AB}=1$. A solution of the Sparling equation for an instanton was first found in \cite{specialsolns}, but their solution is singular over a codimension 1 subset of space-time.

The connection $A$ may then be calculated to be
\begin{equation}\label{eq:1instantonA}
	A_{A\ap CD}= \frac{2}{1+|x|^2} x_{\ap (C} \epsilon_{D)A} = - \epsilon_{A(C} \partial_{D)\ap} \log \phi,
\end{equation}
with $\phi=1+|x|^2$. This is just the standard 1-instanton of width 1 centred at the origin (c.f.\ \cite{Atiyah}, in quaternionic notation.) The curvature of this connection is (evidently ASD)
\[
F_{A\ap B\bp CD}=\frac{2}{(1+|x|^2)^2}\, \epsilon_{\ap\bp} \epsilon_{A(C} \epsilon_{D)B}.
\]

Equation \eqref{eq:1instantonA} may be recognised as the non-singular form of the 't Hooft ansatz for 1-instanton. The general form for an $n$-instanton is (c.f.\ \cite{MasonWoodhouse})
\begin{equation}
	A_{AA'C'D'} = \epsilon_{A(C'} \partial_{D')A} \log \phi,
	\label{eq:thooft}
\end{equation}
where $\phi$ is a solution of the wave equation, usually taken to be
\[
\phi= 1 + \sum_{i=1}^n \frac{\lambda_i^2}{|x-c_i|^2}.
\]
This expression for $A$ is often interpreted as a non-linear superposition of $n$ 1-instantons, centred at $c_i$ and of width\footnote{The `width' of an instanton is only a vague notion, as the field strength is non-zero over the whole of space-time.} $\lambda_i$.

To find the field $a_0$ corresponding to an $n$-instanton, one could solve Equation \eqref{eq:asdGeqn} for $G$, then $a_0= G^{-1} \db_0 G$. However, a non-singular $G$ can only be obtained if a non-singular form of the 't Hooft ansatz is used. This involves a sequence of gauge transformations of the form $(x-c_i)^{AA'}/|x-c_i|$ and the end result is unfortunately rather intractable, even for $n=2$. An alternative approach is to retain the singular form of the potential, find a (singular) $G$ and then look for a gauge transformation that removes the singularities. The necessary calculations are quite long and messy. Since the formulae will not be used further in this work, they are presented in the Appendix. It has to be admitted, however, that our success is only partial --- our solution for $a_0$ is only smooth in an open subset of $\PT'$. Another gauge transformation remains to be found, in order to obtain a completely smooth field.

\subsection{Self-dual solutions}\label{subsec:sd}

Even though SD and ASD fields are related by a simple reversal of space-time orientation, or just an exchange of primed and unprimed indices in spinor notation, the theory on the SD side has traditionally been more difficult (the `googly problem'). The lack of symmetry between ASD and SD bundles is a perennial problem in twistor theory. One approach to address it is to consider the `ambitwistor space' $Q$, which treats right- and left-handed fields on an equal footing. Mason and Skinner \cite{MasonSkinner} have introduced a Lagrangian for Yang-Mills theory on this space, with the aim of reproducing the BCFW recursion relations \cite{BCFW}. As a matter of fact, the Yang-Mills equation received a reformulation in ambitwistor space by Yasskin, Isenberg and Green \cite{MR564452}, and independently Witten \cite{Witten:1978xx} in the late 1970's. For comparison, we outline their idea. They regard (complexified) space-time $M$ as the diagonal in $\C^4\times \C^4$ and consider a system of first-order equations, which encode self-duality in one factor of $\C^4$ and anti-self-duality in the other factor. A connection turns out to satisfy the Yang-Mills equation if it satisfies the first-order equations in the second-order formal neighbourhood of the diagonal $M$. (Concretely, this means that, if $w$ denotes coordinates in the normal direction to $M$, then the connection is quadratic in $w$.) A similar correspondence to Ward's relates a bundle with such a connection to a complex bundle on ambitwistor space $Q$, which can be extended to the third-order neighbourhood of $Q$ in $\P^3\times \P^3$. The focus of this thesis is, however, on twistor space and the ambitwistor space will not be used.

In the present context, we first repeat the argument at the beginning of Section \ref{subsec:asd} with an interchange of primed and unprimed indices. Observe that $\eta^A \eta^B F_{A\ap B\bp}=0$ for an SD field, which allows us to solve
\begin{equation}\label{eq:sdGeqn}
	\eta^A (\partial_{A\ap} G + A_{A\ap} G)=0
\end{equation}
with $G$ a function of $x$ only ($\eta$ is taken to be fixed). Therefore, starting from $a=\pi^\ap A_{A\ap}e^A$ on $pr^*E$, a gauge transformation by $G$ in \eqref{eq:sdGeqn} sets $a_2=\eta^A\pi^\ap A_{A\ap}$ to 0. (Indeed, in the new gauge, $\eta^A A_{A\ap}=0$, so $A_{AA'}=\eta_A A_\ap$ for some function $A_\ap$.) We still keep $a_0=0$, since $G$ is independent of $\pi$. This means that, in order to obtain self-dual solutions on space-time, we should restrict $a_0=0$, and so the holomorphic frame $H$ may be taken to be the identity. The equations of motion become $\db_2 b_0=0$ plus
\begin{equation}\label{eq:sdeom1}
	\db_2 a_1 = - \epsilon \int_{L_x} b_0\> (\pi\cdot \pi')^2\ d^2\pi'
\end{equation}
and
\begin{equation}\label{eq:sdeom2}
	\db_2 b_1 = -\epsilon \int_{L_x\times L_x} b_0^{(1)} b_0^{(2)} \frac{(\pi_1\cdot \pi_2)^3}{(\pi \cdot \pi_1) (\pi_2 \cdot \pi)}\ d^2\pi_1 d^2\pi_2.
\end{equation}
The constraints are
\begin{equation}\label{eq:sdconstraints}
	\db_0 a_1=0 \qquad\text{and}\qquad \db_0 b_1- \db_1 b_0 - [a_1, b_0]=0
\end{equation}

While these are simpler than the full equations (\ref{eq:eom0}-\ref{eq:eom1}), it is by no means clear that there exists any algebraic manipulation similar to those in the previous two sections, which will lead to an expression of $a_1$ and $b_1$ in terms of $a_0$ and $b_0$. In a sense, the self-dual case already incorporates much of the complexity of the full Yang-Mills equation, since a non-zero $a_0$ would not change the appearance of the equations, except for inserting the frame $H$ in certain places. Let us examine the form of the fields in the case of an SD instanton of charge 1, partly to gain an idea for the type of functions involved and partly for comparison with the ASD case in the last subsection.

{\bf Example: 1-instanton} The potential on space-time is, c.f.\ Equation \eqref{eq:1instantonA},
\[
A_{A\ap C'D'}= \frac{2}{1+|x|^2}\> x_{A (C'} \epsilon_{D')\ap},
\]
with corresponding field strength
\[
F_{AA'BB'C'D'}= F_{A'B'C'D'} \epsilon_{AB} = \frac{2}{(1+|x|^2)^2}\> \epsilon_{A'(C'} \epsilon_{D')B'} \epsilon_{AB}.
\]
The solution to Equation \eqref{eq:sdGeqn} is
\[
G^\ap_\bp= \frac{1}{\sqrt{1+|x|^2}}\left(\delta^\ap_\bp + 2 \hat{z}^\ap z_\bp \right),
\]
where $z^\ap=\eta_A x^{A \ap}$ and $\hat{z}^\ap=\hat{\eta}_A x^{A\ap}$. The only non-zero component of $a$ in the new gauge is
\[
a_1 = - \frac{4}{1+|x|^2} \hat{z}_{(C'} \pi_{D')} - \frac{4 \eta \cdot \omega}{(1+|x|^2)^2} \hat{z}_{C'} \hat{z}_{D'}.
\]

Complications arise from the need to work out $b$ and put it in the CSW gauge. Let
\[
W^\cp_{\ D'}= \frac{1}{\epsilon} F^{\quad\ C'}_{A'B'\ D'} \frac{\hpi^\ap \hpi^\bp}{(\pi\cdot \hpi)^2} = - \frac{2}{\epsilon (1+|x|^2)^2} \frac{\hpi^\cp \hpi_{D'}}{(\pi\cdot \hpi)^2}
\]
in the space-time gauge, then the lift of $b$ to twistor space is given by \eqref{eq:bfromB} as
\[
b= 3W e^0 + \frac{\hpi^\ap}{\pi\cdot\hpi} \nabla_{A\ap} W e^A.
\]
From Equation \eqref{eq:gaugefreedom}, the gauge freedom in $b$ is
\begin{equation}
	b\mapsto G^{-1} b G - (e^0 \db_0 + e^A\pi^\ap\nabla_{A\ap}^G) \chi,
	\label{eq:btransformbyGchi}
\end{equation}
where we make the gauge transformation explicit by using the superscript ${}^G$ to denote conjugation by $G$. In order to obtain $b_2=0$, we need to solve
\begin{equation}\label{eq:chieqn}
\frac{1}{\pi\cdot\hpi} \eta^A\hpi^\ap\nabla_{AA'}^G (G^{-1} W G) = \eta^A \pi^\ap \nabla_{AA'}^G \chi.
\end{equation}
Crucially, the gauge transformation $G$ above makes $\eta^A A_{AA'}^G=0$, so that $\eta^A \nabla_{AA'}^G = \eta^A \partial_{AA'}$ in Equation \eqref{eq:chieqn}, which, therefore, can be integrated like an ordinary differential equation. The following shorthand will be used:
\[
u= \eta\cdot\omega,\quad \ubar=\frac{\hat{\eta}\cdot \homega}{\pi\cdot\hpi};\qquad v= -\hat{\eta}\cdot \omega,\quad \vbar=\frac{\eta\cdot\homega}{\pi\cdot\hpi},
\]
so that $\db_2= \eta^A\pi^\ap\partial_{AA'}=\partial/\partial \ubar$, $\frac{1}{\pi\cdot\hpi} \eta^A\hpi^\ap \partial_{AA'}= \partial /\partial v$, etc. Equation \eqref{eq:chieqn} becomes
\[
\frac{\partial}{\partial \ubar} \chi_{C'D'}=\frac{\partial}{\partial v} (G^{-1}  W G)_{C'D'}= \frac{12\vbar}{\epsilon (1+2u\ubar+2v\vbar)^4}\Bigl(\frac{\hpi_\cp}{\pi\cdot\hpi}+ 2 \ubar z_\cp \Bigr) \Bigl({}_{C'}\to {}_{D'} \Bigr).
\]
Letting $\chi_{C'D'}=f \frac{\hpi_\cp \hpi_{D'}}{(\pi\cdot\hpi)^2} + 2g \frac{\hpi_{(C'} z_{D')}}{\pi\cdot\hpi}+ h z_{C'} z_{D'}$, we get three separate equations for $f$, $g$ and $h$, which can be integrated. The difficulty lies in finding \emph{smooth} solutions. As an illustration, the equation for $f$ is
\[
\frac{\partial f}{\partial \ubar} = \frac{12 \vbar}{\epsilon (1+2u\ubar+2v\vbar)^4}
\]
which implies that
\[
f= -\frac{2 \vbar}{\epsilon u(1+2u\ubar+2v\vbar)^3} + \tilde{f}(u,v,\vbar),
\]
where $\tilde{f}$ is any function depending on $u$, $v$ and $\vbar$ only. Therefore, in order for $f$ to be smooth, we have to subtract off the $1/u$ singularity in the first term by choosing $\tilde{f}= 2\vbar/[\epsilon u(1+2v\vbar)^3]$. The other equations for $g$ and $h$ are solved in the same way, but because of higher dependence on $\ubar$, the procedure is rather more complicated.

The final result is
\begin{align*}
\chi_{C'D'} = & \frac{\ubar\vbar}{\epsilon (1+|x|^2)^3 (1+2v\vbar)^3} \left\{ 3 (1+|x|^2)^2 \frac{\hpi_\cp \hpi_{D'}}{(\pi\cdot\hpi)^2}\right. \\
 &\ \left. + \Bigl[4\ubar (1+ 2v\vbar) z_\cp + (2u\ubar + 3 (1+ 2v\vbar) \frac{\hpi_\cp}{\pi\cdot\hpi} \Bigr] \Bigl[{}_{C'}\to {}_{D'} \Bigr] \right\},
\end{align*}
with the freedom of adding on a smooth function of $u$, $v$ or $\vbar$ only.

To find $b$ after transforming by $G$ and $\chi$, every term in Equation \eqref{eq:btransformbyGchi} has to be computed. Written out in full, it is
\[
b= \left( 3 G^{-1}  W G -\db_0 \chi \right) e^0 - \hat{\eta}^A \left( \frac{\hpi^\ap}{\pi \cdot \hpi} \nabla^G_{AA'} (G^{-1}  W G) - \pi^\ap \nabla_{AA'}^G \chi \right) e^1.
\]
The calculations were very tedious, not least because $\nabla= \partial + [A,\ \ ]$ when acting on $W$ and $\chi$. The manipulations were partly implemented in \emph{Mathematica}.\footnote{The spinor algebra was handled by taking $\{\pi_\ap, \hpi_\ap\}$ as a basis and expressing everything as a matrix.} The answer is
\[
b= -\frac{6}{\epsilon (1+2v\vbar)^4} \frac{\hpi_\cp \hpi_{D'}}{(\pi\cdot\hpi)^2} \left(e^0 + \frac{4 \ubar (1+u\ubar+2v\vbar)}{(1+|x|^2)^2}\ e^1 \right).
\]
It is straightforward to calculate that
\[
\db_2 a_1 = \frac{4}{(1+|x|^2)^3} (\pi_\cp - v z_\cp) ({}_{C'}\to {}_{D'})
\]
and
\[
\db_2 b_1 = -\frac{24}{\epsilon (1+|x|^2)^3 (1+2v\vbar)^2} \frac{\hpi_\cp \hpi_{D'}}{(\pi\cdot\hpi)^2}.
\]
Equations \eqref{eq:sdeom1} and \eqref{eq:sdeom2} can be checked by comparing these expressions to the integrals on the right-hand sides. These integrals are performed essentially using the techniques outlined in Appendix \ref{sec:formulae}. The actual computations are not informative and are therefore omitted.

Now, if $b_0$ is known over $\PT'$, then $\db_2 b_1$ is given by performing the integral in Equation \eqref{eq:sdeom2}. The result can be integrated like the $\partial f /\partial \ubar$ equation above, with singularities removed in a similar way. However, analytically continuing initial data may lead to singularities as mentioned earlier and we would like to circumvent this problem by changing $\db_2$ to a real operator. This possibility is considered in the next section.

\section{Minkowskian reformulation}\label{sec:mink}

In the previous section, the classical equations were examined in the cases when the fields have some special feature. The fact that the operator $\db_2$ is complex ($\partial/\partial\bar{z}$ in local coordinates) means that we have to deal with a boundary value problem. For example, to solve a $\db_2$ equation, we had to resort to complicated integral formulae (in Subsection \ref{subsec:abelian}) or to lengthy algebraic manipulations (in Subsection \ref{subsec:sd}). In order to investigate the general case, it would be preferable to have a framework in which $\db_2$ is real, that is, $\partial/\partial t$ in some coordinates. The equations would then be \emph{bona fide} evolution equations and present an initial value problem. Such a framework is provided by a reformulation of the equations in Minkowski space.

One might attempt to carry over the construction of the action in Chapter \ref{chap:action} to the twistor space of Minkowski space. However, the structure of twistor space is more complicated in this case. In particular, it can no longer be viewed simply as a $\P^1$-bundle over space-time and $\PT$ in fact corresponds to the complexified space-time via a `double fibration'. We refer the interested reader to \cite{Huggett} or \cite{WardWells} for a detailed exposition. Suffice it to note that the region in $\PT$ corresponding to real Minkowski space is a 5-dimensional manifold, on which the Chern-Simons form necessarily vanishes. An action of the same form that we have been considering would, therefore, be difficult to construct. This has been a drawback for those who are concerned with the Euclidean signature of space-time in our formulation. It is relevant here to point out that a version of the action could be given for Minkowski space, if we take the total space to be the projective primed spin bundle $\PS$ over Minkowski space. This is a $\P^1$-bundle over $\M$, in a similar fashion to $\PT\to\E$. As before, we use coordinates $(x^{A\ap},\pi_\ap)$, but the real structure on $\PS$ that fixes $\M$, to be described below, is different to the one introduced in Section \ref{sec:twistor}. Analogues of the vector fields $\db_0$, $\db_A$ and forms $e^0$, $e^A$ can be defined and formally, our action may then be transferred, without change, to the new setting. Its equations of motion could be derived as before, but we will demonstrate a more direct approach: we adopt the Yang-Mills equation as our starting point and show that it leads to a set of equations similar to (\ref{eq:eom0}-\ref{eq:eom2}).

\subsection{Field equations}

Going from Euclidean space $\E$ to Minkowski space $\M$ may be regarded as formally letting $(x^0, x^1, x^2, x^3)\mapsto (x^0, ix^1, ix^2,ix^3)$. The matrix $x^{A\ap}$ defined in \eqref{eq:xmutoxaa} will then be Hermitian and preserved by a different conjugation from the $\hat{}$ defined in Section \ref{sec:twistor}. Denoting this conjugation by $\bar{}$, we have
\begin{equation}\label{eq:barconj}
	(\omega^A)=\left(\begin{array}{c}
	\alpha \\ \beta
\end{array}\right)\mapsto (\bar\omega^\ap)=\left(\begin{array}{c}
	\bar\alpha \\ \bar\beta
\end{array}\right),\quad
(\pi_\ap)=\left(\begin{array}{c}
	\gamma \\ \delta
\end{array}\right)\mapsto (\bar\pi_A)=\left(\begin{array}{c}
	\bar\gamma \\ \bar\delta
\end{array}\right).
\end{equation}
Note the interchange of primed and unprimed indices and that $\overline{\overline{\omega}}^A= \omega^A$. It may be readily checked that the action induced on $x^{A\ap}$ is that of Hermitian conjugation and that $x^{A\ap}$ is indeed invariant with respect to it. The `incidence relation' in the Minkowskian case is usually taken to be
\begin{equation}\label{eq:incidence2}
	\omega^A=i x^{A\ap} \pi_\ap.
\end{equation}

It is still possible to define a $\hat{}$ conjugation by choosing a time-like vector $T^{A\ap}$ and setting (c.f.\ \cite{Woodhouse})
\begin{equation}\label{eq:hatbarconj}
	\hpi^\ap=T^{A\ap} \bpi_A.
\end{equation}
It is conventional to choose the normalization $T^2=2$ so that $T^{A\ap} T_{A\bp}= \delta^\ap_\bp$. All the properties of the $\hat{}$ stated in Section \ref{sec:twistor} hold. For definiteness, one could take $T^a=(\sqrt{2},0,0,0)$, then $T^{A\ap}$ will be the identity matrix.

The obvious modification of $\db_2$ to make it real is to let 
\begin{equation}\label{eq:db2defn}
	\db_2=\frac{\bpi^A \pi^\ap}{\pi\cdot \hpi} \partial_{A\ap}.
\end{equation}
This is simply $\partial/\partial t$ if we make the following change of coordinates. First expand $x^{A\ap}$ in terms of $\pi_\ap$,
\[
x^{A\ap}= \frac{1}{\pi\cdot\hpi} (\> t\> \bpi^A \pi^\ap + z\> \bpi^A \hpi^\ap + \bar{z}\> \hat{\bpi}^A \pi^\ap + s\> \hat{\bpi}^A \hpi^\ap),
\]
where $t$ and $s$ are real because $x=\bar{x}$. In terms of the coordinates $(t,z,s,\pi_\ap)$ instead of $(x^{A\ap},\pi_\ap)$, the $\db_2$ in Equation \eqref{eq:db2defn} indeed corresponds to $\partial/\partial t$.

The space $\PS$ may be endowed with an involutive structure by declaring all vector fields in the following distribution to be of type (0,1),
\[
D= \text{span} \left\{\db_0 = -(\pi\cdot \hpi) \hat{\bpi}_A \frac{\partial}{\partial \bpi_A},\quad
\db_1 = \hat{\bpi}^A \pi^\ap \partial_{A\ap},\quad
\db_2 = \frac{\bpi^A \pi^\ap}{\pi\cdot \hpi} \partial_{A\ap} \right\}.
\]
Note that 
\begin{equation}
	[\db_2,\db_0]=\db_1
	\label{eq:db0db2}
\end{equation}
while all the other commutators vanish, so that $[D,D]\subset D$, i.e.\ $D$ is indeed an involutive distribution. The dual (0,1)-forms to $\db_i$ will be denoted by $e^0$, $e^1$ and $e^2$ as before.
\[
e^0=\frac{\bpi^A d\bpi_A}{(\pi\cdot\hpi)^2}, \qquad e^1= \frac{\bpi_A \hpi^\ap dx^A_\ap}{(\pi\cdot\hpi)^2}, \qquad e^2= -\frac{\hat{\bpi}_A \hpi^\ap dx^A_\ap}{\pi\cdot\hpi}\ .
\]

Given a vector bundle $E\to \M$ with a connection $A_{A\ap}$, it can be pulled back to $\PS$, just as in Section \ref{sec:ward}. If we let $B_{\ap\bp}$ denote the self-dual part of the curvature, i.e.
\begin{equation}\label{eq:bdefn}
	B_{\ap\bp}= \partial_{A(\ap} A^A_{\bp)} + A_{A(\ap} A^A_{\bp)},
\end{equation}
then, as we saw in Section \ref{sec:ward}, the Yang-Mills equation is equivalent to
\begin{equation}
	\nabla_{AA'} B^\ap_\bp =0,
\end{equation}
which is the same as
\begin{equation}\label{eq:nablabsymmetry}
	\nabla_{AA'} B_{\bp\cp} = \nabla_{A(A'} B_{\bp\cp)}.
\end{equation}
If we define
\begin{align}
a &= \pi^\ap A_{A\ap} e^A,\notag\\
b &= \frac{\hpi^\ap \hpi^\bp}{(\pi\cdot\hpi)^2} \left( 3 B_{\ap\bp} e^0 + \frac{\hpi^\cp}{\pi\cdot\hpi} \nabla_{A\cp} B_{\ap\bp} e^A \right),\label{eq:abdefns}
\end{align}
Equations \eqref{eq:bdefn} and \eqref{eq:nablabsymmetry} imply that $a$ and $b$ satisfy
\begin{align}
\db_a^2 &= \db a + a\wedge a = \pi^\ap \pi^\bp B_{\ap\bp} e^1\wedge e^2\label{eq:abeqnsinb1}\\
\db_a b &= \frac{\hpi^\ap \hpi^\bp}{(\pi\cdot\hpi)^2} B_{\ap\cp} B^\cp_\bp e^1\wedge e^2.\label{eq:abeqnsinb2}
\end{align}
These can be seen as special cases of Equations \eqref{eq:eomfromb} and \eqref{eq:eomfroma} when $a_0=0$. The latter equations hold equally well in the Minkowski setting, as may be checked directly or using the action principle on $\PS$. They are invariant under the gauge transformations in Equation \eqref{eq:gaugefreedom}, so the forms $a$ and $b$ enjoy enhanced gauge freedom as before. The CSW gauge still makes sense in this context. The gauge condition is $a_2=b_2 =0$, where $a_2$, $b_2$ are now understood to be the components in the $e^2$-direction. We will write out the equations in CSW gauge below.

Expanding $\db_a^2$ into components gives (note $\db_0 e^1 =e^2$)
\[
\db_a^2 = (\db_0 a_1 - \db_1 a_0 + [a_0, a_1]) e^0\wedge e^1 - (\db_2 a_0 - a_1) e^0\wedge e^2 - \db_2 a_1 e^1 \wedge e^2.
\]
Therefore, writing $\db_2$ more suggestively as $\partial_t$, we find that Equation \eqref{eq:eomfromb} becomes, in CSW gauge,
\begin{align}
& \partial_t a_0 = a_1,\notag \\
& \partial_t a_1 = - \int_{L_x} K_{21} b_0^{(1)} K_{12} (\pi_1 \cdot \pi_2)^2\ d^2\pi_1, \label{eq:aeqn}\\
& \db_0 a_1- \db_1 a_0 + [a_0, a_1] =0,\notag
\end{align}
where $H$ is the frame satisfying $(\db_0+a_0)H=0$ and $K_{ij} = H_i H_j^{-1}/ (\pi_i\cdot\pi_j)$ is Green's function for $\db_{a_0}$ as before.

A similar expansion of $\db_a b$ shows that
\begin{align}
& \partial_t b_0 =  b_1,\notag \\
& \partial_t b_1 = -\int_{L_x\times L_x} K_{32}b_0^{(2)} K_{21} b_0^{(1)} K_{13} \> (\pi_1 \cdot \pi_2)^4 \ d^2\pi_1 d^2\pi_2, \label{eq:beqn}\\
& \db_0 b_1 - \db_1 b_0 + [a_0,b_1] - [a_1,b_0] =0.\notag
\end{align}

If we have functions satisfying the equations above, we can recover the Yang-Mills connection on space-time as in Chapter \ref{chap:action}. Concretely at the level of the equations, it works as follows. A gauge transformation by $H$ produces $a'=H^{-1} a H + H^{-1} \db H$, with $a'_0=0$ by the definition of $H$. From the equations for $a$ and $\db_0 H = -a_0 H$,
\begin{align*}
\db_0 a'_1 &= H^{-1} (\db_0 a_1+ [a_0,a_1]) H + H^{-1} a_0 \db_1 H + H^{-1} \db_0 \db_1 H\\
&= \underbrace{H^{-1} (\db_1 a_0) H + H^{-1} a_0 \db_1 H}_{H^{-1} \db_1 (a_0 H)} + H^{-1} \db_0 \db_1 H =0.
\end{align*}
A similar calculation, using $\db_0\db_2=\db_2\db_0 - \db_1$ among other things, gives
\[
\db_0 a'_2= \db_0 (H^{-1} \db_2 H) = -a'_1.
\]
It follows that $\db_0^2 a'_2=0$. On the other hand, the functions $a'_1$ and $a'_2$ have weights 2 and 0 respectively, from the definition of the bases $e^1$ and $e^2$. Hence, they must be of the form
\[
a'_1 = \hat{\bpi}^A \pi^\ap A_{AA'}(x)\quad\text{ and }\quad a'_2 = \frac{\bpi^A \pi^\ap}{\pi\cdot\hpi} A_{AA'}(x).
\]
In other words, $a'_A= \pi^\ap A_{AA'}(x)$, with the form $A_{AA'}(x)$ being the potential on space-time.

\subsection{Iterative solution}

It must be noted that even though $\db_2$ takes a nice form in these coordinates, the equations $\db_2 a_0=\db_2 b_0=0$ unfortunately no longer hold, because of Equation \eqref{eq:db0db2}, or equivalently, $\db_0 e^1 = -e^2$. This means that $a_0$ and $b_0$ evolve non-trivially under $\partial_t$. They cannot be determined straightforwardly from initial data and the solution procedure sketched in Section \ref{sec:eom} is too simple to work. However, we can propose a more sophisticated, iterative version.

Suppose an initial value problem is given for the functions $a_0$, $a_1$, $b_0$ and $b_1$. Initial data for these equations may consist of values of the fields on a space-like hypersurface ${\cal H}$ in $\PS$, e.g.\ $\{t=0\}$ or the past null infinity\footnote{Minkowski space can be conformally compactified by adding a light-cone at infinity. All light rays may be regarded as starting from one part of this light-cone -- the past null infinity ${\cal I}^-$, and ending on another -- the future null infinity ${\cal I}^+$.} ${\cal I}^-$. Note that the last members of Equations \eqref{eq:aeqn} and \eqref{eq:beqn} only involve the initial data and their derivative in directions tangent to ${\cal H}$. As a consistency condition, these equations must, therefore, be satisfied initially. A calculation similar to that in Section \ref{sec:eom} shows that these conditions are preserved by $\partial_t$ and so can be regarded as constraints, which, if satisfied initially, hold for all time.

The iteration procedure is to begin by solving $(\db_0+a_0)H=0$ at time 0, then insert the initial values for the fields (and $H$) into the right-hand sides of the four $\partial_t$ equations and obtain solutions to first order in $t$. After solving $(\db_0+a_0)H=0$ again ($a_0$ now known to order $t$), the first order solutions can be put back into the equations to generate values to second order in $t$, and so on. The constraints should be satisfied at every order in $t$, since they are preserved by $\partial_t$. Formally then, all the fields could be given as power series in $t$. Convergence issues have to be investigated, but will not be dealt with here.

\subsection{Parallel propagator}

An article \cite{Kozameh92} by Kozameh, \emph{et. al.}\ proposes a reformulation of the Yang-Mills equation as a single equation in terms of the `parallel propagator'. In this section, we draw some connection between their work and this thesis.

Given a Yang-Mills connection $A_{AA'}$ on Minkowski space, the parallel propagator $G(x, \pi, \bpi)$ is a gauge group-valued function depending on 6 real variables. As any null direction may be expressed as $l_{AA'}\equiv \bpi_A \pi_\ap/(\pi\cdot\hpi)$, the variable $\pi$ is here regarded as parametrizing the future light-cone of $x$. The function $G$ is defined to be the parallel transport from $x$ to future infinity ${\cal I}^+$ along the null direction given by $\pi$. As a path-ordered integral, it is written,
\begin{equation}
	G(x, \pi, \bpi) = {\cal P} \exp \left( \int_x^\infty A_{A\ap} dx^{A\ap} \right).
\end{equation}
If we parametrize the path from $x$ to ${\cal I}^+$ as $x_{AA'}+t l_{AA'}$, then
\[
G(x, \pi, \bpi) = {\cal P} \exp \left( \int_0^\infty A_{AA'} l^{AA'} dt \right).
\]
It follows from the properties of parallel transport that $\partial_t G = - A_{AA'} l^{AA'} G$, or equivalently
\begin{equation}
	A_{AA'} l^{AA'} = -\partial_t G G^{-1}.
	\label{eq:Al}
\end{equation}
The potential $A$ can be recovered once we note that the knowledge of $\pi^\ap f_\ap(x)$ determines a function $f_\ap(x)$ completely (similarly for $\bpi^A g_A(x)$). This is readily seen by performing an integral as in formula \eqref{eq:piint1}. Alternatively, adopting the approach in \cite{Kozameh92}, we note that
\[
f_\ap = \frac{f\cdot \hpi}{\pi\cdot \hpi} \pi_\ap + \frac{\pi\cdot f}{\pi\cdot \hpi} \hpi_\ap,
\]
where $\pi\cdot f$ is known by assumption. But the first term can also be calculated: let $\partial_0 = -\frac{1}{\pi\cdot \hpi} \hpi_\ap \frac{\partial}{\partial \pi_\ap}$, then
\[
\partial_0 (\pi\cdot f)= \frac{f\cdot \hpi}{\pi\cdot \hpi},
\]
which allows one to reconstruct $f_\ap(x)$. Finding $g_A(x)$ from $\bpi^A g_A(x)$ is entirely the same, except for having to use $\db_0$ instead of $\partial_0$.

Recovering $A(x)$ from Equation \eqref{eq:Al} then involves acting on $-\partial_t G G^{-1}$ with a combination of $\partial_0$ and $\db_0$ and expressing $A_{AA'}$ in terms of the bases $\bpi_A \pi_\ap$, $\hat{\bpi}_A \pi_\ap$, $\bpi_A \hpi_\ap$ and $\hat{\bpi}_A \hpi_\ap$. The final expression, not required here, is given in \cite{Kozameh92}.

Having expressed $A$ entirely in terms of $G$ and its various derivatives, the curvature $F$ can be calculated. The Yang-Mills equation, $\nabla_{BB'} F^{AA'BB'} =0$, is equivalent to 
\begin{equation}
	l_{AA'} \nabla_{BB'} F^{AA'BB'} =0
	\label{eq:ymlF}
\end{equation}
by the argument above. However, to ensure that $-\partial_t G G^{-1}$ really is of the form $ l^{AA'} A_{AA'}(x)$, another condition is needed:
\begin{equation}
	\db^2_0 (\partial_t G G^{-1})=0 \quad\text{ and }\quad \partial^2_0 (\partial_t G G^{-1})=0.
	\label{eq:auxcond}
\end{equation}
This is sufficient to imply that the reconstructed $A$ is truly a function of $x$ only. Equations \eqref{eq:ymlF} and \eqref{eq:auxcond} together allows one to rewrite the Yang-Mills equation in terms of $G$:
\begin{equation}
	\partial_t^3 (\db_0 J) + \partial_t [\partial_t^2 J, \bar{J}] + 2 [\partial_t^2 J, \partial_t \bar{J}] =0,
	\label{eq:kmneqn}
\end{equation}
where $J= G^{-1}\partial_0 G$ and $\bar{J}= G^{-1} \db_0 G$. The paper \cite{Kozameh92} goes on to integrate this equation with respect to $t$ and find an integral equation
\[
\db_0 J - \int_0^\infty dt \left( [\partial_t^2 J, \bar{J}] t + [\partial_t^2 J, \partial_t \bar{J}] t^2 \right) = {\cal C}
\]
where the integration is along $x+tl$ and the `constant' ${\cal C}$ is given by characteristic data on ${\cal I}^+$.

In fact, the parallel propagator appears in our formulation in the guise of $H$ (really $H^{-1}$). Given a Yang-Mills field $A$, the corresponding $a$ and $b$ fields are defined in Equation \eqref{eq:abdefns}. A gauge transformation by the parallel propagator $G$ maps
\[
a_2 \mapsto a'_2 = G^{-1} a_2 G + G^{-1} \db_2 G =0,
\]
thanks to Equation \eqref{eq:Al}. Thus, $a$ is put into the CSW gauge. Recall that the defining property of $H$ is that it provides a transformation from the CSW gauge to space-time gauge (in which $a_0=0$), but $G^{-1}$ evidently accomplishes this. Therefore, in the present context, $G$ and $H^{-1}$ may be identified, up to a change of gauge depending on $x$ only. Note that $a_0$ in the CSW gauge is $a'_0= G^{-1} \db_0 G$, which is the quantity denoted by $\bar{J}$ above.

As the focus of this thesis is on (0,1)-forms, operators like $\partial_0$ ($\partial/\partial z$ in local coordinates) do not appear naturally in our formalism. Therefore, $J$ in Equation \eqref{eq:kmneqn}, in a sense the complex conjugate of $a_0$, has no analogue in our work. Whereas the dependent variables in Equation \eqref{eq:kmneqn} are $J$ and $\bar{J}$ ($=a_0$), those in our equations are $a_0$ and $b_0$ (of which $a_1$ and $b_1$ are derivatives). One difference is that, the auxiliary condition \eqref{eq:auxcond} does not appear to be preserved by $\partial_t$-evolution, unlike the constraints in Equations \eqref{eq:aeqn} and \eqref{eq:beqn}. Therefore, while evolution of $(a_0, b_0)$ from consistent initial conditions leads to solutions of the Yang-Mills equation, as shown previously, it is not known whether Equation \eqref{eq:kmneqn} alone is equivalent to Yang-Mills.

\chapter{Quantum Aspects}\label{chap:quant}

Part of the motivation behind Witten's paper \cite{Witten03} was the simplicity of the tree-level MHV amplitudes. As mentioned in the introduction, it was conjectured from the form of these amplitudes that they may be treated as interaction vertices, joined together by a $1/p^2$ propagator, according to the so-called CSW rules \cite{CSW04}. Since these rules specify how amplitudes are joined up to form amplitudes for larger numbers of external particles, they are called recursion relations. Even simpler recursion relations were discovered by Britto, Cachazo and Feng, which allow one to decompose any gluon tree-level amplitude into combinations of three-point amplitudes. A surprisingly simple proof was given in \cite{BCFW}, using nothing else except properties of a massless field theory. Risager \cite{Risager:2005vk} later showed that CSW rules in fact follow from the BCF relations. However, it is still desirable to obtain an action-based explanation of these rules, not least because the proofs mentioned above only work at tree-level and evidence suggests that the CSW rules also apply to loop amplitudes.

In this chapter, we will examine the quantum implications of our alternative formulation of Yang-Mills theory. We will obtain the Feynman rules that arise from the twistor Lagrangian by finding expressions for the propagators, vertices and external states. These Feynman rules will be seen to lead directly to the CSW rules, when combined with the LSZ reduction formula. Moreover, the `$-++$' amplitude which is not derivable from the MHV rules is present in this framework, in the form of a diagram which apparently vanishes in LSZ reduction, but is in fact non-zero because of special properties of spinor momenta when there are only three external particles.

As we treat Euclidean twistor space $\PT'$ as a $\P^1$-fibration over `physical' space, the quantum theory used is necessarily Euclidean QFT. Despite the absence of a distinguished time coordinate, one justifies the study of scattering processes in Euclidean space by appealing to the analyticity of Green's functions and using Wick rotation to change the signature of space-time. This is uncontroversial in quantum field theory and, for example, Witten's original proposal uses (2,2)-signature. However, from a twistor theorist's point of view, it would be satisfying to obtain a twistor formulation of Yang-Mills in Minkowski space, or even a formulation that does not depend on any particular real structure in twistor space. Whether this is possible is still an open question. In Section \ref{sec:mink}, we mentioned that the action did not make sense in the twistor space over Minkowski space, but it did when taken as living in the primed spin bundle. Therefore, there is a possibility that the appropriate underlying space is the spin bundle rather than twistor space.

\section{Feynman rules}\label{sec:feynman}
\subsection{Propagators}\label{subsec:propagator}

The first ingredient in deriving the Feynman rules is the propagator. Given an action $S[\phi]$ whose quadratic part is $\frac{1}{2}\int \phi D\phi$, with $D$ some self-adjoint operator, the propagator $\Delta(x,y)$ of the theory is the integral kernel of $D^{-1}$. In other words, $\phi(x)=\int \Delta(x,y) J(y)$ solves the equation $D\phi=-J$, or in momentum space, 
\[
\tilde{\phi}(p)=\frac{1}{4\pi^2}\int \tilde{\Delta}(p, -q) \tilde{J}(q) dq.
\]
Accordingly, we add two source terms to the action, $\int_{\PT'} (a\wedge J+b\wedge K)\wedge \Omega$, where $J$ and $K$ are $(0,2)$-forms of weight $-4$ and 0 respectively. The quadratic part of the action plus sources is then, schematically,
\[
\int_{\PT'} b\wedge \db a\wedge\Omega - \frac{\epsilon}{2} \int_{\PT'\times_\E \PT'} b_0^2 + \int_{\PT'} (a\wedge J+b\wedge K)\wedge \Omega.
\]
The propagators will be found by solving
\begin{align}
	\db b &=-J \label{eq:source1}\\
	\db a &=-K+ \epsilon\int_{L_x} b_0 (\pi\cdot\pi')^2 \label{eq:source2}
\end{align}

It will be convenient to express $J$ (and similarly $K$) in components as follows,
\[
J= \frac{1}{2} J_0 e_A \wedge e^A - J_A e^0 \wedge e^A.
\]
This amounts to taking $J$ as the anti-holomorphic Hodge dual of $J_0 e^0+J_A e^A$ with respect to the volume form $\hat{\Omega}$, so that, for example,
\[
a\wedge J= \frac{1}{2}(a_0 J_0+a_A J^A)e^0\wedge e_C \wedge e^C= (a_0 J_0+a_A J^A) \hat{\Omega}.
\]
The expression in the parentheses is of weight 0.

In components, Equation \ref{eq:source1} becomes 
\begin{align}
\db_A b^A &= -J_0 \label{eq:bApropeqn}\\
\db_0 b_A-\db_A b_0 &= J_A.\label{eq:b0propeqn}
\end{align}
If we Fourier transform the $x$ coordinates while noting that $\db J=0$, the equations above may be solved in the CSW gauge, in which $\eta^A b_A=0$. In what follows, the same symbols ($a_0$, $J_0$, etc.) will be used to denote their Fourier transforms.
\begin{align}
b^A &=-\frac{\eta^A}{\eta\pi p} J_0\label{eq:baj}\\
b_0 &=\frac{\eta_A J^A}{\eta\pi p} - \frac{2}{p^2} \delta(\eta\pi p) J_0\label{eq:b0j}
\end{align}
The first line is clear from Equation \eqref{eq:bApropeqn} and the second comes from writing the Fourier transform of Equation \eqref{eq:b0propeqn} as
\begin{align*}
\pi^\ap p_{AA'} b_0 &= -J_A + \db_0 b_A\\ 
&= -J_A -\frac{\eta_A}{\eta\pi p} \db_0J_0 - \eta_A \left(\db_0 \frac{1}{\eta\pi p}\right) J_0\\
&=\underbrace{ -J_A + \frac{\eta_A}{\eta\pi p} \pi^\bp p_{BB'} J^B} -\eta_A \left(\db_0 \frac{1}{\eta\pi p}\right) J_0,\\
&\hspace*{1.3cm} \pi^\ap p_{AA'} \frac{\eta_B J^B}{\eta\pi p}
\end{align*}
where the Fourier transform of $\db_0J_0+\db_AJ^A=0$ and Equation \eqref{eq:antisym} were used; and $\eta\pi p$ is a shorthand for $\eta^A\pi^\ap p_{AA'}$. Finally, to simplify the last term, write $\eta_A$ as
\[
\eta_A=\frac{2(\eta\pi p)}{(\pi\cdot\hpi) p^2}\hpi^\ap p_{AA'}- \frac{2 (\eta\hpi p)}{(\pi\cdot\hpi) p^2} \pi^\ap p_{AA'}.
\]
Then from Equation \eqref{eq:defdeltafn},
\[
\eta_A \left(\db_0 \frac{1}{\eta\pi p}\right) = \pi^\ap p_{AA'} \frac{2}{p^2} \delta(\eta\pi p),
\]
thus leading to Equation \eqref{eq:b0j}.

Equation \eqref{eq:source2} has the same form as Equation \eqref{eq:source1} except for an extra term involving the now known $b_0$. Its solution follows the same pattern as for the $\db b$ equation above.
\begin{align*}
a_A &= \frac{\eta_A}{\eta\pi p}\left(-K_0+ \epsilon\int_{\P^1} \frac{\eta_BJ^B(p,\pi')}{\eta\pi' p} (\pi\cdot\pi')^2 d^2\pi' \right)- \eta_A \frac{2\eta\pi p}{p^2}\ J_0|_{\pi=\eta p},\\
a_0 &= \frac{2}{p^2} \delta(\eta\pi p) \biggl(\qquad \qquad\cdots \quad \cdots \quad \cdots \qquad\qquad \biggr) + \frac{\eta_AK^A}{\eta\pi p}.
\end{align*}

Despite the unwieldy appearance of these formulae, the propagator, written in component form, is fairly simple (all particles taken to be outgoing).
\begin{align}
\langle b_0(p,\pi_1) a_A(q,\pi_2)\rangle &=-\langle b_A(p,\pi_1) a_0(q,\pi_2) \rangle = \delta(p+q) \delta (\pi_1\cdot\pi_2) \frac{\eta_A}{\eta\pi_1 p}\label{eq:prop1}\\
\langle b_0(p,\pi_1) a_0(q,\pi_2)\rangle &= \delta(p+q) \delta (\pi_1\cdot\pi_2) \delta(\eta\pi_1 p) \frac{-2}{p^2}\label{eq:prop2}\\
\langle a_0(p,\pi_1) a_A(q,\pi_2)\rangle &= \epsilon \delta(p+q) \delta(\eta\pi_1 p) \frac{2\eta\pi_2 p}{p^2} \eta_A\label{eq:prop3}\\
\langle a_A(p,\pi_1) a_B(q,\pi_2)\rangle &= \epsilon \delta(p+q) (\pi_1\cdot\pi_2)^2 \frac{\eta_A \eta_B}{(\eta\pi_1 p)(\eta\pi_2 p)}\label{eq:prop4}\\
\langle a_0(p,\pi_1) a_0(q,\pi_2)\rangle &= \langle b_0(p,\pi_1) b_0(q,\pi_2)\rangle = 0 \label{eq:prop5}
\end{align}
As the action depends at most linearly on $b_A$, the propagator $\langle b_A b_B \rangle=0$ trivially. Note that trace and the $\End(E)$ indices on the fields have been suppressed throughout. But they do not play any role in the derivation above and, written out explicitly, $\langle a^\alpha_\beta b^\gamma_\delta\rangle$ will contain a factor of $\delta^\alpha_\delta \delta^\gamma_\beta$.

\begin{figure}[htb]
\begin{align*}
& \epsfig{file=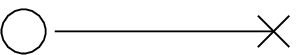} \qquad \delta(\pi_1\cdot\pi_2) \frac{\eta_A}{\eta\pi_1 p}\\
& \epsfig{file=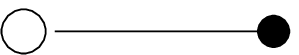} \qquad \delta(\pi_1\cdot\pi_2) \delta(\eta\pi_1 p) \frac{-2}{p^2}\\
& \epsfig{file=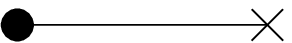} \qquad \epsilon \delta(\eta\pi_1 p) \frac{2\eta\pi_2 p}{p^2} \eta_A \\
& \epsfig{file=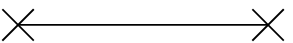} \qquad \epsilon (\pi_1\cdot\pi_2)^2 \frac{\eta_A \eta_B}{(\eta\pi_1 p)(\eta\pi_2 p)}
\end{align*}
	\caption{\emph{Diagrammatic representation of the propagators that arise from the twistor Lagrangian.}}
	\label{fig:prop}
\end{figure}

We will follow convention and propose a pictorial representation of these propagators, akin to internal lines in the usual Feynman diagram. However, because of the mixing of the different fields, we will use lines with different ends instead of the wavy or dashed lines in QCD. In Figure \ref{fig:prop}, the circle represents $b_0$, the cross $a_A$ and the disk $a_0$. When constructing the Feynman rules later, we will require that lines joined together have the same symbols at the end-points. Equations \eqref{eq:prop1}-\eqref{eq:prop4} are then summarised in Figure \ref{fig:prop}, with the delta function $\delta(p+q)$ omitted.

Since we are working in twistor space, isomorphic to $\R^4\times\P^1$, one would expect delta functions $\delta(p+q)$ and $\delta(\pi_1\cdot\pi_2)$, if the theory were local. The $\delta(\pi_1\cdot\pi_2)$ is absent in the $\langle a_0 a_A\rangle$ and $\langle a_A a_B\rangle$ propagators, because of the non-local term $I[a_0,b_0]$. As a result, many integrals over $\P^1$ will have to be performed in computing correlators. Fortunately, many of these are trivial, thanks to delta functions, and there is a trick to simplify the rest. It is to be stressed, however, that the proposed Lagrangian is local in the space-time variables and momentum conservation is not an issue, as borne out by the ubiquity of $\delta(p+q)$ in the propagators.

\subsection{Vertices}\label{subsec:vertices}

Non-quadratic terms in the action come from expanding out $I[a_0,b_0]$, more precisely the part $\int \tr b_1K_{12}b_2K_{21}$. As explained in the Appendix \ref{sec:formulae}, Green's function $K_{12}$ depends non-linearly on the field $a_0$. Its variation with respect to $a_0$ is given by Equation \eqref{eq:varkint}:
\[
\delta K_{12}=-\int_{L_x} K_{13} \delta a_3 K_{32}\ d^2\pi_3,
\]
which enables us to give a series expansion of $K_{12}$ in powers of $a_0$. When $a_0=0$, the bundle is trivial and $K_{12}=\frac{1}{\langle 12\rangle}$, where we use the notation $\langle 12\rangle=\pi_1\cdot\pi_2$. (This will also be used later for contraction between spinor momenta. The context ought to make it clear whether the variable $\pi_\ap$'s or the momenta $\lambda_\ap$'s are involved.) The following shorthand will also be adopted:
\[
\langle i_1 i_2 \cdots i_n\rangle = \langle \pi_{i_1}\cdot\pi_{i_2}\rangle \langle \pi_{i_2}\cdot\pi_{i_3}\rangle \cdots \langle \pi_{i_{n-1}}\cdot \pi_{i_n}\rangle.
\]

The variational formula above implies that
\[
K_{12}=\frac{1}{\langle12\rangle}-\int_{L_x} \frac{1}{\langle 13\rangle} a_3 \frac{1}{\langle 32\rangle} d^2\pi_3 + O(a_0^2).
\]
(Recall that $a_i$ stands for $a_0(x,\pi_i)$, when there is no danger of confusion with the other field $a_A$.) Taking higher variations of Equation \ref{eq:varkint} leads to the expansion
\begin{equation}
K_{12}= \frac{1}{\langle12\rangle} + \sum_{n=3}^\infty (-1)^n \int_{(L_x)^{n}} \frac{a_3\cdots a_n}{\langle 13\cdots n 2\rangle}	\prod_{j=3}^{n} d^2\pi_j.
	\label{eq:kexpansion}
\end{equation}
Inserting this into $\int \tr b_1K_{12}b_2K_{21} (\pi_1\cdot\pi_2)^4$ gives an infinite series of vertices
\begin{align}
(-1)^n  \int_{(L_x)^n} \tr b_1a_3\cdots a_m & b_2 a_{m+1}\cdots a_n\notag\\ 
&\times \frac{\langle 12\rangle^4}{\langle 13\cdots m2(m+1)\cdots n1\rangle} \prod_{k=1}^n d^2\pi_k
\label{eq:vertex}
\end{align}
for $n\geq m\geq 3$. The similarity of these vertices to the MHV amplitude in Equation \ref{eq:MHV} indicates that we are indeed on the right path to deriving the MHV rules of Cachazo, Svrcek and Witten.

Let us represent the $n$-point vertex as follows.
\begin{equation}
	\epsfig{file=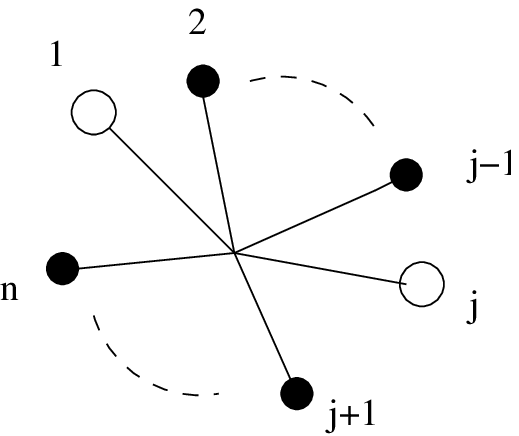, scale=0.8} \quad \displaystyle{\int_{(L_x)^n} \frac{\langle 1j\rangle^4}{\langle 12\cdots n1\rangle}\prod_{k=1}^n d^2\pi_k}
	\label{eq:vertexexp}
\end{equation}

{\bf Note on the coupling constant:} The appearance of the constant $\epsilon$ in the propagators and but not in the vertices is at odds with conventional practice. The remedy is straightforward. To make contact with Yang-Mills theory as presented in physics literature, we let $a\mapsto \epsilon a$ and $S\mapsto S/\epsilon$. As can be readily checked, the effect in the CSW gauge is to remove all appearances of $\epsilon$ from the quadratic terms, while adding a factor of $\epsilon^{n-2}$ to the vertex with $n$ legs\footnote{Strictly speaking, one should scale $a$ to $-\epsilon a$ and get rid of the $(-1)^n$ factor in \eqref{eq:vertex}.}. This constant $\epsilon$ can then be identified with the usual Yang-Mills coupling constant $g$. Thus the expansions in powers of $a$ here are really expansions in powers of the coupling constant. We could have begun with the scaled version of the action, but chose to keep $\epsilon$ as a parameter indicating deviation from the ASD sector, in line with the original exposition in \cite{Mason05}.

\subsection{External states}\label{subsec:extstates}

In order to calculate scattering amplitudes of gluons, one must incorporate the ordinary gauge fields $A_\mu$ in this formalism. The desired relation is provided by Equation \eqref{eq:Ainta},
\begin{equation}
	A_{A\ap}(x) =-2\int_{L_x} H^{-1} (\db_A+ a_A) H\> \frac{\hat{\pi}_\ap}{\pi\cdot\hat{\pi}}\> d^2 \pi.
	\label{eq:Ainta2}
\end{equation}
The aim here, as in the last subsection, is to write $A$ as a power series in $a$. With help from Equation \eqref{eq:hinteqn}, we can expand $H$ in an analogous manner to Equation \eqref{eq:kexpansion},
\[
H(x,\pi_1)= I + \sum_{n=2}^\infty (-1)^{n+1} \int_{(L_x)^{n-1}} \frac{a_2 \cdots a_n}{\langle 12\cdots n\rangle} \frac{\langle o 1\rangle}{\langle o n\rangle} \prod_{k=2}^n d^2 \pi_k,
\]
where, by a slight abuse of notation, $\langle o j\rangle$ stands for $o\cdot \pi_j$. Therefore, the integrand of \eqref{eq:Ainta2} is
\begin{align*}
\left(H^{-1}(\db_A+a_A)H\right)(x,\pi_1) &= \left(I+ \int \frac{a_2}{\langle 12\rangle}\frac{\langle o1\rangle}{\langle o2\rangle}+ \cdots \right)\\
&\quad \times \left[a_A+ (\db_A+a_A) \left(-\int \frac{a_2}{\langle 12\rangle}\frac{\langle o1\rangle}{\langle o2\rangle} + \cdots \right) \right].
\end{align*}
Fourier-transforming to momentum space gives
\begin{align}
	A_{AA'}(p) = & -2 \int a_A(p,\pi_1) \frac{\hat{\pi}_{1\ap}}{\langle 1 \hat{1} \rangle}\> d^2 \pi_1 \notag\\
	&+ 2 \int \pi_1^\bp p_{AB'} \frac{a_0(p,\pi_2)}{\langle 12\rangle} \frac{\langle o1 \rangle}{\langle o2 \rangle} \frac{\hat{\pi}_{1\ap}}{\langle 1 \hat{1} \rangle}\> d^2 \pi_1 d^2 \pi_2\label{eq:Apexpansion}\\
	&-2 \int \frac{[a_0(p',\pi_2),a_A(p-p',\pi_1)]}{\langle 12\rangle} \frac{\langle o1 \rangle}{\langle o2 \rangle} \frac{\hat{\pi}_{1\ap}}{\langle 1 \hat{1} \rangle}\> d^2 \pi_1 d^2 \pi_2 d^4 p'\notag\\
	&+ O(a_0^2, a_A a_0^2)\notag
\end{align}

At this stage, anyone would be excused for balking at the length of the formulae and doubting the utility of our formalism. The amplitude calculations would surely be nothing but a mess? While the vertices are clean and suggestive of the right answer, have we pushed all the complications into the external states? In fact, two ingredients to be used in amplitude calculations will eliminate all but the two leading terms displayed in Equation \eqref{eq:Apexpansion}. The first is a judicious choice of polarization vectors.

Recall, e.g.\ from \cite{Witten03}, how to construct polarization vectors of a given helicity out of spinor variables. Given a null momentum\footnote{The momenta are here taken to be complex, so that $\lambda$ and $\tilde{\lambda}$ are regarded as independent. Formulae specific to Minkowski signature is to be understood as given by analytic continuation.} $p_{AA'}= \lambda_A \tilde{\lambda}_\ap$, the positive and negative helicity polarization vectors can be chosen to be
\begin{equation}
	\varepsilon^+_{AA'}= \frac{\eta_A\tilde{\lambda}_\ap}{\eta\cdot \lambda}\qquad \text{and}\qquad \varepsilon^-_{AA'}= \frac{\lambda_A o_\ap}{\tilde{\lambda}\cdot o}.
	\label{eq:polvec}
\end{equation}
For example, such a $\varepsilon^-$ would produce a field strength proportional to
\[
p_{AA'}\varepsilon^-_{BB'}- p_{BB'}\varepsilon^-_{AA'}=\lambda_A \lambda_B \epsilon_{A'B'},
\]
which is anti-self-dual and so of negative helicity.

The constant spinors $\eta_A$ and $o_\ap$ in Equation \eqref{eq:polvec} can be arbitrary, but it is no coincidence that we have chosen those spinors appearing respectively in Equation \eqref{eq:csw}, the definition of the CSW gauge, and in Equation \eqref{eq:hinteqn}, the definition of $H$. When one computes helicity amplitudes, the states $A^\pm=\varepsilon^\pm_{AA'} A^{AA'}$ will be inserted into the correlators, instead of $A_{AA'}$. The CSW gauge condition $\eta^A a_A=0$ implies that the expansion of $A^+$ contains no term involving $a_A$. On the other hand, the terms depending on $p_{AA'}$ in $A^-$ are contracted with $\lambda_A$ and will therefore vanish, contributing 0 to on-shell scattering amplitudes. Thus we have
\begin{equation}
	A^+(p)= 2\int \langle 1 \tilde{\lambda} \rangle \frac{a_0(p,\pi_2)}{\langle 12 \rangle} \frac{\langle o1\rangle}{\langle o2\rangle} \frac{\langle \hat{1} \tilde{\lambda}\rangle}{\langle 1\hat{1} \rangle} d^2 \pi_1 d^2 \pi_2 + O(a_0^2)
	\label{eq:A+p}
\end{equation}
and
\begin{align}
	A^-(p)= & \ 2\int \frac{\langle o\hat{1} \rangle}{\langle 1 \hat{1} \rangle} \frac{\lambda^A a_A(p,\pi_1)}{\langle \tilde{\lambda} o \rangle} d^2 \pi_1\notag\\
	& +2 \int \frac{[a_0(p',\pi_2),\lambda^A a_A(p-p',\pi_1)]}{\langle 12\rangle \tilde{\lambda}\cdot o} \frac{\langle o1 \rangle}{\langle o2 \rangle} \frac{\langle o \hat{1} \rangle}{\langle 1\hat{1}\rangle} d^2 \pi_1 d^2\pi_2 d^4 p'\notag\\
	& +O(\lambda^A a_A a_0^2).\label{eq:A-p}
\end{align}

Pictorially, these may be represented as

\vspace{0.8cm}
{\parindent=1cm
{\Large $\displaystyle{A^+:}\quad$} \epsfig{file=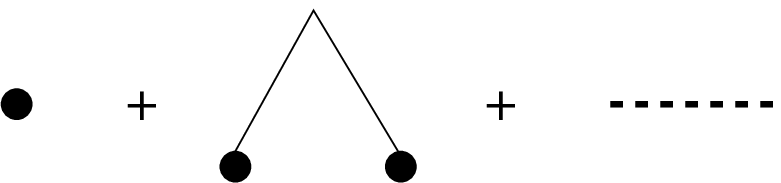}
}
\vspace{0.8cm}

{\parindent=1cm
{\Large $\displaystyle{A^-:}\quad$} \epsfig{file=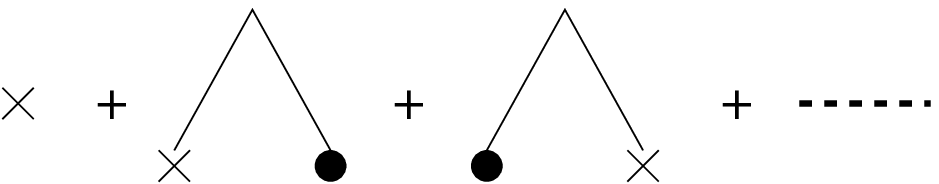}
}
\vspace{0.8cm}

These still involve more terms than necessary for computing amplitudes. The second ingredient that simplifies the form of external states is LSZ reduction, discussed in the next section.

\section{Tree-level scattering}\label{sec:tree}
\subsection{LSZ reduction}

Using results from the last section, we can follow the standard prescription and calculate Green's functions perturbatively by use of Feynman diagrams. However, we are only interested in scattering amplitudes. The link is provided by the well-known Lehmann-Symanzik-Zimmermann (LSZ) reduction formula. The scattering amplitude of $n$ gluons, with momenta $p_i$ and helicities $h_i$, is given by
\begin{equation}
	{\cal A}(p_1,h_1;\cdots;p_n,h_n) = \lim_{\text{All }p_i^2\to 0} \left(\prod_{j=1}^n p_j^2 \right) \langle A_1^{h_1} \cdots A_n^{h_n}\rangle.
	\label{eq:lsz}
\end{equation}
In the usual $\phi^4$-theory, for example, this merely amounts to `amputating the external legs', that is, setting to 1 the propagators connecting external fields to internal vertices. The implications for our formalism are further-reaching. First, those propagators without $p^2$ in the denominator cannot appear in the external legs. More precisely, $\langle b_0 a_A\rangle$ and $\langle a_A a_B \rangle$ do not have $1/p^2$ poles and, when multiplied by $p^2$, will vanish in the limit $p^2\to 0$. Therefore, the two legs \epsfig{file=propagator1.eps, scale=0.8} and \epsfig{file=propagator4.eps, scale=0.8} cannot appear between an external field and an internal vertex.

To be rigorous, we have to consider the possibility that $p^2$ in the numerator is cancelled by a singularity not of the form $1/p^2$. Examining closely the integrals to be evaluated, we see that terms such as $\tilde{\lambda}_i\cdot \tilde{\lambda}_j$ may appear in the denominator. For generic choices of external momenta, these will not vanish, except for the special case when there are only three particles scattering. As will be demonstrated in section \ref{subsec:missing}, momentum conservation may force a term like $\tilde{\lambda}_i\cdot \tilde{\lambda}_j$ to vanish when $p$ goes on shell; and a diagram having \epsfig{file=propagator1.eps, scale=0.8} as an external leg has to be taken into account.

Second, let us consider terms involving more than one $a$ field in the expansion for $A^\pm$, Equations \eqref{eq:A+p} and \eqref{eq:A-p}. A typical diagram involving a second-order term in $A^+$ might look like the following.

\vspace{0.3cm}
\begin{center}
\epsfig{file=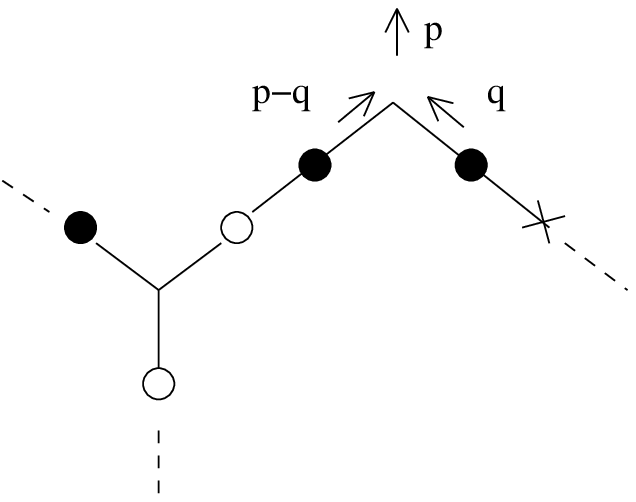}	
\end{center}
\vspace{0.1cm}

Here, $p$ is the external momentum associated with the $A^+$ term and the momentum $q$ is determined by momentum conservation, since we are only considering tree-level scattering. The propagators in the external legs would therefore produce, at most, poles of the form $1/(p-q)^2$. On being multiplied by $p^2$, this again vanishes in the limit that $p$ becomes null. This means that only the first terms in Equations \eqref{eq:A+p} and \eqref{eq:A-p} are necessary for calculations of scattering amplitudes and $A^\pm$ may be represented simply as

\begin{center}
{\large $\displaystyle{A^+:}\quad$} \epsfig{file=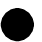}\qquad	and\qquad {\large $\displaystyle{A^-:}\quad$} \epsfig{file=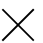}
\end{center}

We are now ready to reproduce the familiar results about gluon scattering at tree-level.

\subsection{Vanishing of $(+\cdots +)$ and $(-\cdots - \pm)$ amplitudes}

The vertices all contain two $b_0$'s and one or more $a_0$'s. Two vertices can only be connected by a \epsfig{file=propagator2.eps,scale=0.7} propagator and, therefore, one can never eliminate $a_0$ from the legs to be joined to an external field. If all the gluons have positive helicity, then inevitably two $a_0$'s will be connected by a propagator. But $\langle a_0 a_0 \rangle=0$. Hence the amplitudes vanish.

The all-minus case is similar. There, a \epsfig{file=propagator1.eps,scale=0.7} propagator will inevitably appear in an external leg, but as explained above, it will have 0 contribution to the amplitude by LSZ reduction.

Since vertices are always joined by \epsfig{file=propagator2.eps,scale=0.7} lines, simple combinatorial considerations show that the number of $b_0$'s, or $\bigcirc$, to be attached to external states is equal to the number of vertices plus 1. When one gluon has positive helicity and the rest negative helicity, it is again unavoidable to have \epsfig{file=propagator1.eps,scale=0.7} in an external leg. Therefore, the $(-\cdots -+)$ amplitude is zero.

\subsection{Parke-Taylor formula}

\begin{figure}[htb]
	\begin{center}
	\epsfig{file=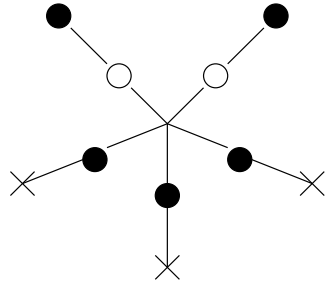}
	\end{center}
	\caption{\emph{The diagram for five-gluon MHV scattering.}}
	\label{fig:mhv5}
\end{figure}

When all but two gluons have negative helicity, we aim to reproduce the well-known formula for MHV amplitudes. For a non-zero contribution, only the two external $A^+$, or $\bullet$, can be joined to a free $b_0$ in a vertex. As just explained, the number of such $b_0$ is the number of vertices plus one. Hence, only one vertex can appear overall and Figure \ref{fig:mhv5} shows a typical diagram.

There are other contributions from vertices with different arrangements of $b_0$ and $a_0$, thus leading to a sum over inequivalent cyclic orderings. The total (tree) amplitude of $n$-gluon scattering will be
\[
{\cal A}_n^\text{tree} = \sum_{\sigma\in S_n\backslash Z_n} \tr \left(T^{a_{\sigma(1)}} \cdots T^{a_{\sigma(n)}} \right) A_n^\text{tree} (p_{\sigma(1)},h_{\sigma(1)};\cdots;p_{\sigma(n)}, h_{\sigma(n)}),
\]
where $p_i$ and $h_j$ are the momenta and helicities respectively, and $\sigma$ is a permutation modulo cyclic ordering. In order to calculate the factors $A_n$, let us evaluate the diagram piece by piece. First, take the $a_0 b_0$ leg, with an outgoing momentum $p$.

\begin{center}
\epsfig{file=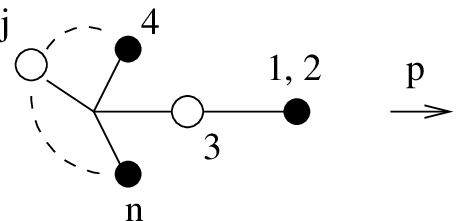, scale=1.5}
\end{center}

Equations \eqref{eq:vertex} and \eqref{eq:A+p} display the precise $\pi$ integral that has to be performed:
\[
\lim_{p^2\to 0}\ p^2 \int \frac{\langle 1\tilde{\lambda} \rangle \langle o1\rangle \langle \hat{1}\tilde{\lambda} \rangle}{\langle 12\rangle \langle o2\rangle \langle 1\hat{1} \rangle}\ \langle a_0(p,\pi_2) b_0(q,\pi_3) \rangle \ \frac{\langle 3j \rangle^4}{\langle 34\cdots n3 \rangle} \prod_{k=1}^n d^2 \pi_k\ d^4 q.
\]
The picture above shows the indices of the dummy $\pi$ variables corresponding to each node. The $p^2$ factor is cancelled by the $1/p^2$ in the propagator, c.f.\ Equation \eqref{eq:prop2}. The delta functions force $\pi_{2A'} =\pi_{3A'} =(\eta\cdot\lambda) \tilde{\lambda}_\ap$ and the integral becomes
\[
-2\int \frac{\langle o1 \rangle \langle \hat{1}\tilde{\lambda} \rangle}{\langle o \tilde{\lambda} \rangle \langle 1 \hat{1} \rangle} \frac{\langle \tilde{\lambda} j \rangle^4}{\langle \tilde{\lambda} 4\cdots n \tilde{\lambda} \rangle} d^2 \pi_1 \prod_{k=4}^n d^2 \pi_k.
\]
Now the formula \eqref{eq:piint1} can be used to see that
\[
-2\int \frac{\langle o1 \rangle \langle \hat{1}\tilde{\lambda} \rangle}{\langle o \tilde{\lambda} \rangle \langle 1 \hat{1} \rangle} d^2 \pi_1 =1.
\]
Thus, the effect of the (amputated) external leg \epsfig{file=propagator2.eps, scale=0.7} is to replace the $\pi$ variable ($\pi_3$ in the example below) associated with the $\bigcirc$ in the vertex with $\tilde{\lambda}$, like a delta function:
\[
\int \frac{\langle 3j \rangle^4}{\langle 34\cdots n3 \rangle} \prod_{k=3}^n d^2 \pi_k \quad\longrightarrow\quad \int \frac{(\tilde{\lambda}\cdot \pi_j)^4}{\langle \tilde{\lambda} 4\cdots n \tilde{\lambda} \rangle}\prod_{k=4}^n d^2 \pi_k.
\]

The case of \epsfig{file=propagator3.eps, scale=0.7} is similar. We label the variables as below.

\begin{center}
\epsfig{file=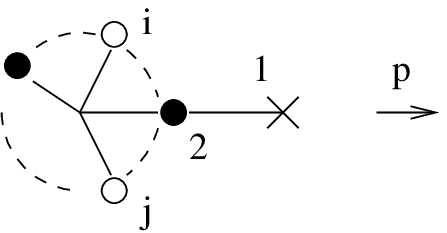, scale=1.5}
\end{center}

The integral is carried out in the same fashion as before.
\begin{align*}
& \lim_{p^2\to 0}\ p^2 \int \frac{\langle o\hat{1} \rangle}{\langle \tilde{\lambda} o \rangle \langle 1 \hat{1} \rangle} \ \lambda^A \langle a_A(p,\pi_1) a_0(q,\pi_2) \rangle \ \frac{\langle ij\rangle^4}{\langle 23\cdots n2\rangle} \prod_{k=1}^n d^2 \pi_k\\
&= -2 \int \frac{\langle o\hat{1} \rangle \langle 1 \tilde{\lambda} \rangle}{\langle \tilde{\lambda} o \rangle \langle 1 \hat{1} \rangle} \ \frac{\langle ij \rangle^4}{\langle \tilde{\lambda} 3\cdots n \tilde{\lambda}\rangle} d^2 \pi_1 \prod_{k=3}^n d^2 \pi_k\\
&= \int \frac{\langle ij\rangle^4}{\langle \tilde{\lambda} 3\cdots n \tilde{\lambda} \rangle} \prod_{k=3}^n d^2 \pi_k.
\end{align*}
We see that the external leg \epsfig{file=propagator3.eps, scale=0.7} also has the same effect as a delta function setting the $\pi$ associated to $\bullet$ to be $\tilde\lambda$.

The final result for the (mostly minus) MHV scattering amplitude is now clear. We recover the Parke-Taylor formula. (See the end of Subsection \ref{subsec:vertices} for discussion about the coupling constant factor.)
\[
A_n^\text{tree} (- \cdots \stackrel{i}{+} \cdots \stackrel{j}{+} \cdots -) = \epsilon^{n-2} \frac{(\tilde\lambda_i \cdot \tilde\lambda_j)^4}{\prod_{k=1}^n (\tilde\lambda_k \cdot \tilde\lambda_{k+1})}.
\]

\subsection{The `missing' $+--$ amplitude}\label{subsec:missing}

One problem with the MHV diagrams has long been the absence of the $+--$ amplitude\footnote{The reader is reminded our helicity assignments are opposite to those commonly found in the articles on twistor strings.}, which is non-zero when external momenta are complex, or in (2,2) signature. Since the twistor and ordinary Yang-Mills Lagrangians have been shown to be classically equivalent, we ought to be able to recover the $+--$ amplitude in our formalism. The diagram below seems to vanish at first sight, as the external leg \epsfig{file=propagator1.eps, scale=0.7} contains no $1/p^2$ pole and so is apparently killed in LSZ reduction.

\begin{figure}[htb]
	\begin{center}
	\epsfig{file=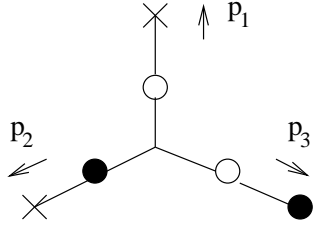}
	\end{center}
	\caption{\emph{The diagram which survives LSZ reduction and reproduces the $+--$ three-point amplitude.}}
	\label{fig:missing}
\end{figure}

However, when there are only three external particles, momentum conservation forces a special relationship between the spinor momenta. Contracting the equation
\begin{equation}
	\lambda_1 \tilde{\lambda}_1+ \lambda_2 \tilde{\lambda}_2 + \lambda_3 \tilde{\lambda}_3 =0
	\label{eq:3momconserv}
\end{equation}
with $\lambda_1$ gives
\begin{equation}
	(\lambda_1 \cdot \lambda_2) \tilde{\lambda}_2 = (\lambda_3\cdot \lambda_1) \tilde{\lambda}_3,
	\label{eq:lam2lam3}
\end{equation}
that is, $\tilde{\lambda}_2$ and $\tilde{\lambda_3}$ are parallel unless both sides of the equation are zero. Similarly, we find
\begin{align}
(\lambda_1 \cdot \lambda_2) \tilde{\lambda}_1 &= (\lambda_2\cdot \lambda_3) \tilde{\lambda}_3,\label{eq:lam1lam3}\\
(\lambda_3 \cdot \lambda_1) \tilde{\lambda}_1 &= (\lambda_2\cdot \lambda_3) \tilde{\lambda}_2.\label{eq:lam1lam2}
\end{align}
Therefore, either all the $\tilde{\lambda}_i$'s or all the $\lambda_i$'s are proportional to each other.

Let us write down the integral which the diagram above represents, recalling that the \epsfig{file=propagator3.eps, scale=0.7} and \epsfig{file=propagator2.eps, scale=0.7} legs have the same effect as delta functions.
\begin{align}
	& \int \frac{\langle 1 \tilde{\lambda}_3\rangle^4}{\langle 1 \tilde{\lambda}_2 \tilde{\lambda}_3 1\rangle} \langle b_0 ( -p_1,\pi_1) a_A(p_1,\pi_2)\rangle \lambda_1^A \frac{\langle o \hat{2}\rangle}{\langle \tilde{\lambda}_1 o \rangle \langle 4 \hat{4}\rangle} d^2\pi_1 d^2\pi_2\notag \\
	=& \int \frac{\langle 1 \tilde{\lambda}_3 \rangle^3}{\langle 1 \tilde{\lambda}_2 \rangle \langle \tilde{\lambda}_2 \tilde{\lambda}_3 \rangle} \frac{1}{\langle 1 \tilde{\lambda}_1 \rangle} \frac{\langle o\hat{1} \rangle}{\langle \tilde{\lambda}_1 o \rangle \langle 1 \hat{1} \rangle} d^2 \pi_1.
	\label{eq:missingint1}
\end{align}
Now, either the $\lambda_i$'s are parallel and the amplitude is zero, or the $\tilde{\lambda}_i$'s are parallel and the term $\langle \tilde{\lambda}_2 \tilde{\lambda}_3 \rangle$ in the denominator represents a singularity in \eqref{eq:missingint1}. Since momentum conservation also implies that
\[
p_1^2 = (p_2+p_3)^2 = 2p_2\cdot p_3= 2(\lambda_2\cdot\lambda_3)(\tilde{\lambda}_2 \cdot \tilde{\lambda}_3),
\]
multiplying \eqref{eq:missingint1} by $p_1^2$ gives
\[
2\frac{\lambda_2\cdot \lambda_3}{\langle \tilde{\lambda}_1 o \rangle} \int \frac{\langle 1 \tilde{\lambda}_3\rangle^3 \langle o \hat{1} \rangle}{\langle 1 \tilde{\lambda}_2 \rangle \langle 1 \tilde{\lambda}_1 \rangle \langle 1 \hat{1} \rangle} d^2 \pi_1.
\]
Using the relations \eqref{eq:lam2lam3}-\eqref{eq:lam1lam2} (and \eqref{eq:piint1} again), we find
\[
2\int \frac{\langle 1 \tilde{\lambda}_1\rangle \langle o \hat{1} \rangle}{\langle \tilde{\lambda}_1 o \rangle \langle 1 \hat{1} \rangle} \frac{(\lambda_1\cdot \lambda_2)^3}{(\lambda_2\cdot \lambda_3) (\lambda_3 \cdot \lambda_1)} d^2 \pi_1 = \frac{(\lambda_1\cdot \lambda_2)^3}{(\lambda_2\cdot \lambda_3) (\lambda_3 \cdot \lambda_1)},
\]
which is precisely the correct form of the $--+$ amplitude.

\subsection{CSW rules}

For a general tree-level scattering amplitude, we can evaluate all the possible Feynman diagrams followed by LSZ reduction. Arguments in the previous sections imply that the process can be made very simple. First of all, only \epsfig{file=propagator3.eps, scale=0.7} and \epsfig{file=propagator2.eps, scale=0.7} in external legs contribute non-trivially to scattering amplitude, and act as delta functions setting the $\pi$ variables to the appropriate spinor momenta. Therefore, the $\pi$ integrals associated with external legs in the expression \eqref{eq:vertexexp} for a vertex are trivial. To see the effect of internal lines, let us look at the following example, involving only one internal propagator.

\begin{figure}[htb]
	\begin{center}
	\epsfig{file=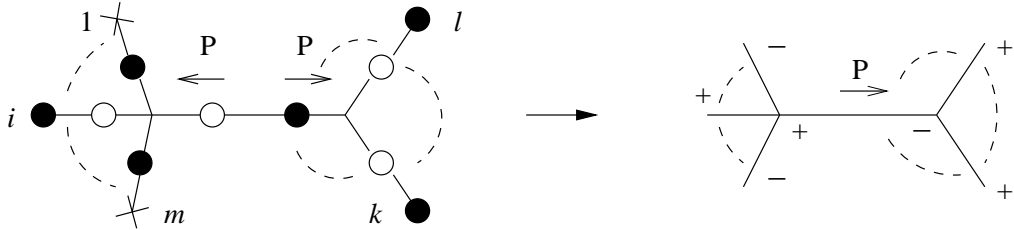}
	\end{center}
	\caption{\emph{A typical Feynman diagram (left) having two internal vertices in our formalism can equivalently be represented by a simplified diagram (right), which is identical to an MHV diagram of Cazhazo, Svrcek and Witten.}}
	\label{fig:csw}
\end{figure}

The left-hand side diagram in Figure \ref{fig:csw} exhibits all the details of a Feynman diagram in our formalism, but since we know the outcome of evaluating all the external legs, we could equally adopt the simplified diagram on the right-hand side, which just uses $+$ and $-$ signs to indicate the fields involved. The internal momentum $P$ is determined by momentum conservation. The integral represented in Figure \ref{fig:csw} is the following (with $i$ standing for $\tilde{\lambda}_i$),
\begin{align*}
& \int \frac{\langle i \pi_1\rangle^4}{\langle \pi_1 1\cdots m \pi_1\rangle} \underbrace{\langle b_0 (-P, \pi_1) a_0 (P,\pi_2) \rangle}_{-2 \delta(\pi_1\cdot \pi_2) \delta(\eta\pi_1 P)/P^2} \frac{\langle k l \rangle^4}{\langle \pi_2 (m+1)\cdots n \pi_2 \rangle} \ d^2\pi_1 d^2\pi_2\\
=& \frac{\langle i (\eta P)\rangle^4}{\langle (\eta P) 1\cdots m (\eta P)\rangle} \frac{-2}{P^2} \frac{\langle k l \rangle^4}{\langle (\eta P) (m+1) \cdots n (\eta P) \rangle},
\end{align*}
where $(\eta P)^\ap$ denotes the spinor $\eta_A P^{AA'}$.

Up to a constant factor\footnote{Unfortunate factors of 2 and $-1$ appear in some formulae, compared to those in the literature. These could be absorbed into re-scalings of $a$ or $\epsilon$.}, the last expression is precisely the CSW prescription for joining up MHV amplitudes into MHV diagrams. When more vertices are involved, it is easily seen that they are joined up in the same way as above. Therefore, we see that the Feynman rules that follow from the twistor Lagrangian reproduce the CSW rules in \cite{CSW04}.

The independence of the amplitudes of $\eta$ is not clear from the original CSW formulation and the authors made non-rigorous arguments, based on integrating over the choice of $\eta$. In our approach, the spinor $\eta$ represents a gauge choice and the gauge invariance of the quantum theory of the twistor Lagrangian ensures that the amplitudes are independent of the choice of $\eta$.

\section{Relation to canonical MHV Lagrangian}\label{sec:mhvlagrangian}

Mansfield initiated an alternative approach to deriving the MHV diagrams from a form of the Yang-Mills Lagrangian \cite{Mansfield:2005yd}. Later developed by him, Morris and others, the idea is to find a field transformation which is canonical and which, by construction, leads to MHV-type vertices \cite{Ettle:2006bw, Ettle:2007qc}. In light-cone gauge, two unphysical degrees of freedom in the gauge field $A$ can be integrated out, leaving two components ${\cal A}$ and $\bar{\cal A}$, which correspond to positive and negative helicity states respectively. The Yang-Mills Lagrangian could be written schematically as
\[
{\cal L}={\cal L}_{+-}+{\cal L}_{++-}+{\cal L}_{--+}+{\cal L}_{++--},
\]
where one $+$ (resp.\ $-$) sign indicates the presence of one factor of ${\cal A}$ (resp.\ $\bar{\cal A}$). The first two terms constitute the self-dual sector and would only produce tree-level diagrams having one negative helicity external state, with the rest positive. Since these vanish on shell, one may reasonably expect to be able to define new fields ${\cal B}$ and $\bar{\cal B}$ such that the self-dual sector becomes free:
\[
{\cal L}_{+-} [{\cal A}, \bar{\cal A}] +{\cal L}_{++-} [{\cal A}, \bar{\cal A}] = \tilde{\cal L}_{+-} [{\cal B}, \bar{\cal B}].
\]
This was achieved and expressions of ${\cal A}$ in terms of $\cal B$ were found (Equations (2.15) and (2.16) in \cite{Ettle:2007qc}). The old field $\cal A$ is a functional purely of $\cal B$ by construction, with $\bar{\cal A}$ a functional of one factor of $\bar{\cal B}$ and any number of $\cal B$. The original Yang-Mills Lagrangian is readily seen to become\footnote{Mansfield and others take MHV to mean `mostly $+$', opposite to the convention adopted in this thesis. The reader is again advised to bear this ${\mathbb{Z}}_2$ symmetry in mind when comparing the two approaches.}
\[
{\cal L}=\tilde{\cal L}_{-+} + \tilde{\cal L}_{--+} + \tilde{\cal L}_{--++} + \tilde{\cal L}_{--+++} + \cdots
\]
in terms of the $B$ fields.

As we saw previously in this chapter, the twistor Lagrangian gives rise to interaction vertices involving two $b_0$ and any number of $a_0$. Perhaps inevitably, given the similarities, there is some relation between the two Lagrangian-based approaches to MHV diagrams. First, the light-cone co-ordinates $(\check{p}, \hat{p}, \tilde{p}, \bar{p})$ in \cite{Ettle:2007qc} are expressed in the spinor notation as follows. Choose two spinors
\[
\eta_A=(0,1),\quad o_\ap=(0,1),
\]
then, letting $(x^0,x^1,x^2,x^3)\to (x^0, ix^1, ix^2, ix^3)$ in formula \eqref{eq:xmutoxaa}, one sees that
\begin{align*}
\check{p}&= \hat{\eta}^A \hat{o}^\ap p_{AA'},\qquad \hat{p}= \eta^A o^\ap p_{AA'}\\
\tilde{p}&= \eta^A \hat{o}^\ap p_{AA'}, \qquad \bar{p}= \hat{\eta}^A o^\ap p_{AA'}.
\end{align*}
In particular, the two fields $\cal A$ and $\bar{\cal A}$ can be written as
\[
{\cal A}= \eta^A \hat{o}^\ap A_{AA'}\quad \text{ and }\quad \bar{\cal A}= \hat{\eta}^A o^\ap A_{AA'}.
\]
The two brackets $(\ ,\ )$ and $\{\ ,\ \}$ defined in \cite{Ettle:2007qc} translate into
\[
(12)= (\eta p_1)\cdot (\eta p_2), \qquad \{12\} = (op_1)\cdot (op_2),
\]
where $(\eta p)$ denotes the spinor $\eta_A p^{AA'}$ (similarly for $(op)$).

Equation \eqref{eq:Apexpansion} displays the expansion of $A_{AA'}(p)$. Each term of the series involves one factor of $a_A$ and any power of $a_0$. In the CSW gauge, $\eta^A a_A=0$, so the component ${\cal A}=\eta^A \hat{o}^\ap A_{AA'}$ will only contain factors of $a_0$. This is the analogy of $\cal A$ being a functional of $\cal B$ only in the canonical Lagrangian formalism. Going into more details, let us recall the expansion of $H$ in powers of $a_0$ given in Section \ref{subsec:extstates}.
\begin{equation}
	H(x,\pi_1)= I + \sum_{n=2}^\infty (-1)^{n+1} \int_{(L_x)^{n-1}} \frac{a_2 \cdots a_n}{\langle 12\cdots n\rangle} \frac{\langle o 1\rangle}{\langle o n\rangle} \prod_{k=2}^n d^2 \pi_k.
	\label{eq:Hexpansion}
\end{equation}
Similarly, we can derive an expansion for $H^{-1}$ from the equation $\db_0 H^{-1}= H^{-1} a_0$.
\begin{equation}
	H^{-1}(x,\pi_1) = I + \sum_{n=2}^\infty \int_{(L_x)^{n-1}} \frac{a_2 \cdots a_n}{\langle 2\cdots n1\rangle} \frac{\langle o 1\rangle}{\langle o 2\rangle} \prod_{k=2}^n d^2 \pi_k.
	\label{eq:Hinvexpansion}
\end{equation}
Note that inverting the series \eqref{eq:Hexpansion} using the binomial theorem would lead to a far messier expression for $H^{-1}$, which, however, would collapse into Equation \eqref{eq:Hinvexpansion} by virtue of spinor algebra identities.

We can now write down the expansion of ${\cal A}=\eta^A \hat{o}^\ap A_{AA'}$ in $a_0$ using Equation \eqref{eq:Ainta2}. To simplify the appearance of the formulae, let $\mu_i^\ap = \eta_A p_i^{AA'}$ and $\langle o 1 \rangle = o\cdot \pi_1$, $\langle 123 \rangle = (\pi_1 \cdot \pi_2)(\pi_2\cdot \pi_3)$, etc.\ as before. Further, denote $a_0(p_k, \pi_l)$ by $a_{k,l}$. After Fourier transform,
\begin{align}
{\cal A}(p_1) &= -2\int H^{-1}(p_2,\pi_1) \langle 1\mu_3\rangle H(p_3,\pi_1) \> \frac{\langle \hat{1} \hat{o}\rangle}{\langle 1 \hat{1}\rangle} \delta(p_1-p_2-p_3) \> d^2 \pi_1 d^4 p_2 d^4 p_3 \notag \\
&= 2 \int \langle 1\mu_1 \rangle \frac{a_{1,2}}{\langle 12\rangle} \frac{\langle o1 \rangle \langle \hat{1} \hat{o} \rangle}{\langle o2\rangle \langle 1 \hat{1}\rangle} d^2\pi_1 d^2 \pi_2 \notag \\
& \quad + 2\int \biggl(\frac{a_{2,2} \langle o1 \rangle}{\langle 12\rangle \langle o2\rangle} \langle 1\mu_3 \rangle \frac{a_{3,3} \langle o1 \rangle}{\langle 13\rangle \langle o3 \rangle} - \langle 1\mu_1 \rangle \frac{a_{2,2} a_{3,3} \langle o1 \rangle}{\langle 123 \rangle \langle o3 \rangle} \biggr) \notag \\
& \hspace{4cm}  \times \frac{\langle \hat{1} \hat{o}\rangle}{\langle 1 \hat{1}\rangle}\ \delta(p_1-p_2-p_3)\ \prod_{j=1}^3 d^2\pi_j\ d^4 p_2 d^4 p_3 \notag \\
& \quad +\ \cdots \cdots.\label{eq:calAexpansion}
\end{align}
Now $a_0(x,\pi)$ satisfies the equation of motion $\db_2 a_0 = \eta^A \pi^\ap \partial_{AA'} a_0 =0$, whose Fourier transform is $(\eta\pi p) a_0(p,\pi)=0$. Therefore, $a_0(p,\pi)$ is of the form $a_0(p) \delta(\eta\pi p)$. (Alternatively, note that any propagator between $a_0$ and another field contains a factor of $\delta(\eta\pi p)$.) Putting this into the formula above, we see that the first term becomes
\[
2\int a_0(p_1) \frac{\langle o1 \rangle \langle \hat{1}\hat{o} \rangle}{\langle o\mu \rangle \langle 1\hat{1} \rangle} d^2\pi_1 = - \frac{a_0(p_1)}{\hat{p}_1},
\]
where $\hat{p}= \langle o\mu \rangle= \eta^A o^\ap p_{AA'}$ in the notation of \cite{Ettle:2007qc}. The second term requires a little algebra first. After the delta functions are integrated out, the terms in the parentheses turn into
\[
\frac{a_0(p_2) a_0(p_3) \langle o1\rangle}{\langle 1\mu_2\rangle \langle o\mu_3 \rangle} \left( \frac{\langle o1 \rangle}{\langle o\mu_2 \rangle} - \frac{\langle 1 \mu_1 \rangle}{\langle \mu_2\mu_3 \rangle} \right) = - \frac{\hat{p}_1 \, a_0(p_2) a_0(p_3)}{\hat{p}_2 \hat{p}_3 \langle \mu_2\mu_3\rangle} \langle o1\rangle,
\]
where the Schouten identity and momentum conservation $\mu_1=\mu_2+\mu_3$ have been used. The $\pi_1$ integral is then easily done and the second term in the expansion \eqref{eq:calAexpansion} becomes, therefore,
\[
\int \frac{\hat{p}_1\, a_0(p_2) a_0(p_3)}{(23) \hat{p}_2 \hat{p_3}}\> \delta(p_1-p_2-p_3) d^4 p_2 d^4 p_3 ,
\]
where $(23)=\langle \mu_2 \mu_3\rangle$ as explained before.

On the other hand, the expansion of $\cal A$ in terms of the newly defined $\cal B$ field is given in \cite{Ettle:2007qc} as
\[
{\cal A}(p_1)= B(p_1) - \int \frac{\hat{p}_1}{(23)} B_2 B_3 \delta(p_1 + p_2 + p_3) d^4 p_2 d^4 p_3 + \cdots.
\]
Comparing the two expansions, it is clear that the field $B(p)$ in the canonical MHV formalism can be identified with $a_0(p)/\hat{p}$ (up to a constant multiple) in our formalism\footnote{Rutger Boels first pointed out this correspondence in an unpublished note.}. Under this identification, the two expansions of $\cal A$ can be checked to agree to all orders. One simply has to apply repeatedly the argument above for the $O(a_0^2)$ term.

It remains to identify the $\bar{\cal B}$ field. Each term in the expansion of $\bar{\cal A} = \hat{\eta}^A o^\ap A_{AA'}$ involves one factor of $\hat{\eta}^A a_A$. Recall that the equation of motion in CSW gauge entails $\eta^A \db_A b_0=0$ and
\[
(\eta^A \db_A) (\hat{\eta}^B a_B) = \int b_0 \, (\pi_1\cdot \pi_2)^2 d^2 \pi_2 + O(a_0 b_0).
\]
Hence, in momentum space, $b_0(p,\pi)= b_0(p) \delta(\eta\pi p)$ and
\[
\hat{\eta}^A a_A = (\eta\pi p) b_0(p) + O(a_0 b_0).
\]
The leading term in the expansion for $\bar{\cal A}$ can then be calculated
\[
2 \int \hat{\eta}^A a_A(p,\pi_1) \frac{\langle o \hat{1}\rangle}{\langle 1\hat{1} \rangle} d^2\pi_1 = \hat{p}\, b_0(p) + O(a_0 b_0).
\]
Since $\bar{\cal A}= \bar{\cal B} + O({\cal B} \bar{\cal B})$ after Mansfield's canonical transformation, we identify $\bar{\cal B}(p)$ with $\hat{p}\, b_0(p)$.

In fact, we could have noted that the propagators $\langle a_0 b_0\rangle$ and $\langle {\cal B} \bar{\cal B} \rangle$ have the same $1/p^2$ form, modulo constants and delta functions. Therefore, since ${\cal B}\leftrightarrow a_0/\hat{p}$, one would expect $\bar{\cal B}\leftrightarrow \hat{p}\, b_0$. From comparing the vertices, one could also see that $\bar{\cal B}$ must be proportional to $b_0$.

\begin{figure}[htb]
	\begin{center}
	\epsfig{file=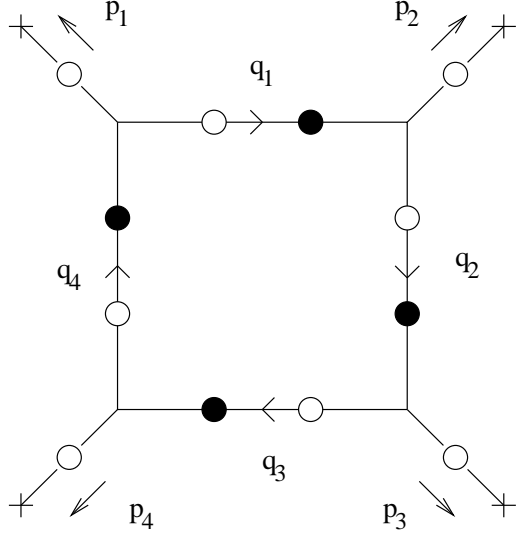}
	\end{center}
	\caption{\emph{The box diagram for the one-loop scattering amplitude with four $-$ helicity gluons.}}
	\label{fig:box}
\end{figure}

The fields in the canonical transformation approach thus correspond, in a fairly simple manner, to the basic fields in the twistor formalism $a_0$, $b_0$. (As to the other twistor fields, $\eta^A a_A=0$ in CSW gauge and $\hat{\eta}^A a_A$ is equivalent to a series in $a_0$ and $b_0$, while the field $b_A$ never appears in amplitude calculations.) The propagators in the two theories can also be checked to agree with each other and the vertices are of the same MHV form. It is, therefore, no surprise that the two theories both reproduce the scattering amplitudes correctly at tree-level. One difference is between the external states that are used. The MHV Lagrangian formalism has the advantage that the field transformation is canonical by construction and allows one to use the new fields in amplitude calculations straightaway, by appealing to the Equivalence Theorem. In the twistor approach, the external fields are expressed as integrals of $a_0$ for $+$ helicity and $\lambda^A a_A$ for $-$ helicity. Since $a_A$ is equivalent to a series in $a_0$ and $b_0$, the $-$ helicity state in the twistor formalism already incorporates some of the `MHV completion vertices' as christened by Mansfield, Morris, \emph{et.\ al.}\ in \cite{Ettle:2007qc}. As a consequence, we did not have to use any sub-leading terms in the expansions for $A^{\pm}$, \eqref{eq:A+p} and \eqref{eq:A-p}, in order to recover the `missing' three-point amplitude. This is in contrast to the solution in \cite{Ettle:2007qc}, which makes use of higher order terms in the expansions for $\cal A$ and $\bar{\cal A}$. The real test, however, lies in the calculations of loop amplitudes. While the one-loop all-plus four-point amplitude has been derived using the canonical MHV Lagrangian, it is not yet clear if the same external states as at tree-level are sufficient to compute loop amplitudes. However, encouraging evidence is offered by preliminary considerations of the all-minus loop (same as all-plus in the other convention) four-point amplitude. The box diagram in Figure \ref{fig:box} does not naively vanish, even though the external propagator \epsfig{file=propagator1.eps, scale=0.7} seems to be killed by LSZ reduction.

The reason that this box diagram survives is the same as for the missing three-point amplitude. Performing the standard quadruple cut amounts to putting all the internal momenta $q_i$ on shell, i.e.\ letting $q_i^2=0$. Each corner of the box is then a replica of Figure \ref{fig:missing} and, letting all the external momenta to become null under LSZ reduction, it contributes the following factor to the final integral over the internal momentum $q_{iAA'}=\mu_A \tilde{\mu}_\ap$,
\[
\left(\frac{\eta\cdot \mu_i}{\eta\cdot\mu_{i-1}} \right)^2 \frac{(\mu_{i-1} \cdot \lambda_i)^3}{(\mu_i\cdot \lambda_i) (\mu_{i-1}\cdot \mu_i)}.
\]
All $\pi$ integrals have been carried out and the remaining momentum integral requires a regularisation procedure. If one adopts dimensional regularisation in the sense of taking $\PT'$ to be $\R^{4-2\epsilon} \times \P^1$, the momentum integral is similar to that obtained in \cite{Ettle:2007qc} (formula 5.1). This suggests that no `completion vertices' are required in the twistor formalism as long as the external states are represented correctly. However, work is still to be done to find a regularisation scheme which is more `natural' for twistor space and which does not rely on space-time methods.

\begin{appendix}

\chapter[Appendix]{}

\section{Some projective spinor formulae}\label{sec:formulae}

Projective spinors are used throughout this thesis. We collect here some of the results needed and indicate their derivation.

Integrals of the form $\int_{\P^1} f(\pi_\ap)$ are often encountered, where $f(\pi_\ap)$ has to be a $(1,1)$-form of total weight 0. They can be evaluated by choosing a basis $\{\lambda_\ap,\hat{\lambda}_\ap\}$, with $\lambda\cdot\hat{\lambda} = 1$, and using scaling to express $\pi_\ap$ as 
\begin{equation}\label{eq:pizi}
\pi_\ap=\lambda_\ap-z\,\hat{\lambda}_\ap.
\end{equation}
The integral will then become one over $\C$. For example, $\pi^\ap d\pi_\ap=dz$ and $\pi\cdot\hat{\pi}=1+|z|^2$. For convenience, we will use the abbreviation
\begin{equation}\label{eq:d2pi}
	d^2 \pi = \frac{1}{2\pi i} \frac{\hpi^\ap d\hpi_\ap\wedge \pi^\bp d\pi_\bp}{(\pi\cdot\hpi)^2}
\end{equation}
for the normalized volume form on $\P^1$. The following formula may be immediately verified.
\begin{equation}\label{eq:piint1}
	\int_{\P^1} d^2 \pi \frac{\pi_\ap \hpi^\bp}{\pi\cdot\hpi} = \frac{1}{2} \delta_\ap^\bp.
\end{equation}

More generally,
\begin{equation}
	\int_{\P^1} d^2 \pi \frac{\pi_\ap \cdots \pi_\bp \hpi^\cp \cdots \hpi^{D'}}{(\pi\cdot\hpi)^n}  = \frac{1}{n+1} \delta^\cp_{(\ap} \cdots \delta^{D'}_{\bp)}.
	\label{eq:multipiint}
\end{equation}
The quickest derivation of this result uses representation theory. The primed spinors $S'$ form  the fundamental representation of $SU(2)$, taken to be matrices $M^\ap_\bp$. The left-hand side of Equation \eqref{eq:multipiint} can be regarded as an equivariant map from $\bigodot^n S'$ to itself, which has to be a multiple of the identity by Schur's Lemma, since $\bigodot^n S'$ is an irreducible representation of $SU(2)$. The constant of proportionality $1/(n+1)$ can be worked out by evaluating both sides after contracting all the indices.

\vspace{0.6 cm}

The same method leads to a representation of Green's function for the $\db$-operator on $\P^1$. It is known that, over $\C$,
\begin{equation}\label{eq:dbarz}
\frac{\partial}{\partial\bar{z}} \frac{1}{z-z'}= 2\pi i \> \delta(z-z'),
\end{equation}
where $\delta(z)$ denotes $\delta(x)\delta(y)$. This follows easily from a standard argument showing that
\begin{equation}\label{eq:intdbz}
\int_\C f(z,\bar{z}) \frac{\partial}{\partial\bar{z}} \frac{1}{z-z'}\ d\bar{z}\wedge dz = 2\pi i f(z',\bar{z}'),
\end{equation}
for $f$ a smooth function, expressible as a power series in $z$ and $\bar{z}$ around $z=z'$.

Over the projective line, we note that 
\[
\db =d\hpi_\ap \frac{\partial}{\partial \hpi_\ap} = d\hpi_\ap \left( \frac{\hpi^\ap\pi_\bp-\pi^\ap\hpi_\bp}{\pi\cdot\hpi}\right) \frac{\partial}{\partial \hpi_\bp}.
\]
The Euler operator $\hpi_\ap (\partial/\partial \hpi_\ap)$ vanishes when acting on functions of zero anti-holomorphic weight. (We will always deal with functions of this type.) Therefore
\begin{equation}
\db	= \frac{\hpi^\ap d\hpi_\ap}{\pi\cdot\hpi} \> \pi_\bp \frac{\partial}{\partial \hpi_\bp}\>=e^0\>\db_0,\label{eq:dbarpi}
\end{equation}
where $e^0=\hpi^\ap d\hpi_\ap /(\pi\cdot\hpi)^2$ and $\db_0=(\pi\cdot\hpi)\pi_\ap (\partial/\partial \hpi_\ap)$. Given a function $g(\pi,\hpi)$ of holomorphic weight $-1$, the generalisation of Equation \eqref{eq:intdbz} is 
\begin{equation}\label{eq:dbargreenfn}
\int_{\P^1} g(\pi,\hpi)\ \db_0 \left(\frac{1}{\pi\cdot\pi'}\right) \ d^2\pi = g(\pi',\hat{\pi}').
\end{equation}
This is easily seen by representing $\pi_\ap$ as $\pi_\ap'-z\hat{\pi}_\ap'$. Hence, we have the delta function of weight 1, in analogy with Equation \eqref{eq:dbarz},
\begin{equation}\label{eq:defdeltafn}
	\delta(\pi\cdot\pi')= \db_0 \left(\frac{1}{\pi\cdot\pi'}\right).
\end{equation}
More generally, given a function $g_k$ of weight $k$, we can multiply it by $\left(\frac{\pi'\cdot o}{\pi\cdot o}\right)^{k+1}$, where $o_\ap$ is any spinor not proportional to $\pi'$, and obtain a function of weight $-1$. Inserting this into the integral \eqref{eq:dbargreenfn} evidently gives $g_k(\pi',\hat{\pi}')$, independent of the choice of $o_\ap$.

In this thesis, an important role is played by frames which are holomorphic up the $\P^1$-fibres over $\E$, i.e.\ we demand a matrix function on $\PT'$ satisfying the following equation
\begin{equation}\label{eq:hdiffeqn}
(\db_0+a_0)H=0.
\end{equation}
The integral form of this equation is 
\begin{equation}\label{eq:hinteqn}
H(x,\pi)=I-\int_{L_x} a_0 H(x,\pi')\> \frac{\pi\cdot o}{\pi'\cdot o}\> \frac{1}{\pi\cdot\pi'} \> d^2\pi',
\end{equation}
as can be seen using Equation \eqref{eq:dbargreenfn}. Here the spinor $o$ has the interpretation of a selected point in $L_x$, at which $H$ is fixed to be the identity matrix. Note that Equation \eqref{eq:hdiffeqn} implies that, for any $\End(E)$-valued form $\chi$,
\begin{equation}\label{eq:hconjdba0}
	H^{-1} \db_{a_0} \chi H= \db_0 (H^{-1} \chi H).
\end{equation}

To compute the variation of $H$ with respect to $a_0$, we vary Equation \eqref{eq:hdiffeqn}, i.e.\ replace $a_0$ by $a_0+\delta a_0$ and $H$ by $H+\delta H$. This leads to
\[
\db_0(H^{-1}\delta H)=-H^{-1}\delta a_0 H
\]
and its integral form
\begin{equation}\label{eq:varh}
	H^{-1}\delta H=-\int_{L_x} H^{-1}\delta a_0 H \> \frac{\pi\cdot o}{\pi'\cdot o}\> \frac{1}{\pi\cdot\pi'} \> d^2\pi'.
\end{equation}

Equation \eqref{eq:defdeltafn} may be interpreted as saying that $1/(\pi_1\cdot\pi_2)$ is Green's function for the operator $\db_0$ (with respect to $\pi_1$) on $\P^1$. Equation \eqref{eq:hdiffeqn} implies that 
\begin{equation}\label{eq:defk12}
	K_{12}=\frac{H_1 H_2^{-1}}{\pi_1\cdot\pi_2} \equiv H(x,\pi_1) \frac{1}{\pi_1\cdot\pi_2} H^{-1}(x,\pi_2)
\end{equation}
may be regarded as Green's function for $\db_0+a_0$, that is, 
\begin{equation}\label{eq:dagreenfn}
\db_{a_0} K_{12}=\delta(\pi_1\cdot \pi_2),
\end{equation}
where the operator acts on $\pi_1$.

After some calculations, using Equation \eqref{eq:varh} and the identity \eqref{eq:antisym}, we find that the variation of $K_{12}$ with respect to $a_0$ is
\begin{equation}\label{eq:varkint}
	\delta K_{12}=-\int_{L_x} K_{13} \delta a_3 K_{32}\ d^2\pi_3,
\end{equation}
where $\delta a_3$ denote the value of $\delta a_0$ at $\pi_3$. Hence, displaying the matrix indices explicitly, we have
\begin{equation}\label{eq:varkdiff}
	\frac{\delta K_{12}{}^A_{\ B}}{\delta a_3 {}^C_{\ D}} = -K_{13}{}^A_{\ C} K_{32}{}^D_{\ B}\ .
\end{equation}

\section{ASD multi-instantons in CSW gauge}\label{sec:multiinstanton}

It is described in Section \ref{subsec:asd} that the equation
\begin{equation}
	\pi^\ap (\partial_{A\ap} G + A_{A\ap} G)=0
	\label{eq:Geqn}
\end{equation}
can be solved if the potential $A$ has ASD curvature; and the solution $G$ provides the gauge transformation from space-time gauge to CSW gauge. In the CSW gauge, only the field $a_0(\omega, \pi, \hpi)$ is significant and we would like to find its form for an ASD instanton of charge $n$. The t' Hooft ansatz \eqref{eq:thooft} gives a simple formula for the potentials of certain $n$-instantons,
\[
A_{AA'C'D'} = \epsilon_{A(C'} \partial_{D')A} \log \phi,\quad \text{ where }\ \phi= 1 + \sum_{i=1}^n \frac{\lambda_i^2}{|x-c_i|^2}.
\]
However, owing to the singularities at $c_i$, no smooth solutions could be found for Equation \eqref{eq:Geqn}. As a matter of fact, this difficulty prevented us from finding a smooth solution for the 1-instanton for some time, until it was realised that the singularities had to be gauge away. Taking the 2-instanton as an example, we can achieve this as follows. Let $x_i=x-c_i$ and $P=c_1-c_2$. Transforming $A$ by $x_{1A}^\ap P^A_\bp x_{2B}^\bp/ |x_1||P||x_2|$ leads to, after a couple of pages of calculations,
\begin{align*}
A_{AA'CD} = & \frac{2}{\lambda_1^2 |x_2|^2+ \lambda_2^2 |x_1|^2 + |x_1|^2|x_2|^2}\\ 
&\times \biggl[ \frac{2}{P^2}(|x_2|^2 x_{1CA'} P_{AE'} x_{1D}^{E'} - |x_1|^2 x_{2CA'} P_{AE'} x^{E'}_{2D})\\ &+ (\lambda_1^2 + |x_1|^2) x_{2A'C} \epsilon_{DA} + (\lambda_2^2+ |x_2|^2) x_{1A'C} \epsilon_{DA} \biggr]_{\text{symmetrize in } C,\>D}
\end{align*}
The case of the instanton charge $n=1$ is fairly simple and the solution obtained is presented in Section \ref{subsec:asd}. In contrast, the $n=2$ case is evidently much more complex and, despite considerable effort, Equation \eqref{eq:Geqn} could not be solved in this case.

Nevertheless, if one were only interested in the form of $a_0$, one could adopt an alternative approach. It is possible to retain the singular form of $A$, find a singular solution and then apply a gauge transformation to make $G$ smooth. If we let
\[
N_{C'D'} = \sqrt{\phi} G_{C'D'},
\]
then Equation \eqref{eq:Geqn} becomes
\[
\phi \db_A N_{C'D'} + \pi_\ap N^\ap_{D'} \partial_{C'A} \phi =0.
\]
Contracting this equation with $\pi^\cp$ shows that $\pi_\ap N^\ap_{D'} = \phi\, f_{D'}$, for some function $f_{D'}(\omega,\pi,\hpi)$ independent of $\homega$. It is not too difficult to guess a solution
\[
G^\ap_{B'} = \frac{1}{\sqrt{\phi}} N^\ap_\bp= \frac{1}{\sqrt{\phi}} \left( \delta^\ap_\bp + \sum_{i=1}^n \frac{\lambda_i^2 \eta_A x_i^{AA'} \pi_\bp}{|x_i|^2 (\eta\cdot \omega_i)} \right).
\]
As $a_0=G^{-1} \db_0 G$ and the solution above is ostensibly independent of $\hpi$, we seem to have transformed away $a$ entirely. This is only an illusion, because of the $(\eta\cdot \omega_i)$ singularities in the formula. Given a constant $\eta$, for any $x$, there is a value of $\pi$ such that $\eta\cdot \omega= \eta_A \pi_\ap x^{AA'}=0$. In the light of the last section, $a_0=G^{-1} \db_0 G$ might be presented as a sum of delta functions. We could also use the observation that Equation \eqref{eq:Geqn} is unchanged if $G$ is replaced by $G\gamma$, for any matrix $\gamma(\omega,\pi,\hpi)$. If an suitable $\gamma$ that cancels the singularities in $G$ could be found, then $a_0=\gamma^{-1} G^{-1} \db_0 (G\gamma)= \gamma^{-1} \db_0 \gamma$, since $G$ is formally annihilated by $\db_0$.

We will attempt to find an appropriate $\gamma$ in the case $n=2$. The solution in the case $n=1$ suggests that the $(\eta\cdot \omega_1)$ singularity is removed by choosing 
\[
\gamma_1= \frac{\hpi^\ap \omega_{1A}}{\lambda_1 \pi\cdot \hpi} + \frac{\lambda_1 \pi^\ap \eta_A}{2\eta\cdot \omega_1}.
\]
Indeed, using Equation \eqref{eq:antisym}, which implies that $\eta_A x_1^{AA'} \omega_{1B}= (\eta\cdot\omega_1) x_{1B}^\ap - \frac{1}{2} |x_1|^2 \pi^\ap \eta_B$,
\[
G\gamma_1 = \frac{\hpi^\ap \omega_{1B}}{\lambda_1 \pi\cdot\hpi} + \frac{\lambda_1 x^\ap_{1B}}{|x_1|^2} + \frac{\lambda_2^2 \eta_A x_2^{AA'} \omega_{1B}}{\lambda_1 |x_2|^2 (\eta\cdot\omega_2)}.
\]
To cancel the other singularity, $(\eta\cdot\omega_2)$, we considered combinations of $\omega$, $\pi$ and $\hpi$ that are of weight 0. The best candidate that we found was 
\[
\gamma_2 = \frac{\lambda_2 \omega_1^A \eta_B}{\lambda_1 \eta\cdot\omega_2} + \frac{\lambda_1 \hat{o}^A \omega_{2B}}{\lambda_2 \omega_1\cdot \hat{o}},
\]
where $o^A=v^{AA'}\pi_\ap$, for some vector $v\in\R^4\backslash \{0\}$. The matrix $G\gamma_1 \gamma_2$ is free from $(\eta\cdot\omega_i)$ singularities, but is still not smooth where $\omega_1\cdot \hat{o}=0$. This happens when $x_1\cdot v=0$. (To see this, note that
\begin{align*}
\omega_1 \cdot \hat{o} &= \pi_\ap \hpi_\bp x^\ap_{1A} v^{AB'}\\
 &= \frac{1}{2} (\pi\cdot\hpi) (x_1\cdot v) + \pi_\ap \hpi_\bp x_{1A}^{(A'} v^{B')A}.
\end{align*}
Since $x$ and $v$ are both real under the conjugation $\hat{}$, the first term in the last line is real, while the second is imaginary. Hence, $\omega_1\cdot \hat{o}=0$ only if $x_1\cdot v=0$.)

The inverses of $\gamma_1$ and $\gamma_2$ are
\[
\gamma_1^{-1}= -2 \left( \frac{\omega_1^A \hpi_\ap}{\lambda_1 \pi\cdot\hpi}+ \frac{\lambda_1 \eta^A \pi_\ap}{2 \eta\cdot \omega_1} \right) \quad\text{and}\quad \gamma_2^{-1}= -\left(\frac{\lambda_2 \eta^A \omega_{1B}}{\lambda_1 \eta\cdot \omega_2} + \frac{\lambda_1 \omega_2^A \hat{o}_B}{\lambda_2 \omega\cdot \hat{o}} \right).
\]
The resulting $a_0$ can then be calculated to be
\[
a_0 = \gamma_2^{-1} \gamma_1^{-1} \db_0 (\gamma_1 \gamma_2) = \frac{2}{\lambda_2^2} \left(1 + \left(\frac{|v| \lambda_1 (\pi \cdot \hpi)}{2\>\omega_1 \cdot \hat{o}} \right)^2 \right) \omega_2^A \omega_{2B}.
\]
When $\lambda_1=0$, i.e.\ when the instanton really has charge $n=1$, the formula above evidently reduces to the solution we found for the 1-instanton.

From the discussion above, the function $a_0$ as given above is only smooth in an open set in twistor space, namely where $x_1\cdot v\neq 0$. Unsuccessful attempts were made to find a gauge transformation $h(\omega,\pi,\hpi)$ such that $h^{-1}a_0h+ h^{-1}\db_0 h$ is smooth everywhere. As the formula for $a_0$ for a general multi-instanton would not be used further in this thesis, we have not pursued this problem further. It remains to observe that for $n>2$, it is necessary to apply more transformations like $\gamma_{1,2}$ and the eventual expression for $a_0$ can be seen to be $\omega_n^A \omega_{nB}$ times a polynomial in terms like $(\pi\cdot\hpi)/(\omega_i\cdot \hat{o})$, of degree $2n$.

\end{appendix}

\addcontentsline{toc}{chapter}{Bibliography}
\bibliography{references}
\bibliographystyle{plain}

\end{document}